 \documentclass[final,1p,times]{elsarticle}

\usepackage{amssymb}

\usepackage{latexsym,amsfonts,amsmath,theorem,amssymb}
\usepackage{stmaryrd,array,tabularx,bbm}
\usepackage{pstricks,graphicx}
\usepackage{a4wide}
\usepackage{color}
\usepackage{caption}
\usepackage{theorem}
\usepackage{framed}
\usepackage{blindtext}
\usepackage{float}
\usepackage[caption=false]{subfig}
\usepackage{algorithm}
\usepackage{algorithmicx}
\usepackage[noend]{algpseudocode}
\usepackage{tikz}
\usetikzlibrary{arrows}
\usetikzlibrary{decorations.pathreplacing,angles,quotes}

\algrenewcommand\algorithmicrequire{\textbf{Input:}}
\algrenewcommand\algorithmicensure{\textbf{Output:}}
 
\newtheorem{theorem}{Theorem}[section]

\newtheorem{proposition}[theorem]{Proposition}
\newtheorem{assumptions}[theorem]{Assumptions}

{\theoremstyle{plain} \theorembodyfont{\rmfamily}
\newtheorem{example}[theorem]{Example}
}

\numberwithin{equation}{section}
\numberwithin{figure}{section}
\numberwithin{table}{section}

\newcommand{\cv}{\mathrm{cv}}

\newcommand{\TMP}{T}
\newcommand{\BDG}{B}

\journal{J. Comp. Phys.}

\begin{document}

\begin{frontmatter}



\title{Multilevel Sequential${}^2$ Monte Carlo for Bayesian Inverse Problems}


\author[Ma]{Jonas Latz}
\ead{jonas.latz@ma.tum.de}
\author[Era]{Iason Papaioannou}
\ead{iason.papaioannou@tum.de}
\author[Ma]{Elisabeth Ullmann}
\ead{elisabeth.ullmann@ma.tum.de}
\address[Ma]{Chair of Numerical Analysis, TU M\"{u}nchen, Boltzmannstr. 3, 85748 Garching b.M., Germany}
\address[Era]{Engineering Risk Analysis Group, TU M\"{u}nchen, Theresienstr. 90, 80333 M\"{u}nchen, Germany}

\begin{abstract}
The identification of parameters in mathematical models using noisy observations is a common task in uncertainty quantification. We employ the framework of Bayesian inversion: we combine monitoring and observational data with prior information to estimate the posterior distribution of a parameter. Specifically, we are interested in the distribution of a diffusion coefficient of an elliptic PDE. In this setting, the sample space is high-dimensional, and each sample of the PDE solution is expensive. To address these issues we propose and analyse a novel Sequential Monte Carlo (SMC) sampler for the approximation of the posterior distribution. Classical, single-level SMC constructs a sequence of measures, starting with the prior distribution, and finishing with the posterior distribution. The intermediate measures arise from a tempering of the likelihood, or, equivalently, a rescaling of the noise. The resolution of the PDE discretisation is fixed. In contrast, our estimator employs a hierarchy 
of PDE discretisations to decrease the computational cost. We construct a sequence of intermediate measures by decreasing the temperature or by increasing the discretisation level at the same time. This idea builds on and generalises the multi-resolution sampler proposed in [P.S. Koutsourelakis, J. Comput. Phys., 228 (2009), pp. 6184-6211] where a bridging scheme is used to transfer samples from coarse to fine discretisation levels. Importantly, our choice between tempering and bridging is fully adaptive. We present numerical experiments in 2D space, comparing our estimator to single-level SMC and the multi-resolution sampler.

\end{abstract}

\begin{keyword}
uncertainty quantification \sep partial differential equation \sep finite element method \sep particle filter \sep sequential importance sampling \sep tempering


\MSC 35R60 \sep 62F15 \sep 65C05 \sep 65C35 \sep 65N21 \sep 65N30 

\end{keyword}

\end{frontmatter}


\section{Introduction}\label{sec:intro}

In science and engineering we use mathematical models to simulate and understand physical processes.
These models require input parameters.
Once the parameters are specified we can solve the so-called forward problem to obtain output quantities of interest.
In this work we focus on models that involve partial differential equations (PDEs).
To date approximate forward solvers are available for many PDE-based models, and output quantities of interest can be approximated efficiently.
In contrast, the identification of input parameters (the inverse problem) is more challenging.
Often the physical process is only given implicitly by observations (data, measurements).
These measurements are typically noisy and/or sparse, and do not contain sufficient information on the underlying parameter or are disturbed in such a way that the true parameter cannot be recovered at all.
The inverse problem is ill-posed.

A classical example is the simulation of steady-state groundwater flow to assess the safety of proposed long-term radioactive waste repositories. 
The quantity of interest is the travel time of radioactive particles to the boundary of a safety zone. 
The simulation requires the hydraulic conductivity of the ground; it can be observed implicitly by pumping tests, and by pressure measurements.
The objective of the groundwater flow inverse problem is the identification of the conductivity.
In this example, the mathematical model involves an elliptic PDE.
The groundwater flow inverse problem is well known, see e.g. \cite{Dashti2011UncertaintyProblem,Dashti2017TheProblems,Marzouk2009DimensionalityProblems,Richter1981AnEquation,Schillings2017AnalysisProblems}.

In contrast to deterministic regularisation techniques, the Bayesian approach to inverse problems uses the probabilistic framework of Bayesian inference.
Bayesian inference is built on \textit{Bayes' Formula} {   in the formulation given by Laplace \cite[II.1]{Laplace1812TheorieProbabilites}}.
We remark that other formulations are possible, see e.g. the work by Matthies et al. \cite{Matthies2016InverseSetting}.
We make use of the mathematical framework for Bayesian Inverse Problems (BIPs) given by Stuart \cite{Stuart2010InversePerspective}.  
Under weak assumptions -- which we will give below -- one can show that the BIP is well-posed.
The solution of the BIP is the conditional probability measure of the unknown parameter given the observations.

The Bayesian framework is very general and can handle different types of forward models.
However, in this work we consider PDE-based forward models, and in particular an elliptic PDE.
The exact solution of the associated BIP is often inaccessible for two reasons: $(i)$ there is no closed form expression for the posterior measure, and $(ii)$ the underlying PDE cannot be solved analytically. 
We focus on $(i)$, and study efficient approximations to the \textit{full} posterior measure.
Alternatively, one could also only approximate the expectation of output quantities of interest with respect to the posterior measure, or estimate the model evidence, the normalization constant of the posterior measure.

 {Typically, BIPs are approached with sampling based methods, such as Markov Chain Monte Carlo (MCMC) or Importance Sampling. 
 	Classical MCMC samplers are the algorithms suggested by Metropolis et al. \cite{Metropolis1953EquationMachines} and the generalisation by Hastings \cite{Hastings1970MonteApplications}. 
Advanced MCMC methods for BIP settings are Hamiltonian Monte Carlo \cite{Bui-Thanh2014SolvingCarlo} and preconditioned Crank-Nicholson MCMC \cite{Beskos2008AnBridges,Cotter2013MCMCFaster}. 
A disadvantage of MCMC samplers is the fact that it is often difficult to assess their convergence after an initial burn-in phase.
Importance Sampling \cite{Agapiou2015ImportanceCost} on the other hand does not require burn-in. 
However, Importance Sampling is inefficient if the sampling density differs significantly from the target density. }
For these reasons we employ Sequential Monte Carlo (SMC) \cite{Chopin2002AModels,DelMoral2006SequentialSamplers,Neal2001AnnealedSampling} to approximate the posterior measure.
SMC {was initially developed} to approximate sequences of measures which arise from time-dependent estimation problems in data assimilation.
In our setting, since the elliptic PDE models a steady-state process the SMC sequences are constructed artificially {such that, starting from the prior measure, they gradually approach the posterior measure}.
Artificial sequences of measures arise also in simulated annealing \cite{DelMoral2006SequentialSamplers}, the estimation of rare events \cite{Papaioannou2016SequentialAnalysis}, model selection \cite{Zhou2016TowardApproach}, and bridging \cite{Gelman1998SimulatingSampling}.

In some situations it is convenient to determine the artificial sequences ``on the fly'' during the execution of the algorithm. 
The associated method is termed \textit{adaptive SMC}; see \cite{Doucet2011ALater,Jasra2011InferenceCarlo} for a discussion, and \cite{Beskos2016OnMethods} for a careful analysis. 
A well-known drawback is the fact that adaptive SMC returns a biased model evidence estimate, however, the model evidence is not the major focus of our work.
The estimation of the model evidence with non-adaptive SMC is discussed in \cite{Gelman1998SimulatingSampling,Neal2005EstimatingSampling}.

{The major advantage of SMC is its dimension-independent convergence which is often observed in practise and which can be proved e.g. for uniformly bounded update densities \cite{Beskos2015SequentialProblems}.	
Thus SMC can be used in high- and infinite dimensional settings.
See \cite{Rebeschini2015CanDimensionality} for a discussion of this point.}
Similar results are also known for the Ensemble Kalman Filter (EnKF) applied to linear inverse problems with a finite number of particles \cite{Schillings2017AnalysisProblems}.
The EnKF is a linearised version of SMC and has been applied to linear and nonlinear inverse problems (see \cite{Iglesias2013EnsembleProblems}).


SMC has already been used to solve BIPs where the forward model is an elliptic \cite{Beskos2015SequentialProblems} or Navier-Stokes equation \cite{Kantas2014SequentialEquations}.
The computational challenge is that PDE-based forward solves are in general very expensive.
Thus every sample, required by standard solvers such as MCMC or SMC, is expensive.
The total computational budget might allow only a few samples and thus the sample error can be considerably large.
We handle this problem by constructing a \textit{multilevel} SMC sampler. 
To do this we assume that the PDE can be discretised with multiple levels of accuracy.
In our work these levels are associated with different mesh sizes in a spatial domain.
However, it is also possible to consider e.g. different time step sizes, or target accuracies of Newton's method.

Multilevel samplers enjoy considerable attention at the moment, and are available for various tasks in uncertainty quantification.
Multilevel Monte Carlo is widely used in forward uncertainty quantification; see \cite{Giles2015MultilevelMethods} for an overview. 
In the pioneering work by Giles \cite{Giles2008MultilevelSimulation} the multilevel idea is combined with standard Monte Carlo in a forward setting.
However, it can be used with other samplers such as MCMC, SMC, and the EnKF, {  for the estimation of rare} events, for filtering problems in data assimilation, and to solve Bayesian Inverse Problems.
For example, multilevel Ensemble Kalman Filters have been proposed in \cite{Chernov2016MultilevelModels, Hoel2016MultilevelFiltering}.
The authors in \cite{Hoel2016MultilevelFiltering} consider continuous-time data assimilation with multiple time-step discretisations.
In contrast, the work in \cite{Chernov2016MultilevelModels} considers data assimilation for spatially extended models e.g. time dependent stochastic partial differential equation.
{The multilevel estimation of rare events with an SMC type method has been proposed in \cite{Ullmann2015MultilevelEvents}.}

For Bayesian Inverse Problems a multilevel MCMC method has been introduced in \cite{Dodwell2015AFlow}.
Multilevel Sequential Monte Carlo is introduced in \cite{Beskos2017MultilevelSamplers} and further discussed in \cite{Beskos2017MultilevelProposals,DelMoral2017MultilevelConstants,DelMoral2017MultilevelConditions}.
Both the multilevel MCMC and multilevel SMC use coarse PDE discretisations for variance reduction with the help of a telescoping sum expansion.
Moreover, these multilevel samplers are built to integrate output quantities of interest with respect to the posterior.
In contrast, in our work we do not rely on a telescoping sum, and we construct approximations to the full posterior measure.

We build on and generalise the work by Koutsourelakis in \cite{Koutsourelakis2009AParameters}. 
As in \cite{Koutsourelakis2009AParameters} we combine SMC with tempering on a fixed PDE discretisation level, and bridging between two consecutive discretisation levels.
The major novel contribution of our work is a fully adaptive algorithm to decide when to increase the discretisation accuracy (bridging) and when to proceed with SMC (tempering). 
In numerical experiments we show that our sampler -- termed multilevel Sequential$^2$ Monte Carlo -- gives an approximation accuracy similar to single-level SMC while decreasing the computational cost substantially.
Our method works also consistently in the small noise limit.
We note that a similar SMC based multilevel method has been proposed in \cite{Calvetti2017IterativeInversion}; in this work the model error is given as part of the measurement noise and is updated iteratively.

The remainder of this paper is organised as follows.
In \S\ref{sec:Problem} we formulate the Bayesian inverse problem and discuss its discretisation.
Moreover, we review the basic idea of sequential Monte Carlo.
In \S\ref{sec:measures} we give an overview of classical constructions for the intermediate SMC measures, in particular, tempering, bridging, and multilevel bridging.
In \S\ref{Sec:InverseTemp} we discuss adaptive SMC samplers where the inverse temperature is selected to bound the effective sample size or, equivalently, the coefficient of variation of the sample weights.
The major contribution of this work is presented in \S\ref{sec:ML} where we introduce the Multilevel Sequential$^2$ Monte Carlo sampler.
We discuss its computational cost, and suggest an adaptive update scheme for the combination of bridging and tempering.
This update scheme is consistent with the adaptive bridging and tempering discussed in \S\ref{Sec:InverseTemp}.
Finally, in \S\ref{sec:numerics} we present numerical experiments for a test problem in 2D space.
In \S\ref{sec:concl} we give a summary and an outlook of our work.

\section{Background}\label{sec:Problem}

\subsection{Bayesian Inverse Problem} \label{subsec:Problemsetting}

We consider engineering systems or models subject to an uncertain parameter $\theta \in X$. 
The \textit{parameter space} $X$ is a separable Banach space. 
The \textit{forward response operator} $\mathcal{G}: X \rightarrow Y$ maps the parameter to the (finite-dimensional) data space $Y := \mathbb{R}^{N_{\mathrm{obs}}}$.
We are interested in models where $\mathcal{G} = \mathcal{O} \circ G$ is the composition of an \textit{observation operator} $\mathcal{O}$, and a \textit{solution operator} $G$ of a  partial differential equation.
In addition, we consider a \textit{quantity of interest} $Q\colon X \rightarrow \mathbb{R}$ which depends on the parameter $\theta \in X$.

Observations $y \in Y$ are often noisy. 
We model this by assuming that $y$ is a realisation of $\mathcal{G}(\theta_{\mathrm{true}}) + \eta$ where $\theta_{\mathrm{true}} \in X$ is the true model parameter and $\eta \sim \mathrm{N}(0, \Gamma)$ is mean-zero Gaussian noise with non-singular covariance matrix $\Gamma$. 
Hence, using the data vector $y$, we wish to identify $\theta_{\mathrm{true}}$ via the equation
\begin{equation*}
\mathcal{G}(\theta_{\mathrm{true}}) +\eta = y.
\end{equation*}
This (inverse) problem is in general ill-posed in the sense of Hadamard \cite{Hadamard1902SurPhysique} since often $\mathcal{G}(X) \not\ni y$ or $\dim Y \ll \dim X$. 
For this reason we employ the framework of Bayesian inverse problems.
Instead of identifying the deterministic parameter $\theta_{\mathrm{true}}$ we assume that $\theta \sim \mu_0$ is a square-integrable $X$-valued random variable distributed according to a \textit{prior measure} $\mu_0$.
Moreover, we assume that $\theta$ is independent of the noise $\eta$.
Then, the solution of the \textit{Bayesian Inverse Problem} is the \textit{posterior measure} $\mu^y$ of $\theta$,
\begin{equation*}
\mu^y := \mathbb{P}(\theta \in \cdot | \mathcal{G}(\theta) + \eta = y).
\end{equation*}
 Under Assumptions \ref{Assumptions_Prior} and \ref{Main_Assumptions_Stuart2.6} it can be proved that $\mu^y$ exists and that $\mu_0\text{-almost surely}$ it holds
\begin{equation}
\frac{\mathrm{d}\mu^y}{\mathrm{d}\mu_0}(\theta) = \frac{1}{Z_y} \exp(-\Phi(\theta;y)). \ \ \label{BayesBIP}
\end{equation}
In \eqref{BayesBIP} the term $\exp(-\Phi(\theta;y))$ is called \textit{likelihood}, 
\begin{equation}\label{phi}
\Phi(\theta;y) := \frac{1}{2}\|\Gamma^{-\frac{1}{2}}(y - \mathcal{G}(\theta))\|^2_{Y} 
\end{equation}
is a so-called \textit{potential} (the negative log-likelihood), and 
\begin{equation}\label{evidence}
Z_y := \int_X \exp(-\Phi(\theta;y)) \mathrm{d}\mu_0(\theta)
\end{equation}
 denotes the normalising constant of $\mu^y$, or so-called \textit{model evidence}.
The proof of the existence of the posterior measure is given in \cite{Stuart2010InversePerspective} for Gaussian prior measures $\mu_0$. 
It can also be proved that the posterior measure $\mu^y$ is Lipschitz continuous with respect to (w.r.t.) the data space $Y$. 
In this sense the BIP is well-posed. 
Now we state the assumptions on the prior measure.

\begin{assumptions}[Prior measure]\label{Assumptions_Prior} 
Let $m_0 \in X$, and let $H$ be a Hilbert space with $H\supseteq X$.
Let $C_0: H \rightarrow H$ be a trace-class, positive definite and self-adjoint linear operator on $H$. 
Furthermore, let $m_0$ and $C_0$ be chosen such that $\mathrm{N}(m_0, C_0)(X) = 1$. 
Moreover, the prior measure $\mu_0$ is absolutely continuous with respect to $\mathrm{N}(m_0, C_0)$ and its Radon-Nikodym-derivative is $\mathrm{N}(m_0, C_0)$-a.s. given by
\begin{equation*}
\frac{\mathrm{d}\mu_0}{\mathrm{d}\mathrm{N}(m_0, C_0)} \propto \exp(- \Phi_{0}),
\end{equation*}
where $\Phi_{0}:X \times Y \rightarrow \mathbb{R}$ is a  potential.
\end{assumptions}
Note that if $C_0$ satisfies \cite[Assumptions 2.9]{Stuart2010InversePerspective}, then $\mathrm{N}(m_0, C_0)(X) = 1$ holds.
We also remark that Assumptions~\ref{Assumptions_Prior} allow for {certain} non-Gaussian priors.
Note that $C_0$ can be given in terms of the so-called \textit{precision} $C_0^{-1}$.
Then it can happen that $C_0$ is only densely defined on $H$, however, Assumptions~\ref{Assumptions_Prior} are also satisfied in this case.

In addition, for any \textit{potential} $\Phi^\dagger$  we consider the following assumptions.
\begin{assumptions}[Potential] \label{Main_Assumptions_Stuart2.6}
A potential  $\Phi^\dagger: X \times Y \rightarrow \mathbb{R}$ satisfies the following conditions:
\begin{enumerate}
\item For every $\varepsilon, r > 0$ there is an $M(\varepsilon,r) \in \mathbb{R}$ such that 
\begin{equation*}
\Phi^\dagger(\theta; y) \geq M(\varepsilon,r)- \varepsilon\|\theta\|_X^2. \ \  (\theta \in X, y \in Y, \text{ where }\|y\|_Y < r)
\end{equation*}
\item For every $r > 0$  there is a $K(r) > 0$ such that
\begin{equation*}
\Phi^\dagger(\theta;y) \leq K(r). \ \ (\theta \in X, y \in Y, \text{ where } \max\{ \|\theta\|_X, \|y\|_Y\}<r)
\end{equation*}
\item For every $r > 0$ there is an $L(r) > 0$ such that 
\begin{align*}
|\Phi^\dagger(\theta_1;y)-\Phi^\dagger(\theta_2;y)| < L(r)  \|\theta_1 - \theta_2\|_X. \ \ &( \theta_1, \theta_2 \in X, y \in Y, \text{ where } \\ &\max\{\|\theta_1\|_X,\|\theta_2\|_X,\|y\|_Y\} < r)
\end{align*}
\item For every $\varepsilon, r > 0$ there is a $C(\varepsilon, r) \in \mathbb{R}$ such that
\begin{align*}
|\Phi^\dagger(\theta;y_1)-\Phi^\dagger(\theta; y_2)|\leq \exp(\varepsilon\|\theta\|_X^2 + C( \varepsilon, r)) \|y_1- y_2\|_Y. \ \  &(\theta \in X, y_1, y_2 \in Y, \text{ where } \\  &\max\{\|y_1\|_Y,\|y_2\|_Y\} < r)
\end{align*}
\end{enumerate}

\end{assumptions}
The potential $\Phi$ in \eqref{phi} is a typical example in our setting with Gaussian noise. If we are not particularly interested in the data dependence, we sometimes drop this dependence and set $\Phi(\cdot) := \Phi(\cdot;y)$ for a specific $y \in Y$.
In this case, Assumptions~\ref{Main_Assumptions_Stuart2.6} are satisfied if $G$ is the solution operator of an elliptic BVP (see \S\ref{sec:numerics}), and the observation operator $\mathcal{O}$ is linear.
In general, one can also consider non-Gaussian noise, e.g. lognormal (multiplicative) noise \cite[\S 3.2.2]{Kaipio2005StatisticalProblems}, or other PDE operators, e.g. Navier-Stokes \cite{Iglesias2013EnsembleProblems,Kantas2014SequentialEquations}.

Note that if two potentials $\Phi_{0}$ and $\Phi$ satisfy Assumptions \ref{Main_Assumptions_Stuart2.6} then the sum $\Phi+\Phi_{0}$ does as well. 
Thus, we can also consider (posterior) measures with a $\mathrm{N}(m_0, C_0)$-density that is proportional to $\exp(-(\Phi+\Phi_{0}))$. 
This situation occurs in our setting since the sequential Monte Carlo estimator approximates a posterior measure which is then used as prior measure in the next step of the estimation.

\subsection{Discretisation}
In most applications it is not possible to evaluate $\mathcal{G}$, $Q$ or $\mu^y$ analytically. 
Furthermore, the parameter space $X$ is often infinite-dimensional.
This motivates the need to study approximations to the solution of the BIP.

We begin by discretising the physical space associated with the PDE solution operator $\mathcal{G}$.
Let $\mathcal{G}_h$ denote an approximation of $\mathcal{G}$.
Here, $h>0$ refers to a characteristic finite element mesh size. 
The associated approximate potential $\Phi_h$ is given by 
\begin{equation*}
\Phi_h := {\tfrac12}\| \Gamma^{-\frac{1}{2}}(y- \mathcal{G}_h)\|^2_{Y}.
\end{equation*}
Analogously, we approximate the quantity of interest $Q$ by $Q_h$.

Let ${N_{\mathrm{sto}}} \in \mathbb{N}$. 
We approximate the parameter space $X$ by the $N_{\mathrm{sto}}$-dimensional space $X_{N_{\mathrm{sto}}}$. 
For example, if the prior measure is Gaussian, then the parameter space approximation can be constructed using a truncated Karhunen-Lo\`{e}ve (KL) expansion. See \cite{Ghosal2017FundamentalsInference} for details.
In this case $N_{\mathrm{sto}}$ is the number of terms retained in the KL expansion.

Finally, we construct \textit{particle-based} approximations $\widehat{\mu}^y$ to the posterior measure $\mu^y$. 
For particles $(\theta^{(j)} : j = 1,\dots,J) \in X^{J}$ we define
\begin{equation*}
\widehat{\mu}^y := \sum_{j = 1}^Jw^{(j)}\delta_{\theta^{(j)}},
\end{equation*} 
where $w^{(j)}$ is the weight associated with $\theta^{(j)}$. 
The sum of the weights $(w^{(j)}: j = 1,\dots,J)$ is equal to one. 
Typically, the particles $(\theta^{(j)} : j = 1,\dots,J)$ are random samples.
We obtain the particles by either deterministic or non-deterministic transformations of i.i.d. samples from the prior measure. 
If the approximation of the posterior measure involves the discretised potential $\Phi_h$, then we write $\widehat{\mu}^y_h$ in place of $\widehat{\mu}^y$. 

\subsection{Importance Sampling}\label{sec:IS}
Let $\nu_0, \nu_1$ denote probability measures on the measurable space $(X, \mathcal{F})$ and $\nu_1 \ll \nu_0$. 
importance sampling approximates expected values with respect to $\nu_1$ given samples from  $\nu_0$. 
Let $Q: X\rightarrow \mathbb{R}$ denote a quantity of interest that is integrable w.r.t. $\nu_1$.  
According to  the Radon-Nikodym Theorem \cite[Cor. 7.34]{Klenke2014ProbabilityCourse} the expected value of $Q$ with respect to $\nu_1$ can be written as
\begin{equation} \label{EqIS1}
\mathbb{E}_{\nu_1}[Q] =   \mathbb{E}_{\nu_0}\left[\frac{\mathrm{d}\nu_1}{\mathrm{d}\nu_0} Q\right].
\end{equation}
The right-hand side of (\ref{EqIS1}) is an integral with respect to $\nu_0$. 
We approximate this integral by standard Monte Carlo with independent samples $(\theta^{(j)} : j = 1,\dots,J)$ with measure $\nu_0$. 
This gives the \textit{importance sampling estimator} for $\mathbb{E}_{\nu_1}[Q]$,
\begin{align}\label{EqIS2}
\widehat{Q}^{| \mathrm{IS}}_{J}(\theta) &:= \sum_{j=1}^J w^{(j)} Q(\theta^{(j)}), \\
w^{(j)} &:=  J^{-1} \cdot\frac{\mathrm{d}\nu_1}{\mathrm{d}\nu_0}(\theta^{(j)}).\nonumber
\end{align}
{  Often}, the Radon-Nikodym derivative $\gamma_1 \propto \frac{\mathrm{d}\nu_1}{\mathrm{d}\nu_0}$ is known only up to a normalising constant. 
In this case, we use the normalized weights
\begin{equation}
w^{(j)} := \frac{\gamma_1(\theta^{(j)})}{\sum_{l=1}^J\gamma_1(\theta^{(\ell)})}.\nonumber
\end{equation}
Finally, the (normalized) weights $w^{(1)},\dots,w^{(J)}$ can be used to approximate $\nu_1$,
\begin{equation} \label{IS_Prob_Dist_Estim}
\widehat{\nu}^{|\mathrm{IS}}_1 := \sum_{j=1}^J w^{(j)}\delta_{\theta^{(j)}} .
\end{equation}
If the variance of the importance sampling estimator is finite, then the Strong Law of Large Numbers implies that the estimator in \eqref{EqIS1} converges (a.s.) to the desired expected value as $J\rightarrow\infty$.
Sufficient conditions for a finite variance of this estimator are discussed in \cite[\S 3.3.2]{Robert2004MonteMethods}. 
{In the setting of BIPs, $\nu_1$ is the posterior. 
	A straighforward way to apply importance sampling is to choose $\nu_0$ as the prior.
	In this case $\gamma_1$ is the unnormalized likelihood.} 
It is easy to see that if $Q$ is bounded, then the variance of the estimator is finite. 
If the posterior is concentrated in a small area of the prior, then nearly all importance sampling weights are close to zero.
In this situation the estimator is extremely inaccurate given a fixed sample budget, or a large number of samples is required to obtain a desired accuracy.
Sequential Monte Carlo overcomes this problem by using a sequence, not only a pair, of appropriate intermediate measures.

\subsection{Sequential Monte Carlo}\label{sec:SMC_Temp}

Consider a finite sequence of probability measures $\nu_0, \nu_1, \ldots, \nu_{N_{\mathrm{seq}}}$ on $(X, \mathcal{F})$, where $\nu_{k} \ll \nu_{\ell}$, $\ell,k=0,\ldots,{N_{\mathrm{seq}}}.$ 
Sequential Monte Carlo approximates each measure in the sequence with weighted particles; these are constructed sequentially with (variants of) importance sampling.
We denote the Radon-Nikodym derivatives of all measures w.r.t. $\nu_0$ by
\begin{equation}
\gamma_k := Z_k \frac{\mathrm{d}\nu_k}{\mathrm{d}\nu_{0}}, \ \ \ k=1,\dots,{N_{\mathrm{seq}}},
\end{equation}
where $\gamma_k$ is $\nu_{k}$-almost surely (a.s.) positive and $Z_k := \int \gamma_k \mathrm{d}\nu_0 \in (0, \infty)$ is the normalising constant associated with $\gamma_k$. 
By the Radon-Nikodym Theorem it follows that
\begin{equation}\label{EQ:SMC_sequence}
\frac{\mathrm{d}\nu_k}{\mathrm{d}\nu_{k-1}} = \frac{\mathrm{d}\nu_k}{\mathrm{d}\nu_{0}}  \cdot \frac{\mathrm{d}\nu_{0}}{\mathrm{d}\nu_{k-1}} \propto \frac{\gamma_k}{\gamma_{k-1}}  \ \ \ k=1,\ldots,{N_{\mathrm{seq}}}.
\end{equation}
We assume that we can generate independent samples distributed according to $\nu_0$.
Then, we apply importance sampling sequentially to update $\nu_{k-1} \mapsto \nu_k$. 
The measures $\nu_1,\dots,\nu_{N_{\mathrm{seq}}}$ are approximated as in (\ref{IS_Prob_Dist_Estim}).  
In practice, this can be inefficient, especially if $\nu_0$ and $\nu_k$ have a different mass concentration for $k \gg 0$. 
In this case, the approximation of $\nu_k$ would still rely on $\nu_0$-distributed samples. 
Therefore, it is a good idea to apply a Markov kernel that is stationary with respect to $\nu_k$ to the $\nu_k$-distributed particles. 
This moves the particles into the high-probability areas of the measure $\nu_k$.
{Before applying the Markov move,} the particles are \textit{resampled}; this eliminates particles with small weights. 

\section{Construction of intermediate measures} \label{sec:measures}

\subsection{Tempering} \label{Subsec:Tempering}
In BIPs the posterior measure is often concentrated in a small area of the high-dimensional parameter space $X$.
\textit{Tempering} (T) is a widely-used method to approximate such measures. 
The fundamental idea -- borrowed from Statistical Physics -- is to adjust the temperature $\mathcal{T}$ in the Boltzmann distribution.\footnote{The Boltzmann distribution is a discrete probability measure on the set of energy states $S$ of some system of particles. 
Its $\#$-density is proportional to  \[S \ni s \mapsto \exp\left(-\frac{E_s}{\mathcal{T} \cdot k_{\mathrm{Boltz}}}\right),\] where $E_s$ is the energy of state $s$ and $k_{\mathrm{Boltz}}$ is the Boltzmann constant. 
A large temperature $\mathcal{T}$ allows the particles to move faster. 
See \cite[Chapter VIII]{Gibbs1902ElementaryMechanics}, \cite[\S1.1]{Liu2004MonteComputing} and \cite{Metropolis1953EquationMachines} for details.} 
In a Monte Carlo setting tempering is the systematic raising of a density to some power $\beta \in (0,1]$.
Looking at the Boltzmann distribution this means that $\mathcal{T} \in [1,\infty)$. 
If a probability measure is unimodal, increasing the temperature increases the variance of the measure.
This makes it easier to approximate the measure by importance sampling.

We apply tempering in combination with an SMC sampler with ${N_{\mathrm{\TMP}}} \in \mathbb{N}$ intermediate steps. 
We start with the prior $\nu_0 := \mu_0$; this is equivalent to an infinite temperature $\mathcal{T} = \infty$ or an inverse temperature $\beta_0 = \mathcal{T}^{-1} = 0$. 
In the subsequent steps we scale down the temperature $\mathcal{T}$ successively until $\beta_{N_{\mathrm{\TMP}}} = \mathcal{T}^{-1} = 1$, and we have arrived at the posterior $\nu_{N_{\mathrm{\TMP}}} = \mu^y$.
Formally, we define a finite, strictly increasing sequence of inverse temperatures $(\beta_k : k=0,\dots,{N_{\mathrm{\TMP}}})$, where $\beta_0 = 0$ and $\beta_{N_{\mathrm{\TMP}}} = 1$. 
The SMC sequence of probability measures $(\nu_k : k=0,\ldots,{N_{\mathrm{\TMP}}})$ is then given by
\begin{equation*}\label{SMC_RN1}
\frac{\mathrm{d}\nu_k}{\mathrm{d}\nu_0} \propto \gamma_k :=  \exp(-\Phi)^{\beta_k} =  \exp\left(-\tfrac{1}{2}\|(\beta^{-1}_k\Gamma)^{-\frac{1}{2}}(y - \mathcal{G})\|^2_{Y}\right), \quad k \neq 0. 
\end{equation*}
The last term on the right-hand side above tells us that 
\begin{equation*}
\nu_k = \mathbb{P}(\theta \in \cdot | \mathcal{G}(\theta) + \beta^{-1}_k\eta = y), \quad k \neq 0.
\end{equation*}
Hence, an upscaling of the temperature $\mathcal{T} = \beta_1^{-1},...,\beta_{N_{\mathrm{\TMP}}}^{-1}$ is equivalent to an upscaling of the noise level in BIPs. 
Moreover $\beta_0 = 0$ corresponds to an infinitely large noise level, where the likelihood does not contain any information.
Hence, $\nu_0 = \mu_0$ is consistent.

The densities $(\gamma_k : k=1,\dots,{N_{\mathrm{\TMP}}})$ are strictly positive. 
Hence, the intermediate densities in the SMC sampler (see \ref{EQ:SMC_sequence}) are given by
\begin{equation*} \label{EQ:TMP_sequence}
\frac{\mathrm{d}\nu_k}{\mathrm{d}\nu_{k-1}} \propto \frac{\gamma_k}{\gamma_{k-1}} = \exp(-(\beta_k-\beta_{k-1})\Phi)^{}.
\end{equation*}
{We refer to this method as either \textit{SMC with Tempering} or simply \textit{single-level SMC}.

\subsection{Standard Bridging} \label{Subsec:Bridging}

\textit{Bridging} (\BDG) is an SMC type method, where the sequence of probability measures represents a smooth transition from one probability measure $\nu$ to another probability  measure $\nu^{\ast}$. 
We assume that both these probability measures are defined on a common measurable space $(X, \mathcal{F})$, that $\nu\ll \nu^\ast$ and $\nu^\ast\ll\nu$. 
We also assume that $\nu$ and $\nu^\ast$ are absolutely continuous with respect to a $\sigma$-finite measure $\overline{\nu}$ on $(X, \mathcal{F})$. 
Then, the Radon-Nikodym Theorem tells us that ${\mathrm{d}\nu}/{\mathrm{d}\overline{\nu}}$ and ${\mathrm{d}\nu^{\ast}}/{\mathrm{d}\overline{\nu}}$ exist and are unique $\overline{\nu}$-almost everywhere.
Moreover, these densities are strictly positive almost everywhere on the support of $\nu$ and $\nu^{\ast}$.

Now, let $\nu$ and $\nu^{\ast}$ be based on functions $f, f^{\ast}: X \rightarrow \mathbb{R}$ which are proportional to the Radon-Nikodym derivatives given above.
That is,
\begin{align*}
f \propto \frac{\mathrm{d}\nu}{\mathrm{d}\overline{\nu}}, \quad \text{and} \quad   f^{\ast} \propto \frac{\mathrm{d}\nu^{\ast}}{\mathrm{d}\overline{\nu}}.
\end{align*}
Let ${N_{\mathrm{\BDG}}} \in \mathbb{N}$ and  $(\zeta_k : k = 0,\ldots,{N_{\mathrm{\BDG}}}) \in [0,1]^{({N_{\mathrm{\BDG}}}+1)}$ be a strictly increasing finite sequence, where $\zeta_0 = 0$ and ${\zeta}_{N_{\mathrm{\BDG}}} = 1$. 
Then, the bridging sequence of measures $(\nu_k : k = 0,\ldots,{N_{\mathrm{\BDG}}})$ is defined as
\begin{equation*}
\frac{\mathrm{d}\nu_k}{\mathrm{d}\overline{\nu}} \propto \gamma_k :=  f^{(1-\zeta_k)}\cdot (f^{\ast})^{\zeta_k},
\end{equation*}
or, equivalently,
\begin{equation*}\label{Bridging_Sequence}
\frac{\mathrm{d}\nu_k}{\mathrm{d}\nu_{k-1}} \propto \frac{\gamma_k}{\gamma_{k-1}} :=  f^{(\zeta_{k-1}-\zeta_k)}\cdot (f^{\ast})^{(\zeta_{k}-\zeta_{k-1})}.
\end{equation*}
Note that $\nu_0 = \nu$ and $\nu_{{N_{\mathrm{\BDG}}}} = \nu^{\ast}$.

Now, we consider specific functions $f$, $f^\ast$ associated with BIPs.
We assume that
\begin{align*}
f := \exp(-\Phi), \ \ \ \ f^{\ast} := \exp(-\Phi^{\ast}),
\end{align*}
where $\Phi$, $\Phi^{\ast}: X \rightarrow \mathbb{R}$ are (bounded) potentials which satisfy Assumption~\ref{Main_Assumptions_Stuart2.6}. 
Moreover, we assume that $\overline{\nu}$ satisfies the same conditions as the prior measure $\mu_0$ in Assumption~\ref{Assumptions_Prior}. 
Then, the bridging sequence is given by
\begin{equation*} \label{Eq:Bridging_Seq}
\frac{\mathrm{d}\nu_{k}}{\mathrm{d}\nu_{k-1}} \propto \frac{\gamma_k}{\gamma_{k-1}} = \exp(-(\zeta_{k}-\zeta_{k-1})(\Phi^{\ast}-\Phi))
\end{equation*}
with well-defined probability measures $\nu_0,...,\nu_{{N_{\mathrm{\BDG}}}}$. 
Indeed, 
\begin{equation*}
\frac{\mathrm{d}\nu_k}{\mathrm{d}\overline{\nu}} \propto \exp(-[(1-\zeta_k) \Phi + \zeta_k\Phi^{\ast}]) =: \exp(-\Phi_k).
\end{equation*}
The intermediate bridging measures $(\nu_k : k = 1,...,{N_{\mathrm{\BDG}}})$ are given in terms of potentials $(\Phi_k : k = 1,...,{N_{\mathrm{\BDG}}})$ which satisfy Assumptions~\ref{Main_Assumptions_Stuart2.6}. 
Thus, the existence of $(\nu_k : k = 1,...,{N_{\mathrm{\BDG}}})$ is equivalent to the existence of the posterior measure in a BIP (see \S \ref{BayesBIP} for details).

\subsection{Multilevel Bridging} \label{SubSubSec:BridgingMLKoutsourelakis}
It is possible to generalize the idea of standard bridging to a setting where the probability measures $\nu$ and $\nu^\ast$ depend on discretisation parameters $h$, $h^{\ast}$.
In BIPs this is the case if the forward response operator $\mathcal{G}$ is discretised using two different mesh sizes $h, h^{\ast} > 0$. 
Here, $h^{\ast}$ refers to a more accurate yet computationally more expensive PDE solve compared to $h$. 

Suppose that the BIP has been solved on a coarse mesh $h$, and that we wish to obtain a more accurate solution with $h^{\ast} < h$.
This means that $\mu_h^y$ shall be refined to $\mu_{h^{\ast}}^y$.  
In \cite{Koutsourelakis2009AParameters} the author proposes to bridge between the two probability  measures, that is, apply standard bridging for $\nu=\mu_h^y$ and $\nu^\ast=\mu_{h^{\ast}}^y$.
In fact, this idea is carried out in a multilevel way by bridging between a hierarchy of probability  measures associated with a sequence of decreasing mesh sizes.
For this reason we refer to the method as \textit{Multilevel Bridging} (MLB).
We briefly summarize the ideas given in \cite{Koutsourelakis2009AParameters}. 

Let ${N_{\mathrm{L}}} \in \mathbb{N}$ and $(h_{\ell} : {\ell} =1,\ldots,{N_{\mathrm{L}}}) \in (0, \infty)^{N_{\mathrm{L}}}$ denote the hierarchy of mesh sizes, where $h_{N_{\mathrm{L}}}$ is the desired final mesh size and $h_1,...,h_{{N_{\mathrm{L}}}-1}$ are the intermediate mesh sizes. The sequence $(h_{\ell} : {\ell} =1,\ldots,{N_{\mathrm{L}}})$ is strictly decreasing.
Starting with the prior, we first use tempering to compute the posterior measure $\mu_{h_1}^y$ associated with the forward response operator $\mathcal{G}_{h_1}$.
This step is based on the following densities:
\begin{align*}
\frac{\mathrm{d}\nu_k^{\mathrm{\TMP}}}{\mathrm{d}\mu_0} \propto \gamma_k^{\mathrm{\TMP}} := \exp(-\Phi_{h_1})^{\beta_k},
\end{align*}
where $\nu_0^{\mathrm{\TMP}} := \mu_0$ is the prior measure and $(\beta_k : k=0,\ldots,N_{\mathrm{T}})$ is the vector of inverse temperatures.
Then, we proceed iteratively by bridging $\mu^y_{h_{\ell-1}} \mapsto \mu^y_{h_{{\ell}}}$ for each ${\ell} =2,\ldots,{N_{\mathrm{L}}}$. 
In every bridging update we use ${N_{\mathrm{B}}^{({\ell})}}$ intermediate steps based on the (bridging) inverse temperatures $(\zeta^{({\ell})}_k : k = 0,\ldots,{N_{\mathrm{B}}^{({\ell})}})$. 
In particular,
\begin{align*} 
\frac{\mathrm{d}\nu_{{\ell},k}^{\mathrm{\BDG}}}{\mathrm{d}\mu_0} &\propto \gamma_{{\ell},k}^{\mathrm{\BDG}} := \exp(-[{\zeta_{k}^{({\ell})}}\Phi_{h_{\ell}}+(1-{\zeta_{k}^{({\ell})}})\Phi_{h_{{\ell}-1}}]), {\ell}=2,\ldots,{N_{\mathrm{L}}},
\end{align*}
where $\nu_{{\ell}+1,0}^{\mathrm{\BDG}} := \nu_{{\ell},{N_{\mathrm{B}}^{({\ell})}}}^{\mathrm{\BDG}} \text{ and } \nu_{2,0}^{\mathrm{\BDG}} := \nu_{N_{\mathrm{T}}}^{\mathrm{\TMP}}$.

\subsection{Adaptive Sequential Monte Carlo}\label{Sec:InverseTemp}
The accuracy and computational cost of all SMC samplers such as tempering, standard and multilevel bridging, depend crucially on the number of intermediate probability measures $N_{\mathrm{seq}}$ ($\in \{N_{\mathrm{B}},  N_{\mathrm{T}},N_{\mathrm{B}}^{(1)},...,N_{\mathrm{B}}^{(N_{\mathrm{L}})}\}$, respectively) and the choice of the inverse temperatures $(\beta_k : k = 0,\ldots,N_{\mathrm{seq}})$. 
Up to now we assumed that $N_{\mathrm{seq}}$ and $(\beta_k : k = 0,\ldots,N_{\mathrm{seq}})$ are given a priori. 
However, we can also determine the inverse temperatures and associated intermediate probability  measures adaptively ``on the fly''.
In the literature, several strategies for adapting the inverse temperatures are known.
We review (and implement) methods based on the coefficient of variation of the update weights. 
{In the remainder of this text} we do not formally distinguish between SMC with fixed intermediate probability  measures (as in \S\ref{sec:SMC_Temp}) and \textit{adaptive} SMC as it is often done in the literature.
Moreover, adaptivity refers only to the choice of the inverse temperatures.
We do not consider adaptive schemes for the Markov kernel in the MCMC step.
 
To simplify the notation we drop the subscript $k$ and consider the SMC update  $\nu \mapsto \nu^{\ast}$ in the remainder of this section.
Let $w^\ast$ denote the density of $\nu^\ast$ with respect to $\nu$.
The probability measures $\nu$ and $\nu^\ast$ are approximated by $\widehat{\nu}$ and $\widehat{\nu}^{\ast}$, respectively, and are based on $J$ particles each.
Then, the \textit{effective sample size (ESS)} for the SMC update step is defined by
\begin{equation} \label{EQ:ESS:SMC_Update}
\mathrm{ESS} := \frac{J}{1+\mathrm{cv}^2_{\widehat{\nu}}(w^\ast)},
\end{equation}
where
\begin{align*}
\mathrm{cv}_{\widehat{\nu}}(w^\ast) 
&:= \frac{\mathrm{StD}_{\widehat{\nu}}(w^\ast)}{\mathbb{E}_{\widehat{\nu}}[w^\ast]} 
:= \frac{\sqrt{\mathrm{Var}_{\widehat{\nu}}(w^\ast)}}{\mathbb{E}_{\widehat{\nu}}[w^\ast]}
\end{align*}
is the \textit{coefficient of variation} of $w^\ast$. 
In general, one would consider the standard deviation $\mathrm{StD}_{\widehat{\nu}}(w^\ast)$ in place of the coefficient of variation $\mathrm{cv}_{\widehat{\nu}}(w^\ast)$. However, in the SMC setting this allows us to work with \textit{unnormalized} weights, since $\mathbb{E}_{\widehat{\nu}}[w^\ast] = 1$ if $w^\ast$ is normalised. 
Hence $\mathrm{cv}_{\widehat{\nu}}(w^\ast)$ is equal to the standard deviation of the normalized weights.

Now, the inverse temperature $\beta$ associated with $\nu$ is known. 
Our task is to define the inverse temperature $\beta^\ast$ associated with $\nu^{\ast}$.
Clearly, the density of $\nu^{\ast}$ with respect to $\nu$ depends on $\beta^{\ast}$. 
For this reason we write $w^{\ast} = w^{\ast}(\beta^{\ast})$. 
Then, the ESS of the SMC update also depends on $\beta^{\ast}$ and we write
\begin{align*}
\mathrm{ESS}(\beta^{\ast}) &:= \frac{J}{1+{\mathrm{cv}}_{\widehat{\nu}}^2\left(w^\ast(\beta^{\ast})\right)}. 
\end{align*}
Note that $\mathrm{ESS}(\beta^{\ast})$ can be computed without further evaluations of the (expensive) forward response operator, for any $\beta^{\ast} \in (\beta, 1]$. 
In our implementation we choose $\beta^{\ast}$ such that $\mathrm{ESS}(\beta^{\ast})$ is equal to some predefined target value $\tau_{\mathrm{ESS}}>0$.
In practice, we would like to avoid inverse temperature choices that meet the target ESS, that is, $\mathrm{ESS}(\beta^{\ast}) = \tau_{\mathrm{ESS}}$, but do not increase the inverse temperature by at least some $\varepsilon = \beta^{\ast} - \beta > 0$. 
Thus, we define  
\begin{equation} \label{Eq_ESS_opt}
\beta^{\ast} := \underset{{\beta' \in [\min\{\beta+\varepsilon, 1\},1]}}{\mathrm{argmin}}\left({\mathrm{ESS}}(\beta') - \tau_{\mathrm{ESS}}\right)^2.
\end{equation}
Note that the optimisation problem in (\ref{Eq_ESS_opt}) is equivalent to the following problem:
\begin{equation}\label{Eq_cv_opt}
\beta^{\ast} = \underset{{\beta' \in [\min\{\beta+\varepsilon, 1\},1]}}{\mathrm{argmin}}\left({\mathrm{cv}}_{\widehat{\nu}}\left(w^\ast(\beta')\right) - \tau^\ast\right)^2,
\end{equation}
where $\tau^\ast:= \sqrt{({J-\tau_{\mathrm{ESS}}})/{\tau_{\mathrm{ESS}}}}$.
Hence the fitting of the effective sample size is equivalent to a fitting of the coefficient of variation of the weights.

\section{Multilevel Sequential${}^2$ Monte Carlo}\label{sec:ML}
In this section we generalize Multilevel Bridging and propose the \textit{Multilevel Sequential${}^2$ Monte Carlo} (MLS$^{2}$MC) sampler.  
We explain the advantages of this generalisation in \S\ref{MLBoptimal}, but before we do this, we introduce the sampler formally in \S\ref{MLS2MC:Formal_Introduction} and discuss its accuracy and computational cost in \S\ref{MLS2MC:ComputationalCost}.

MLS$^{2}$MC is a Sequential Monte Carlo method which combines Tempering and Multilevel Bridging. 
Sequential$^{2}$ refers to two individual sequences in a Sequential Monte Carlo sampler, namely a sequence of inverse temperatures $(\beta_k : k = 0,\ldots,{N_{\mathrm{T}}})$ and a sequence of discretisation levels $(h_\ell : \ell = 1,\ldots,N_{\mathrm{L}})$. 
Starting with the prior measure $\mu_0$ and discretisation level $\ell=1$, the MLS${}^2$MC update either increases the discretisation resolution $h_\ell \mapsto h_{\ell+1}$ ($\ell = 1,\ldots,{N_{\mathrm{L}}-1}$) or the inverse temperature $\beta_k \mapsto \beta_{k+1}$ ($k = 1,\ldots,{N_{\mathrm{L}}-1}$). 
This process is repeated until we arrive at the inverse temperature  $\beta_{N_{\mathrm{T}}}= 1$ and maximal discretisation level $N_{\mathrm{L}}$.
See Figure \ref{Figure_ComparisonPlot} for an illustration.
 
\begin{figure}[thb]
\centering
\begin{tikzpicture}[scale=0.67]

\draw			(0,-0.15) node[anchor=north] {1}
		(1.5,-0.15) node[anchor=north] {2}
        (3,-0.15) node[anchor=north] {3}
        (5.25,-0.15) node[anchor=north] {$\cdots$}
		
		(7.5,-0.15) node[anchor=north] {$N_{\mathrm{L}}-1$}
        (9,-0.15) node[anchor=north] {$N_{\mathrm{L}}$};
\draw   (-0.15,1.5) node[anchor=east] {$\beta_{1}$}
		(-0.15,3) node[anchor=east] {$\beta_{2}$}
        (-0.15,5.25) node[anchor=east] {$\vdots$}
		(-0.15,7.5) node[anchor=east] {$\beta_{N_{\mathrm{T}}-1}$}
		(-0.15,9) node[anchor=east] {$1=\beta_{N_{\mathrm{T}}}$};

\draw[color=teal,line width=2] (0.1,0) -- (0.1,3);
\draw[color=teal,line width=2,dashed] (0.1,3) -- (0.1,6);
\draw[color=teal,line width=2] (0.1,6) -- (0.1,9);
\draw[color=teal,line width=2] (0.1,9) -- (3,9);
\draw[color=teal,line width=2,dashed] (3,9) -- (6,9);
\draw[color=teal,line width=2] (6,9) -- (9,9);
\draw   (0.1,9) node[anchor= north west, text width=5cm] {\textcolor{teal}{\mbox{MLB} (\cite{Koutsourelakis2009AParameters} and \S\ref{SubSubSec:BridgingMLKoutsourelakis})}};

\draw[color = magenta,line width=2] (9,0) -- (9,3);
\draw[color = magenta,line width=2,dashed] (9,3) -- (9,6);
\draw[color = magenta,line width=2] (9,6) -- (9,9);
\draw (9.5,2.5) node[anchor = west, text width=3cm] {\textcolor{magenta}{single-level SMC \mbox{(\cite{Beskos2015SequentialProblems} and \S\ref{Subsec:Tempering})}}};

\draw[color = blue,line width=2] (0.09,0) -- (0.09,1.5) -- (1.5,1.5) -- (1.5,3)--(3,3);
\draw (1.5,2.25) node[anchor = west,text width=6cm] {\textcolor{blue}{MLS${}^2$MC}};
\draw[color = blue, line width=2, dashed] (3,3) --(3,4.5) -- (4.5,4.5) -- (4.5,6)  --  (6,6);
\draw[color = blue,line width=2] (6,6) -- (6,7.5) -- (7.5,7.5)--(7.5,8.96)--(9,8.96);

\draw[line width=1.5,->] (0,0) -- (0,9.5) node[anchor=south,text width=3.5cm] {\textbf{Inv. Temp.}};

\draw[line width=1.5,->] (0,0) -- (9.5,0) node[anchor=west,text width=2.5cm] {\textbf{Discr. lvl.}};

\draw[line width=1] (1.5,0) -- (1.5,-0.15)
					 (3,0) -- (3,-0.15)
                
                        (7.5,0) -- (7.5,-0.15)
                        (9,0) -- (9,-0.15)
                        (-0.15,1.5) -- (0,1.5)
					 (-0.15,3) -- (0,3)
                  
                        (0,7.5) -- (-0.15,7.5)
                        (0,9) -- (-0.15,9);

\fill[black] (8.975,8.975) circle (3.5pt);
\draw	(9,9) node[anchor=south,text width=3.5cm]  {\textcolor{black}{(Target distr. $\mu^y_{h_{N_{\mathrm{L}}}}$)}};
\fill[black] (0.025,0.025) circle (3.5pt); 
\draw	(0.1,-0.2) node[anchor=north east]  {\textcolor{black}{(Prior distr. $\mu_0$) }};
\draw (0,0) node[anchor=east]{ $0=\beta_0$};
\end{tikzpicture}

\caption{The update schemes associated with Multilevel Bridging, single-level SMC, and MLS${}^2$MC.} 
\label{Figure_ComparisonPlot}
\end{figure}
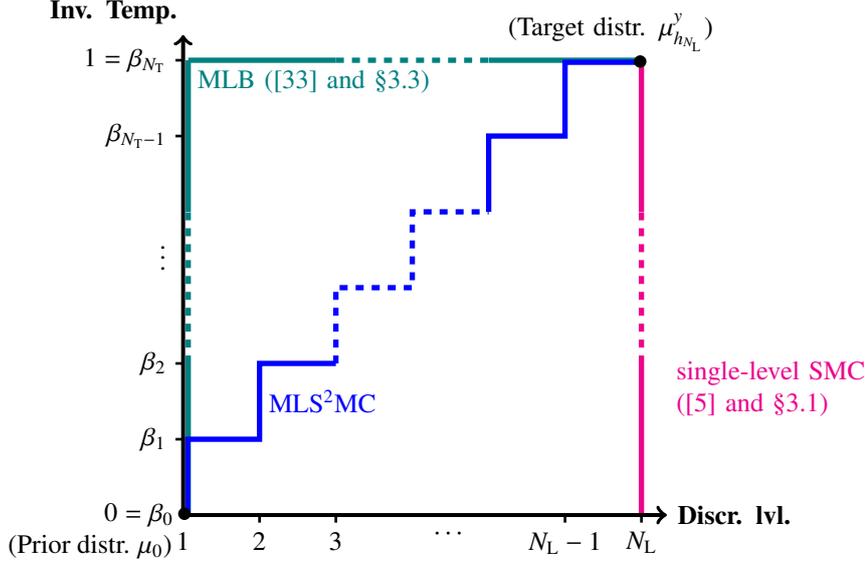

\subsection{Formal Introduction}\label{MLS2MC:Formal_Introduction}
We introduce a general framework to describe MLS$^{2}$MC update strategies. 
Let $N_{\mathrm{S}^2} = N_{\mathrm{\TMP}}+N_{\mathrm{L}}$ denote the total number of bridging steps and inverse temperature updates. 
Let $u : \{0,...,N_{\mathrm{S}^2}\} \rightarrow \{0,..., N_{\mathrm{\TMP}}\} \times \{1,...,N_{\mathrm{L}}\}$ denote a function, where
\begin{align}
u_i(s) &= u_i(s-1) \Leftrightarrow  u_j(s) = u_j(s-1) + 1, \ \ (i, j = 1,{2}, i \neq j) \\
u(0) &= (0,1),\\
u(N_{\mathrm{S}^2}) &= (N_{\mathrm{\TMP}},N_{\mathrm{L}}).
\end{align}
We refer to $u$ as \textit{update scheme}. 
In each step $s=0,\dots,N_{\mathrm{S}^2}$ of the algorithm $u_1(s)=k$ refers to the inverse temperature and $u_2(s)=\ell$ refers to the  discretisation level. 
The update function $u$ is convenient for the discussion and analysis of various update schemes. 
If we consider only a single update scheme $u$, we define $u_1(s) =: {\mathrm{\TMP}(s)}$ and $u_2(s) =: {\mathrm{\BDG}(s)}$. 
Furthermore, if it is clear whether $s$ refers to  ${\mathrm{\TMP}(s)}$ or ${\mathrm{\BDG}(s)}$ or to both, then we use the notation
\begin{align*}
\Phi_s &:= \Phi_{h_{{\mathrm{\BDG}(s)}}}, \ \ \mathcal{G}_s := \mathcal{G}_{h_{{\mathrm{\BDG}(s)}}}, \ \ \beta_s := \beta_{\mathrm{\TMP}(s)}, \quad (s = 0,\ldots,{N_{\mathrm{S}^2}}).
\end{align*}
Before we present the formal definition of MLS${}^2$MC we give two examples for alternative update schemes.
See Figure \ref{Figure_ComparisonPlot} for an illustration.

\begin{example} \label{example_SMC}
Let ${N_{\mathrm{L}}} = 1$ and define the update scheme $u :\{0,\dots,{N_{\mathrm{S}^2}}\} \rightarrow \{0,\dots,{N_{\mathrm{T}}}\} \times \{1\}$, where $s \mapsto  (s,1)$.
Then, the associated sampler is equivalent to single-level SMC.
\end{example}

\begin{example} \label{example_MLB}
Let $u : \{0,\dots,{N_{\mathrm{S}^2}}\} \rightarrow \{0,\dots,{N_{\mathrm{T}}}\} \times \{1,\dots,{N_{\mathrm{L}}}\}$, where
\begin{equation*}
s \mapsto  \begin{cases} (s,1), &\text{if } s \leq {N_{\mathrm{T}}},\\ 
({N_{\mathrm{T}}}, s-{N_{\mathrm{T}}}+1), &\text{otherwise}.
\end{cases}
\end{equation*}
The corresponding sampler is equivalent to Multilevel Bridging.
\end{example}

Now, we define MLS${}^2$MC as a Sequential Monte Carlo sampler (see \S\ref{sec:SMC_Temp}). 
Hence, we construct a sequence of probability measures $(\mu_{u(s)} : s = 0,\dots,{N_{\mathrm{S}^2}})$, where $\mu_{u(0)} = \mu_0$ and $\mu_{u(N_{\mathrm{S}^2})} = \mu^y_{h_{N_{\mathrm{L}}}}$. The intermediate probability  measures are based on the update scheme $u$ and are given once again by the Radon-Nikodym Theorem:
\begin{equation*}
\frac{\mathrm{d}\mu_{u(s)}}{\mathrm{d}\mu_0}(\theta) \propto \exp\left(- \beta_{\mathrm{\TMP}(s)} \Phi_{h_{{\mathrm{\BDG}(s)}}}(\theta)\right) \ \ (s = 1,\ldots,{N_{\mathrm{S}^2}},\ \theta \in X).
\end{equation*}
In the MLS${}^2$MC sampler we distinguish two update types.
Let $s = 1,\dots,{N_{\mathrm{S}^2}}$. 
If $ {\mathrm{\TMP}(s)} = {\mathrm{\TMP}(s-1)}+1 $, then
\begin{equation*}
\frac{\mathrm{d}\mu_{u(s)}}{\mathrm{d}\mu_{u(s-1)}}(\theta) \propto \exp\left(- (\beta_{\mathrm{\TMP}(s)}-\beta_{\mathrm{\TMP}(s-1)}) \Phi_{h_{{\mathrm{\BDG}(s)}}}(\theta)\right) \ \ (s = 1,\ldots,{N_{\mathrm{S}^2}}, \theta \in X).
\end{equation*}
We refer to this update as \textit{inverse temperature update (ITU)}.
Otherwise, if ${\mathrm{\BDG}(s)} = {\mathrm{\BDG}(s-1)}+1$, then the update is given by
\begin{equation*}
\frac{\mathrm{d}\mu_{u(s)}}{\mathrm{d}\mu_{u(s-1)}}(\theta) \propto \exp\left(- \beta_{{\mathrm{\TMP}(s)}} (\Phi_{h_{{\mathrm{\BDG}(s)}}}(\theta)-\Phi_{h_{{\mathrm{\BDG}(s-1)}}}(\theta))\right) \ \ (s = 1,\ldots,{N_{\mathrm{S}^2}},  \theta \in X).
\end{equation*}
We refer to this update as \textit{level update (LU)}.
However, we usually perform more than one Bridging step from one discretisation level to the next (see \S \ref{Subsec:Bridging}). 
We can redefine the update by the following (telescoping) product of $N_{\mathrm{B}}^{({{\mathrm{\BDG}(s)}})} =: {N_{\mathrm{B}}^{(s)}} \in \mathbb{N}$ densities, each of which reflects a particular intermediate bridging measure that is based on {bridging inverse temperatures} $(\zeta_m^{(s)} : m = 1,\ldots,{N_{\mathrm{B}}^{(s)}})$:
\begin{align*}
\frac{\mathrm{d}\mu_{u(s)}}{\mathrm{d}\mu_{u(s-1)}}(\theta) \propto \prod_{m=1}^{{N_{\mathrm{B}}^{(s)}}} \exp\left(-\beta_{ {\mathrm{\TMP}(s)}}(\zeta_m^{(s)} - \zeta_{m-1}^{(s)}) (\Phi_{h_{{\mathrm{\BDG}(s)}}}(\theta)-\Phi_{{h_{{\mathrm{\BDG}(s-1)}}}}(\theta))\right) \nonumber \\ (s = 1,\ldots,{N_{\mathrm{S}^2}}, \theta \in X). \nonumber
\end{align*}
For clarity of presentation we do not include the intermediate bridging measures in the update scheme $u$. 
Furthermore, if it is clear which update scheme is used, we write $\mu_s := \mu_{u(s)}$.

\subsection{Computational cost and accuracy} \label{MLS2MC:ComputationalCost}
Before we propose an efficient update scheme for the MLS${}^2$MC sampler we briefly discuss its computational cost and accuracy.
Let $\mathcal{C}_{\ell} \in (0, \infty)$ denote the computational cost of one evaluation of $\Phi_{h_{\ell}}$ (for ${\ell} = 1,\ldots,{{N_{\mathrm{L}}}})$.
Moreover, we denote the total cost of the MLS${}^2$MC sampler with associated update scheme $u$ by $\mathrm{Cost}(u)$.
We typically measure $\mathcal{C}_\ell$ in terms of the number of floating point operations that are required to evaluate $\mathcal{G}_{h_\ell}$.  
One could also think of estimating the elapsed time of model evaluations or e.g. the memory requirement.

\begin{example} \label{Examples_CostEvaluations}
Let $\mathcal{G}:=\mathcal{O}\circ G$ denote a forward response operator, where $G$ is the solution operator of an elliptic boundary value problem in $d$-dimensional space ($d = 1,2,3$). 
Furthermore, let $h_{\ell} = 2^{-\ell}h_0, \ell \in \mathbb{N}, h_0 > 0$, denote the mesh size of the discretised model $\mathcal{G}_{h_{\ell}}$, respectively the discretised potential $\Phi_{h_{\ell}}$. 
Then, {the ratio of the computational cost associated with two consecutive levels} in terms of floating point operations is 
\begin{equation*}
\frac{\mathcal{C}_{\ell+1}}{\mathcal{C}_{\ell}} = 2^d, \quad \ell \in \mathbb{N}.
\end{equation*}
Given a maximal level ${N_{\mathrm{L}}} \in \mathbb{N}$, we normalize the values such that $\mathcal{C}_{N_{\mathrm{L}}} = 1$. 
We arrive at
\begin{equation*}
{\mathcal{C}_{\ell}} := 2^{d(\ell-N_{\mathrm{L}})}, \quad \ell = 1,\dots,N_{\mathrm{L}}.
\end{equation*}
\end{example}
In the following we discuss the computational cost of MLS${}^2$MC in terms of the update scheme $u$ and the costs $(\mathcal{C}_{\ell} : \ell=1,...,{N_{\mathrm{L}}})$. 
To begin, we consider inverse temperature updates. 
If the Markov kernel update is performed by a Metropolis-Hastings scheme, then one PDE solve for each of the $J$ particles is required, to evaluate the acceptance probability.
The acceptance step also requires the model evaluations of the current particles. 
This however should remain in the memory, until the particles are updated. 
Hence, the computational cost of the inverse temperature updates is given by
\begin{equation*}
 \sum_{\substack{s=1 \\  s \text{ is an ITU}}}^{N_{\mathrm{S}^2}}J \mathcal{C}_{\mathrm{\BDG}(s)}.
\end{equation*}
In Bridging, we also perform a Markov kernel step for each of the ${N_{\mathrm{B}}^{(s)}}$ intermediate Bridging steps and each of the $J$ particles. 
Here, the evaluation of the {  Markov} update density requires two model evaluations in total, namely one on each discretisation level $\mathrm{B}(s-1)$ and $\mathrm{B}(s)$, respectively.
Thus, we perform $N_{\mathrm{B}}^{(s)} \cdot J$ evaluations of the model on the two levels. 
In addition, we have to consider the first intermediate Bridging step. 
As opposed to the inverse temperature update, we do not yet have the model evaluation of the current particles on level $\mathrm{B}(s)$. 
Thus, we need to add the cost of $J\cdot \mathcal{C}_{\mathrm{\BDG}(s)}$ to each of the level updates. 
In summary, the computational cost for a level update is given by
\begin{equation*}
\sum_{\substack{s=1 \\ s \text{ is an LU}}}^{N_{\mathrm{S}^2}}J\left( \mathcal{C}_{\mathrm{\BDG}(s)} + ({N_{\mathrm{B}}^{(s)}})( \mathcal{C}_{\mathrm{\BDG}(s)} + \mathcal{C}_{{\mathrm{\BDG}(s-1)}})\right).
\end{equation*}
Adding the costs for bridging and inverse temperature updates, respectively, we arrive at the following total cost.
\begin{proposition} \label{Cost_Proposition} 
Let the Markov kernels in the MLS${}^2$MC sampler be given in terms of a single Metropolis-Hastings MCMC update. 
Then, the total computational cost of the Multilevel Sequential${}^2$ Monte Carlo sampler is given by
\begin{equation*}
\mathrm{Cost}(u) = \sum_{\substack{s=1 \\ s \text{ is an ITU}}}^{N_{\mathrm{S}^2}} J\mathcal{C}_{\mathrm{\BDG}(s)} + \sum_{\substack{s=1 \\ s \text{ is an LU}}}^{N_{\mathrm{S}^2}}J\left( \mathcal{C}_{\mathrm{\BDG}(s)} + ({N_{\mathrm{\BDG}}^{(s)}})( \mathcal{C}_{\mathrm{\BDG}(s)} + \mathcal{C}_{{\mathrm{\BDG}(s-1)}})\right).
\end{equation*}
\end{proposition}

Next we discuss the accuracy of the MLS${}^2$MC sampler in terms of the following root mean square error type metric
\begin{equation*}
\sup_{\|q\|_{\infty} = 1}\sqrt{\int\left(\mathbb{E}_{\widehat{\mu}^y_{(\theta)}}[q]-\mathbb{E}_{\mu^y}[q]\right)^2\mathrm{d}\mathbb{P}(\theta)},
\end{equation*}
where $\widehat{\mu}^y_{(\theta)}$ is the particle  based MLS$^2$MC approximation of $\mu^y$. Note that  $\widehat{\mu}^y_{(\theta)}$ is a random measure.
We make use of the following observation:
In every MLS$^2$MC update we perform a Monte Carlo estimation with weighted samples. 
Hence, in each update the approximation accuracy measured in terms of the root mean square error is of order
\[
\mathcal{O}(\mathrm{ESS}^{-1/2}; \mathrm{ESS} \rightarrow \infty).
\] 
Here, $\mathrm{ESS}$ is the effective sample size defined in \eqref{EQ:ESS:SMC_Update}. 
We refer to \cite{Agapiou2015ImportanceCost,Beskos2016OnMethods,Beskos2015SequentialProblems,Rebeschini2015CanDimensionality}  for details on the approximation accuracy of SMC type samplers.
Recall that we choose the update steps adaptively (see \S\ref{Sec:InverseTemp}).
Thus, the $\mathrm{ESS}$ is constant in every step.
Hence, every Bridging and Tempering step has the same influence on the accuracy.
Thus, the total accuracy is bounded by the sum of the individual accuracies associated with the update steps.
For this reason, we can maximize the accuracy of the MLS${}^2$MC approximation by minimizing the total number of MLS$^2$MC update steps.
The latter is given by
\[
\#\mathrm{Upd}(u) = N_{\mathrm{\TMP}} + \sum_{\substack{s=1 \\ s \text{ is an LU}}}^{N_{\mathrm{S}^2}}N_{\mathrm{\BDG}}^{(s)}.
\]
In summary, we wish to design an update scheme which minimizes both $\#\mathrm{Upd}(\cdot)$ and $\mathrm{Cost}(\cdot)$.

\subsection{Is Multilevel Bridging optimal?} \label{MLBoptimal}
Now we discuss the computational cost of Multilevel Bridging (see \S\ref{SubSubSec:BridgingMLKoutsourelakis} for details). 
We do this to motivate our generalisation, the MLS$^2$MC sampler.
{First, we state two assumptions on the inverse temperatures and number of intermediate bridging steps.}
\begin{assumptions}\label{Assumptions_MLB_optimal}
In the MLS$^2$MC sampler,
\begin{enumerate}
\item[(a)] the inverse temperature $\beta_{\mathrm{\TMP}(s)}$ is independent of the discretisation level $\mathrm{\BDG}(s-1)$, for any $s = 1,\dots,N_{\mathrm{S}^2}$, where $s$ refers to an ITU, and 
\item[(b)] the number of intermediate bridging steps $N^{(s)}_{\mathrm{\BDG}}$ is independent of the inverse temperature $\beta_{\mathrm{\TMP}(s-1)}$, for any $s = 1,\dots,N_{\mathrm{S}^2}$, where $s$ refers to an LU.
\end{enumerate}
\end{assumptions}
Given these assumptions, $\#\mathrm{Upd}(u)$ is constant for every possible update scheme $u$. 
Hence, we expect the same accuracy for any MLS${}^2$MC sampler independently of $u$. 
One can argue analogously for the cost of the bridging steps: 
Due to the Assumption \ref{Assumptions_MLB_optimal}(b) the number of Bridging steps is fixed throughout all feasible update schemes. 
Thus, the crucial factor contributing to the total cost is the tempering. 
In MLB the tempering is performed completely on level $\ell = 1$ which requires the least computational effort. 
We summarize this paragraph in the following proposition.

\begin{proposition}\label{Proposition_MLB_optimal} 
Let $u$ be the Multilevel Bridging update scheme defined in Example \ref{example_MLB}. 
If Assumptions \ref{Assumptions_MLB_optimal} are satisfied, then $u$ minimizes both $\#\mathrm{Upd}(\cdot)$ and $\mathrm{Cost}(\cdot)$.
\end{proposition}

We now comment on Assumptions \ref{Assumptions_MLB_optimal}, starting with (a). 
The major reason for performing the tempering is the concentrated support of the posterior in the small noise limit. 
The width of this concentrated support is associated with the posterior variance which in turn reflects the certainty in the considered parameter. 
This certainty in the parameter is based on the precision $\Gamma^{-1}$ of the data  which we define a priori in the likelihood. 
Since $\Gamma^{-1}$ is chosen independently of the discretisation resolution $h$, Assumption \ref{Assumptions_MLB_optimal}(a) is likely satisfied.

In contrast, Assumption \ref{Assumptions_MLB_optimal}(b) is not always justified.
If $\Phi_{h_{\ell-1}}$ is a good approximation to $\Phi_{h_{\ell}}$, then also $\exp(-\Phi_{h_{\ell-1}})^{\beta} \approx \exp(-\Phi_{h_{\ell}})^{\beta}$, independently of the inverse temperature $\beta$.
Hence, the support of the associated posterior measures differs only in a small area of the parameter space, and a small number of intermediate bridging steps is required from $\ell-1 \rightarrow \ell$.
However, on two consecutive coarse discretisation levels the discrepancy of $\Phi_{h_{\ell-1}}$ and $\Phi_{h_{\ell}}$ might be large.
This is not a big problem for small inverse temperatures associated with a larger noise level in the likelihood; it is still possible that there is a substantial overlap of the support of the associated probability  measures, and hence a moderate number of intermediate bridging steps is required.
However, for large inverse temperatures and a small noise level the associated probability  measures are likely highly concentrated, and their supports might have a small intersection.
Thus, either a large number of intermediate bridging steps is required, or bridging might not be possible at all.
In any case, Assumption \ref{Assumptions_MLB_optimal}(b) is hardly justified.
Note that a large number of intermediate bridging steps also reduces the overall accuracy of the SMC sampler. 

When is bridging practically impossible? Let $s$ refer to an LU. 
Given a fixed number of particles $J$, it is possible that the update density $\mathrm{d}\mu_s/\mathrm{d}\mu_{s-1}$ is numerically zero for all particles. 
Then, we refer to $\mu_{s-1}$ and $\mu_{s}$ as \textit{numerically singular}. 
Importantly, we are not able to carry out an MLS$^2$MC update from $\mu_{s-1}$ to $\mu_{s}$ in this case.

Now, in MLB all level updates are performed with the untempered likelihood, i.e. $\beta = 1$, even for coarse discretisations.
As explained above, this might result in an inaccurate or expensive estimate, or the estimation might not be possible at all.
In the numerical experiments in \S\ref{sec:numerics} we will illustrate these issues. 

Of course, these problems can be cured by starting the MLB on a fine discretisation level where Assumption \ref{Assumptions_MLB_optimal}(b) is satisfied.
If the model $\mathcal{G}_{h_{(\cdot)}}$ is well understood it might even be possible to define a suitable minimal starting level.
In this case, the cost of MLB might often be cheaper than the cost of the adaptive update scheme that we propose in the next section. 
However, in most cases the model $\mathcal{G}_{h_{(\cdot)}}$ is not well understood or even only given in a black box sense. 
In this case, determining a sufficiently fine starting level for MLB is not possible.
This motivates us to introduce an efficient, parameter-free, adaptive update scheme which does not require a priori information on the model resolution. 

\subsection{An efficient update scheme} \label{Subsec:Adaptive_Update_Schemes}

Now we discuss the major component in the proposed MLS$^2$MC sampler, namely, the choice of the inverse temperature and level updates, respectively.
Balancing these updates with the computational cost is a nontrivial task.
If we increase the discretisation level too early in the update scheme, then many inverse temperature updates on an expensive level are required. 
Increasing the discretisation level too late could result in the undesirable situation that many intermediate bridging steps might be required later on (see the discussion in \S\ref{MLBoptimal}). 
To simplify the derivation of the computational cost we work under Assumption \ref{Assumptions_MLB_optimal}(a) which is likely satisfied for a large class of relevant BIPs.
In this case, to obtain a good accuracy of the MLS$^2$MC approximation, we aim at minimizing the number of bridging steps.
However, we also need to consider the computational cost associated with the proposed path since MLS$^2$MC should operate with minimal cost.

Suppose we are in the update step from $\mu_{s-1}$ to $\mu_s$, where $s \in \{2,\dots,N_{\mathrm{S}^2}\}$ and $u(s-1) =: (k-1,\ell-1)$. 
Under Assumption \ref{Assumptions_MLB_optimal}(a) we study the following \textit{decision problem:}  
Do we update the discretisation level $\ell-1 \mapsto \ell$ or the temperature $k-1 \mapsto k$?

To account for the full impact of this decision we consider the cost and the loss in accuracy of \textit{all} future update steps.  
We split the future path into two parts, namely, from $\mu_{s-1}$  to $\mu_s$ and from $\mu_s$ to $\mu_{N_{\mathrm{S}^2}}$.
For simplification we suppose that for the second part both Assumptions \ref{Assumptions_MLB_optimal}(a) and (b) are satisfied. 
By Proposition \ref{Proposition_MLB_optimal} the optimal strategy for the second part starting in $s$ is to first increase the inverse temperature to $\beta = 1$ (in multiple steps) and then to bridge to the maximal level $N_{\mathrm{L}}$. 
This is equivalent to carrying out the Multilevel Bridging with initial probability  measure  $\mu_s$.
Hence the second part of the path is determined, and we only need to decide on the path from $\mu_{s-1}$ to $\mu_s$.
We now investigate this.

Let $s_{N_{\mathrm{T}}} = \min\{s \in \{1,\dots,N_{\mathrm{S}^2}\} : \beta_s = 1\}$.
In Figure~\ref{Figure_Decisionproblem} we show the two options:
\begin{itemize}
\item Path $w$: Update the level $\ell-1 \mapsto \ell$, in step $s-1 \mapsto s$, then proceed as in MLB,
\item Path $v$: Update the inverse temperature $k-1 \mapsto k$ in step $s-1 \mapsto s$,  then proceed as in MLB,
\end{itemize}
where $w(s-1) = v(s-1) = (k-1,\ell-1)$, $w(s) = (k-1,\ell)$ and $v(s) = (k-1,\ell)$. 
Note that $s_{N_{\mathrm{T}}}$ differs for the paths $v$ and $w$.  
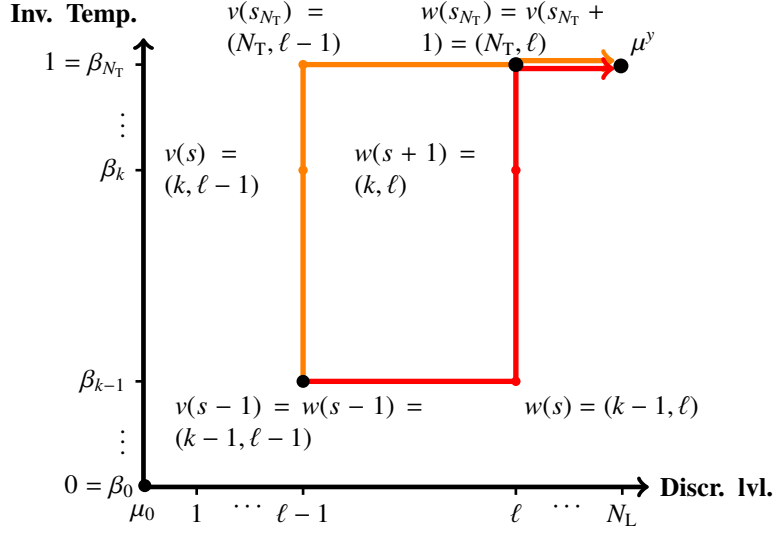
\begin{figure}[htb]
\centering
\begin{tikzpicture}[scale=0.70]

\draw			(1,-0.15) node[anchor=north] {1}
		(2,-0.15) node[anchor=north] {$\cdots$}
        (3,-0.15) node[anchor=north] {$\ell-1$}
		(7,-0.15) node[anchor=north] {$\ell$}
		(8,-0.15) node[anchor=north] {$\cdots$}
        (9,-0.15) node[anchor=north] {${N_{\mathrm{L}}}$};
\draw   (-0.15,1) node[anchor=east] {$\vdots$}
		(-0.15,2) node[anchor=east] {$\beta_{k-1}$}
        (-0.15,4.5) node[anchor=east] {}
		(-0.15,6) node[anchor=east] {$\beta_{k}$}
		(-0.15,7) node[anchor=east] {$\vdots$}
		(-0.15,8) node[anchor=east] {$1=\beta_{{N_{\mathrm{T}}}}$};

\draw[line width=1.75,->] (0,0) -- (0,8.5) node[anchor=south,text width=3.5cm] {\textbf{Inv. Temp.}}; 

\draw[line width=1.75,->] (0,0) -- (9.5,0) node[anchor=west,text width=2.5cm] {\textbf{Discr. lvl.}};
s


\draw[line width=1] (1,0) -- (1,-0.15)
					 (3,0) -- (3,-0.15)
                
                       (7,0) -- (7,-0.15)
                        (9,0) -- (9,-0.15)
					 (-0.15,2) -- (0,2)
                  
                       (0,6) -- (-0.15,6)
                        (0,8) -- (-0.15,8);
                        
 \draw[line width = 2, color = orange,-](3,2) -- (3,8) 
 								(3,8) -- (7,8);
 \draw[line width = 2, color = red,-](3,2) -- (7,2) 
 								(7,2) -- (7,8);
 \draw[line width = 2, color = orange,->](7,8.075) -- (8.85,8.075);
 \draw[line width = 2, color = red,->](7,7.925) --(8.85,7.925);
\fill[red] (7,2) circle (2.5pt);
\draw	(8.8,1.9) node[anchor=north]  {{{$w(s)=(k-1,{\ell})$}}};

\fill[red] (7,6) circle (2.5pt);
\draw	(7,6) node[anchor=east,text width=2cm]  {{$w(s+1)=(k,{\ell})$}};

\fill[black] (7,8) circle (4pt);
\draw	(7,8) node[anchor=south,text width=2.5cm]  {\textcolor{black}{$w(s_{N_{\mathrm{T}}}) = v(s_{N_{\mathrm{T}}}+1) =  ({N_{\mathrm{T}}},{\ell})$}};

\fill[orange] (3,8) circle (2.5pt);
\draw	(3,8) node[anchor=south,text width=2cm]  {\textcolor{black}{{$v(s_{N_{\mathrm{T}}})=({N_{\mathrm{T}}},{\ell}-1)$}}};

\fill[orange] (3,6) circle (2.5pt);
\draw	(3,6) node[anchor=east,text width=1.7cm]  {\textcolor{black}{{$v(s)=(k,{\ell}-1)$}}};

\fill[black] (3,2) circle (3.5pt);
\draw	(3.2,1.9) node[anchor=north,text width=3.65cm]  {\textcolor{black}{$v(s-1)=w(s-1)=(k-1,{\ell}-1)$}};

\fill[black] (8.975,7.975) circle (4pt);
\draw	(9,8) node[anchor=south west]  {\textcolor{black}{$\mu^y$}};
\fill[black] (0.025,0.025) circle (3.5pt); 
\draw	(0,-0.15) node[anchor=north]  {\textcolor{black}{$\mu_0$}};
\draw (0,0) node[anchor=east]{$0=\beta_0$};

\end{tikzpicture}
\caption{Decision problem in MLS${}^2$MC: Which path is cost-optimal? First level, then inverse temperature update (${w}$, red)  or first inverse temperature, then level update ($v$, orange).}
\label{Figure_Decisionproblem}
\end{figure}

We assume that ${N_{\mathrm{B}}^{(s)}} \leq {N_{\mathrm{B}}^{(s+1)}}$.
This is reasonable since the probability  measures that are bridged in path $v$ contain a smaller noise level and are thus more concentrated.
The computational costs associated with paths $v$ and $w$ between $\mu_{s-1}$ and $\mu_{(N_{\mathrm{T}}, \ell)}$ are given by 
\begin{align}
\mathrm{Cost}\left(w|_{\{s-1 ,\dots, s_{N_{\mathrm{T}}}\}} \right)&= \underbrace{J{N_{\mathrm{B}}^{(s)}}(\mathcal{C}_{\ell-1} + \mathcal{C}_{\ell}) + J\mathcal{C}_{{\ell}-1}}_{\text{LU }{\ell}-1 \mapsto {\ell}} + \underbrace{J ({N_{\mathrm{T}}}-k+1) \mathcal{C}_{\ell}}_{\text{ITU }k-1 \mapsto {N_{\mathrm{T}}}}, \label{CostLUthenITU}\\
\mathrm{Cost}\left(v|_{\{s-1 ,\dots, s_{N_{\mathrm{T}}}+1\}} \right)&=  \underbrace{J ({N_{\mathrm{T}}}-k+1) \mathcal{C}_{{\ell}-1}}_{\text{ITU }k-1 \mapsto {N_{\mathrm{T}}}} + \underbrace{J {N_{\mathrm{B}}^{(s+1)}}(\mathcal{C}_{{\ell}-1}+ \mathcal{C}_{\ell})+ J\mathcal{C}_{{\ell}-1}}_{\text{LU }{\ell}-1 \mapsto {\ell}} . \label{CostITUthenLU}
\end{align}
Note that we do not consider the cost of Bridging from $\mu_{(N_{\mathrm{T}}, \ell)}$ to $\mu_{(N_{\mathrm{T}}, N_{\mathrm{L}})} = \mu_{N_{\mathrm{S}^2}}$ since this cost is identical for both paths.
Given our assumptions we can reformulate the decision problem in terms of computational cost as follows: 

\textit{Is the number of additional bridging steps ${N_{\mathrm{B}}^{(s+1)}}$ needed in comparison to ${N_{\mathrm{B}}^{(s)}}$ more expensive than the increased computational cost of the inverse temperature update on level $\ell$ compared to level $\ell-1$?}

This question corresponds directly to the expressions in \eqref{CostLUthenITU} and \eqref{CostITUthenLU}. 
However, we need to minimise both the computational cost and the number of updates $\# \mathrm{Upd}$. 
If bridging and tempering are performed non-adaptively, then all quantities in \eqref{CostLUthenITU} and \eqref{CostITUthenLU} are known, as are $\# \mathrm{Upd}(v)$ and $\# \mathrm{Upd}(w)$.
Hence, we can simply compare the costs and the number of update steps and choose the path that is more appropriate.
However, this is not the focus of our paper.
From now on we consider adaptive tempering and bridging only.

Without loss of generality we assume that $\tau^{\ast} > 0$ is the target value for the coefficient of variation of the weights in \textit{every} tempering and bridging update.
Unfortunately, there are in general no simple analytic expressions for the interdependency of $\tau^{\ast}$, $\beta_s$ and ${N_{\mathrm{B}}^{(s)}}$.
Furthermore, given the probability  measure $\mu_{s-1}$, it is difficult to estimate how many intermediate bridging steps ${N_{\mathrm{B}}^{(s)}}$ are required for the bridging $\ell-1 \mapsto \ell$.
To make progress we continue as follows.
We select a small proportion $\tilde{J}$ of the $J$ samples and estimate the coefficient of variation associated with a bridging update using ${N_{\mathrm{B}}^{(s)}} = 1$ steps based on these $\tilde{J}$ samples.
We obtain
\begin{equation*}
\cv_{\mu_{s-1}} \left[ \exp\left(- \beta_{ s} (\Phi_{s}(\theta)-\Phi_{s-1}(\theta))\right)\right] =: \cv_{s}^{\text{LU}}.
\end{equation*}
This estimation requires $\tilde{J}$ additional evaluations of $\mathcal{G}_{h_{\ell}}$.  
If we update the discretisation level immediately afterwards, then these evaluations can be re-used for the bridging update. 
If this is not the case, then the additional samples are discarded. 
In \S\ref{sec:numerics} we consider various  proportions $\tilde{J}/J$.

To continue we make the following observation. 
If $\cv_{s}^{\text{LU}} < \tau_{\mathrm{LU}}$, where $\tau_{\mathrm{LU}} \in (0, {\tau^\ast}]$, then the bridging can be performed with only one intermediate step. 
We use this observation as a measure of the accuracy of the approximation $\mathcal{G}_{h_{\ell-1}} \approx\mathcal{G}_{h_{\ell}}$. 
\begin{itemize}
\item 
If the accuracy is small (i.e. $\cv_{s}^{\text{LU}} > \tau_{\mathrm{LU}}$), then we bridge immediately, since we would otherwise propagate the inaccurate model to an inverse temperature that is unreasonably small. 
\item
If the accuracy is high (i.e. $\cv_{s}^{\text{LU}} < \tau_{\mathrm{LU}}$), we know that ${N_{\mathrm{B}}^{(s)}} = 1$.
Moreover, we define ${N_{\mathrm{B}}^{\ast}} := {N_{\mathrm{B}}^{(s+1)}}$ and ${N_{\mathrm{T}}^{\ast}} := {N_{\mathrm{T}}}-k+1.$
Based on comparing the costs in \eqref{CostLUthenITU} and \eqref{CostITUthenLU} we perform an inverse temperature update from $s-1 \mapsto s$ if the condition
\begin{equation} \label{ITU_Condition}
J(\mathcal{C}_{\ell-1} + \mathcal{C}_{\ell}) +J {N_{\mathrm{T}}^{\ast}}\mathcal{C}_{\ell}\geq  
J  {N_{\mathrm{T}}^{\ast}}\mathcal{C}_{{\ell}-1}+ J {N_{\mathrm{B}}^{\ast}}(\mathcal{C}_{{\ell}-1}+ \mathcal{C}_{\ell})
\end{equation}
is satisfied (since then the ITU cost is cheaper than the LU cost).
If \eqref{ITU_Condition} is not satisfied, then we perform a level update.
\end{itemize}
Note that the condition in (\ref{ITU_Condition}) is equivalent to
\begin{equation} \label{ITU_Condition_Short}
\frac{\mathcal{C}_\ell}{\mathcal{C}_{\ell-1}} \geq \frac{{N_{\mathrm{T}}^{\ast}}+{N_{\mathrm{B}}^{\ast}}-1}{{N_{\mathrm{T}}^{\ast}}-{N_{\mathrm{B}}^{\ast}}+1},
\end{equation}
where we define $\frac{1}{0}:= \infty$. 
We visualize the condition in \eqref{ITU_Condition_Short} in Figure~\ref{Figure_ITULU_cond} where we show which combinations of ${N_{\mathrm{B}}^{\ast}}$ and ${N_{\mathrm{T}}^{\ast}}$ satisfy \eqref{ITU_Condition_Short} for $\mathcal{C}_{\ell}/\mathcal{C}_{l-1} \in \{2,4,8\}$.
These three cases refer to solves of elliptic PDEs in 1D, 2D and 3D (see Example \ref{Examples_CostEvaluations}). 
We see that condition (\ref{ITU_Condition}) {in the 3D case} implies ${N_{\mathrm{B}}^{\ast}}+1\approx {N_{\mathrm{T}}^{\ast}}$.
\begin{figure}[htb]
\centering
\includegraphics[scale=0.65]{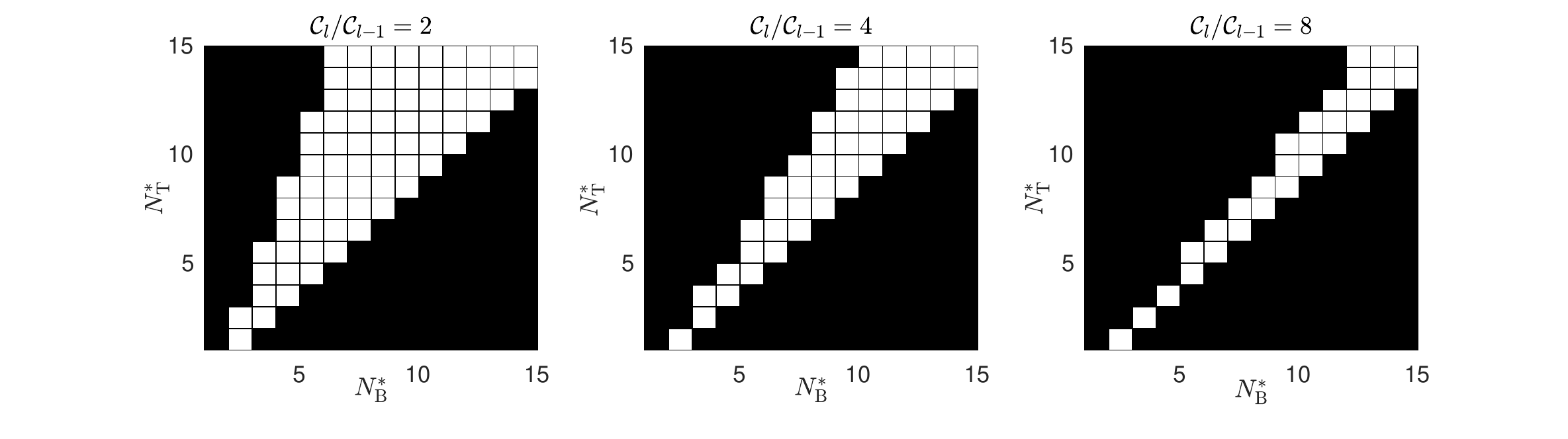}
\caption{Visualisation of the combinations of ${N_{\mathrm{T}}^{\ast}}$ and ${N_{\mathrm{B}}^{\ast}}$ satisfying condition (\ref{ITU_Condition_Short}) (black squares).} \label{Figure_ITULU_cond}
\end{figure}

Of course, the evaluation of \eqref{ITU_Condition_Short} requires ${N_{\mathrm{T}}^{\ast}}$ and ${N_{\mathrm{B}}^{\ast}}$.
However, these quantities are (still) not known a priori, since bridging and tempering are performed adaptively.
By consideration of certain special cases we obtain the approximations
\begin{equation}
{N_{\mathrm{B}}^{\ast}} \approx \left\lceil{\beta_{s+1}}/{\beta_{s}}\right\rceil, \ \ {N_{\mathrm{T}}^{\ast}} \approx \max\left\lbrace\left\lceil{\|\Gamma^{-1}\|_2} \right\rceil-(k+1),1\right\rbrace. \label{M*K*approx}
\end{equation}
However, in practice we observed that these approximations are quite inaccurate. 
Furthermore, if the increase of the computational cost $\mathcal{C}_{\ell}/\mathcal{C}_{\ell-1}$ is large (as is the case in realistic applications), then most combinations of $({N_{\mathrm{B}}^{\ast}}, {N_{\mathrm{T}}^{\ast}})$ satisfy (\ref{ITU_Condition_Short}), see Figure~\ref{Figure_ITULU_cond}.
Thus, if ${N_{\mathrm{B}}^{(s)}} = 1$, one might as well skip checking condition \eqref{ITU_Condition_Short} and always perform an inverse temperature update. 
We follow this strategy from now on. 

The noise in the BIP can be understood as a combination of observational noise and model (discretisation) error. 
This point of view fits very well with our MLS${}^2$MC framework.
Indeed, we reduce the noise level while increasing the accuracy of our model evaluation. 
See also the method presented in \cite{Calvetti2017IterativeInversion} for a further discussion of this idea. 
Suppose now that in the update scheme the inverse temperature has not yet reached its maximum $\beta_{(\cdot)} = 1$.
Given the argument above it is a good idea to increase the inverse temperature after every level update (non-adaptively). 
This reduces the total computational cost, since we save $\tilde{J}$ model evaluations in situations where a level update is very unlikely.
We implement both these ideas in our algorithm.
That is, we do not check the condition \eqref{ITU_Condition_Short}, and we update the inverse temperature automatically after every level update. 
This update scheme is given by the formula
\begin{equation} \label{eq_updatescheme_final}
u(s) = \begin{cases}
(\mathrm{T}(s-1)+1,\mathrm{B}(s-1)), &\text{if } B(s-1) = B(s-2)+1, \\  &\text{\underline{or} }\cv_{s}^{\text{LU}} < \tau_{\mathrm{LU}}, \\ 
(\mathrm{T}(s-1),\mathrm{B}(s-1)+1), &\text{otherwise,}
\end{cases}
\end{equation}
for any $s = 1,...,N_{\mathrm{S}^2}$, where $u(0):=(0,1)$ and $u(-1):=(0,0)$.
Note that the update scheme in \eqref{eq_updatescheme_final} is independent of $\mathcal{C}_{\ell}/\mathcal{C}_{\ell-1}$. 
Clearly, its effectiveness depends on the computational costs at each level, but the cost does not determine the adaptive choice between level update and inverse temperature update.  
Instead, the algorithm reduces the total number of intermediate bridging steps ${N_{\mathrm{B}}^{\ast}}$, and thus also increases the accuracy.
The numerical results in \S \ref{sec:numerics} show that the update scheme in \eqref{eq_updatescheme_final} implements a compromise between computational cost and accuracy.

In summary, our proposed update scheme uses a heuristic backtracking type method to find a suitable inverse temperature for the bridging from discretisation level $\ell-1$ to $\ell$. 
This means that we bridge on the smallest of the adaptively determined inverse temperatures where bridging is necessary. 
Let $s-1 \mapsto s$ refer to a level update. 
The approximation $\mu_{s-1} \approx \mu_{s}$ is accurate, if $\ell$ is sufficiently large, or if $\beta_k$ is sufficiently small. 
Hence, we deduce that our method leads to a small number of required intermediate bridging updates ${N_{\mathrm{B}}^{(\ell)}}$. 
This in turn implies a small total computational cost and a good accuracy of the measure approximation.

\subsection{Maximum level $N_{\mathrm{L}}^{\max}$} \label{Final level updates} 
A natural question in the context of the adaptive update scheme \eqref{eq_updatescheme_final} is: Do we need to go to the level $\ell = N_{\mathrm{L}}$ or can we stop earlier? 
To address this question we proceed as follows.
Let $F:= \{t = 1,\ldots,{N_{\mathrm{S}^2}} :  {\mathrm{\TMP}(t)} = N_{\mathrm{T}}, {\mathrm{\BDG}(t)} \leq {N_{\mathrm{L}}}\}$ denote the subset of the domain of the update scheme $u$ where the maximal inverse temperature $\beta_{N_{\mathrm{T}}} = 1$ is reached and some Bridging steps remain. 
Note that $F$ can be the empty set. 
If this is not the case we refer to $F$ as the set of \textit{final level updates}. 
We reformulate the question above as follows.
Is there an $s \in F$, such that the intermediate probability measure $\mu_s$ is a sufficiently accurate approximation to the target posterior measure $\mu^y_{h_{N_{\mathrm{L}}}}$?

We assess the necessity of updating the discretisation level in terms of the information gain associated with the update. 
If the information gain  of the level update is smaller than a certain threshold, then the  algorithm terminates. 
To be consistent with the update scheme \eqref{eq_updatescheme_final} we measure the information gain in terms of $\cv_{s}^{\text{LU}}$. 
This coefficient of variation gives an upper bound for the Kullback-Leibler divergence from $\mu_{s-1}$ to $\mu_s$. 
See \cite{Agapiou2015ImportanceCost} for details.
Let $\tau_{\min} > 0$, where $\tau_{\min} \ll \tau_{\mathrm{LU}}$, denote a threshold parameter.
The modified update scheme $u'$ reads as follows:

\begin{equation} \label{eq_updatescheme_final_Lmax}
u'(s) = \begin{cases}
u'(s-1) \text{ and terminate}, &\text{{if }} s-1 \in F \text{ \underline{and} }\cv_{s}^{\text{LU}} < \tau_{\min},\\
(\mathrm{T}(s-1)+1,\mathrm{B}(s-1)), &\text{if } B(s-1) = B(s-2)+1, \\  &\text{\underline{or} }\cv_{s}^{\text{LU}} < \tau_{\mathrm{LU}}, \\ 
(\mathrm{T}(s-1),\mathrm{B}(s-1)+1), &\text{otherwise,}
\end{cases}
\end{equation}
for any $s = 1,...,N_{\mathrm{S}^2}$, where $u(0):=(0,1)$ and $u(-1):=(0,0)$. 
If the algorithm terminates for $s < N_{\mathrm{S}^2}$, we define  $N_{\mathrm{S}^2}:=N_{\mathrm{T}}+N_{\mathrm{L}}^{\max}$ and $N_{\mathrm{L}}^{\max} := \mathrm{B}(s-1)$. 
Otherwise, we let $N_{\mathrm{L}}^{\max} := N_{\mathrm{L}}$.
We test the performance the modified update scheme $u'$ in \S\ref{sec:numerics}.

\section{Numerical experiments}\label{sec:numerics}
We consider a steady-state groundwater flow problem on the unit square domain $D = (0,1)^2$. 
The permeability $\kappa(\theta)$ and the hydrostatic pressure $p$ are coupled via the elliptic PDE
\begin{align*}
-\nabla \cdot\left({\kappa(\theta(x)) }\nabla p(x)\right) &= f(x) &(x \in D).
\end{align*}
The source term $f$ and the boundary conditions are specified below.
We observe the pressure at $N_{\text{obs}}$ points $(d_n : n = 1,\ldots,N_{\text{obs}})$ in the domain $D$. 
Thus the observation operator $\mathcal{O}$ maps $p \mapsto (p(d_n) :  n = 1,\ldots,N_{\text{obs}})$. 

This inverse problem is well studied in the literature, see e.g. \cite{Beskos2015SequentialProblems,Dashti2011UncertaintyProblem,Dashti2017TheProblems,Marzouk2009DimensionalityProblems,Richter1981AnEquation,Schillings2017AnalysisProblems}. 
Moreover, in \cite[\S3.7]{Stuart2010InversePerspective} it is proved that Assumptions \ref{Main_Assumptions_Stuart2.6} on the potential are satisfied for this problem.

 {The parameter $\kappa(\theta)$ is a $\log$-normal random field.
In particular, we set $\kappa(\cdot) := \exp(\cdot)$ and assume that the prior distribution of $\theta$ is a Gaussian random field with mean and covariance operator specified below. 
This Gaussian random field is discretised by a truncated KL expansion, which takes the form
\begin{equation}\label{truncKL}
\theta \approx {\theta}_{N_{\mathrm{sto}}} := m_0(x) + \sum_{n=1}^{N_{\mathrm{sto}}} m_n(x)\theta^{\mathrm{KL}}_n,
\end{equation}
where $\theta^{\mathrm{KL}}_1, \dots, \theta^{\mathrm{KL}}_{N_{\mathrm{sto}}}$ denote standard Gaussian random variables.
We generate the true parameter by sampling from the discretised prior random field. 
The observations $y$ are given by the model evaluation of the true parameter plus (additive) Gaussian measurement noise $\eta \sim \mathrm{N}(0, 0.01 \cdot \mathrm{Id})$.}

We consider three estimation problems.

\begin{example} \label{Example1}
Here the pressure on the boundary of $D$ is zero,
\begin{equation*}
p(x) = 0 \ \  (x \in \partial D).
\end{equation*}
The source term $f$ models nine smoothed point sources that are distributed uniformly over the domain:
\begin{equation*}
f(x) = \sum_{n,m=1}^3 \mathrm{N}\left(x_1;0.25n,0.001\right) \mathrm{N}\left(x_2;0.25m,0.001\right),
\end{equation*}
where $\mathrm{N}( \cdot ;E,V)$ is the probability density function of the one-dimensional Gaussian measure with mean $E$ and variance $V$.
The prior random field $\theta \sim \mu_0 = \mathrm{N}(m_0, C_0)$, where $m_0 \equiv 0$ and $C_0$ is the Mat\'{e}rn covariance operator   with correlation length $\lambda=0.65$, smoothness parameter $\nu = 1.5$, and variance $\sigma^2 = 1$. See \cite{Matern1986SpatialVariation,Minasny2005TheVariograms} for details.
The random field $\theta$ is discretised by a truncated KL expansion using the $N_{\mathrm{sto}} := 10$ leading terms which capture $94.5\%$ of the variance. 
Note that for the prior field these random variables are uncorrelated. 
For the posterior field this is not necessarily the case.
However, we only consider the marginals of the posterior distribution.
The action of the operator $G$ is approximated by piecewise linear, continuous finite elements on uniform meshes with {$2\cdot 8^2, 2 \cdot 16^2, 2 \cdot 32^2, 2 \cdot 64^2$ and $2 \cdot 128^2$ triangles}. 
The observation operator $\mathcal{O}$ returns the pressure at 25 points in the spatial domain. 
The 25 points are shown in Figure~\ref{Figure: Measurement-Loc} along with the actual pressure given the true underlying permeability.   
Finally, the covariance operator of the noise is given by the matrix $\Gamma = 0.07^2 \cdot \mathrm{Id}$. 
\end{example}

\begin{example} \label{Example1b}
The inverse problem and its discretisation is the same as in Example~\ref{Example1} but with noise covariance matrix $\Gamma = 0.035^2 \cdot \mathrm{Id}$. 
\end{example}

\begin{example} \label{Example2}
We consider a flow cell problem on $D=(0,1)^2$.
We have flow in the $x_1$-direction and no-flow boundaries along the $x_2$-direction,
\begin{align*}
p(x) &= 0 \ \  &(x \in \{0\}\times[0,1]), \\
p(x) &= 1 \ \  &(x \in \{1\}\times[0,1]), \\
\frac{\partial p}{\partial \vec{n}}(x) &= 0 &(x \in (0,1) \times \{0,1\}).
\end{align*}
Furthermore, the source term $f \equiv 0$.
The prior random field is $\theta \sim \mu_0 = \mathrm{N}(m_0', C_0')$, where $m_0' \equiv 2$ and $C_0'$ is the Mat\'{e}rn covariance operator with correlation length $\lambda=0.1$, smoothness parameter $\nu = 1.5$, and variance $\sigma^2 = 1$.
The random field $\theta$ is discretised by a truncated KL expansion of the form \eqref{truncKL} using the leading $N_{\mathrm{sto}} := 320$ terms which capture $95\%$ of the variance. 
The action of the operator $G$ is approximated by piecewise linear, continuous finite elements on uniform meshes with {$2 \cdot 16^2, 2 \cdot 32^2, 2 \cdot 64^2, 2 \cdot 128^2$ and $2 \cdot 256^2$} triangles.
The measurement locations are uniformly distributed as in Example~\ref{Example1}, however, we use 49 measurements (see Figure~\ref{Figure: Measurement-Loc}).
\end{example} 

\begin{figure}
\centering
\includegraphics[width=0.8\textwidth]{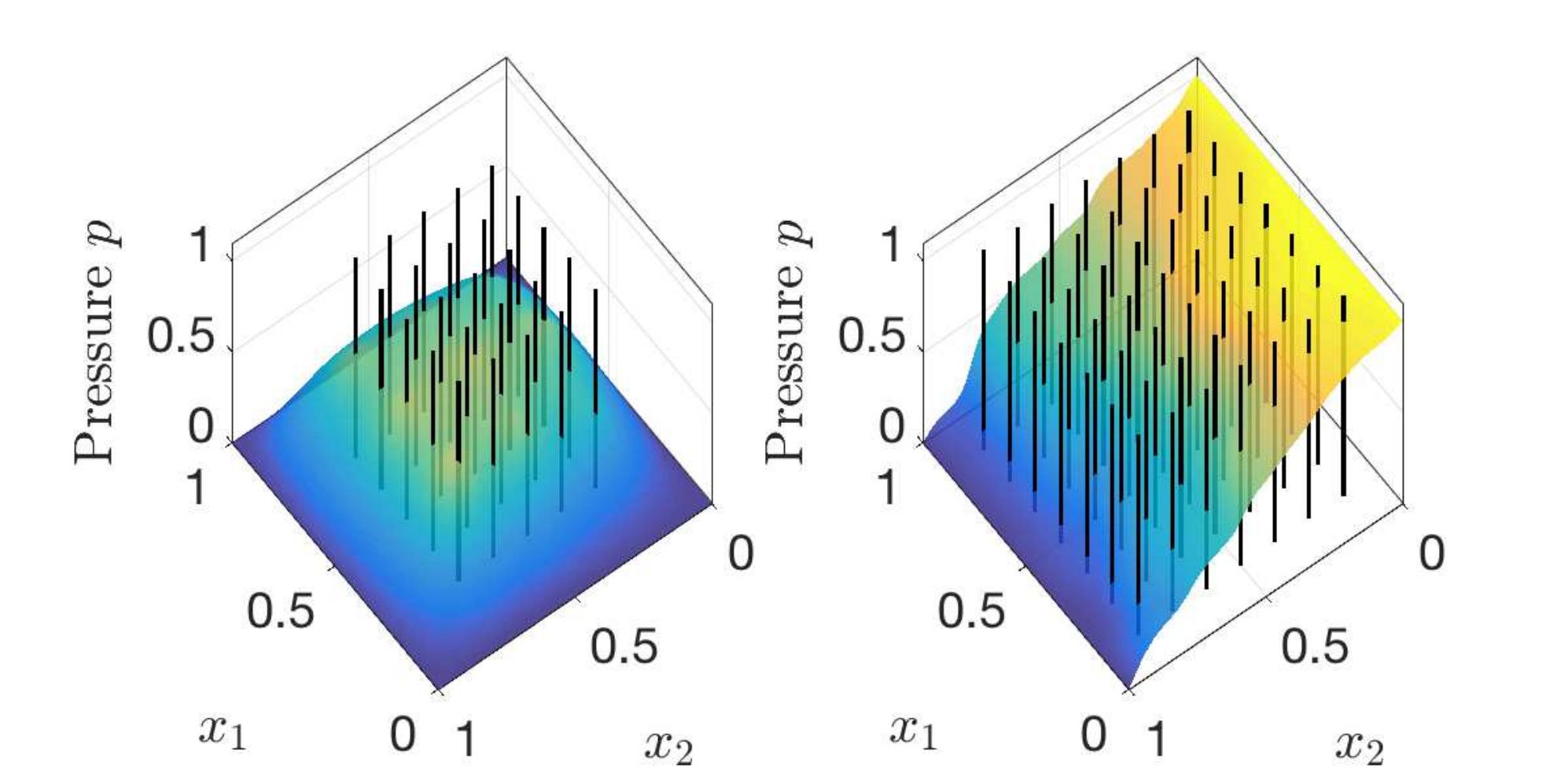}
\caption{Measurement locations and pressure.
The surface plots show the hydrostatic pressure given the true permeability. 
The vertical lines indicate the measurement points. 
On the left: Examples \ref{Example1} and \ref{Example1b}. 
On the right: Example \ref{Example2}.} 
\label{Figure: Measurement-Loc}
\end{figure}

In all examples we test the performance of single-level SMC on the finest mesh (from now on simply `SMC') as well as MLB and MLS${}^2$MC on the given mesh hierarchy.
We observe that in Example~\ref{Example2} the adaptive update scheme of MLS${}^2$MC is identical to the MLB update scheme. 
Interestingly, in Example \ref{Example1b} it is impossible to perform the update $\ell = 1$ to $\ell =2$ with MLB since the probability  measures $\mu^y_{h_1}$ and $\mu^y_{h_2}$ are numerically singular.
We anticipated this situation in \S\ref{MLBoptimal}.

For each of the tests above we consider different numbers of particles and different target values $\tau^\ast$ for the coefficient of variation in the adaptive bridging and tempering updates. 
Furthermore, we choose the maximal discretisation level $N^{\max}_{\mathrm{L}}$ adaptively in Example \ref{Example2}, using the modified update scheme $u'$ in \eqref{eq_updatescheme_final_Lmax}.
The simulation setups are summarised in Table~\ref{Sim_setup_table}.

 \begin{table}[htb]
\centering
\begin{tabular}{|l|ccc|}
\hline
Example   & \ref{Example1}                              & \multicolumn{1}{c}{\ref{Example1b}}          & \multicolumn{1}{c|}{\ref{Example2}}     \\ \hline
$\#$ runs & \multicolumn{3}{c|}{50 per setup}                                                                                                                    \\ \cline{2-4} 
$J$       & \multicolumn{2}{c|}{156, 312, 625, 1250, 2500}                                                 & \multicolumn{1}{c|}{250, 500, 1000, 2000} \\ \cline{2-4} 
$\tau^\ast$    & \multicolumn{2}{c|}{0.5, 1}                                                                    & 1                                         \\ \cline{2-4} 
$\tau_{\mathrm{LU}}$ & \multicolumn{3}{c|}{$\tau^\ast$}                                                                                                                    \\ \cline{2-4} 
$N_{\mathrm{sto}}$    & \multicolumn{2}{c|}{10}                                                                    & 320                                        \\\cline{2-4} 
$h^{-1}$    & \multicolumn{2}{c|}{(8, 16, 32, 64, 128)}                                                                    & (16, 32, 64, 128, 256)                                         \\ \cline{2-4} 
Update scheme    & \multicolumn{2}{c|}{$u$ in \eqref{eq_updatescheme_final}}                                                                    & $u'$ in \eqref{eq_updatescheme_final_Lmax};    $\tau_{\min} = 0.001$                                  \\  \cline{2-4} 
$\Gamma$  & \multicolumn{1}{c|}{$0.07^2 \cdot \mathrm{Id}$} & \multicolumn{1}{c|}{$0.035^2 \cdot \mathrm{Id}$} & $0.045^2 \cdot \mathrm{Id}$                 \\ \hline
\end{tabular}
\caption{Simulation settings}
\label{Sim_setup_table}
\end{table}

All SMC samplers use a Markov kernel.
We choose a single step of a Random Walk Metropolis MCMC sampler with Gaussian proposal density. 
 {The covariance operator of this proposal density is given by $C^{\mathrm{prop}} = \frac{2.38^2}{N_{\mathrm{sto}}}\mathrm{Id}$.}
It remains unchanged for all intermediate measures.
In high dimensions it would be a good idea to employ the preconditioned Crank-Nicholson MCMC sampler, however, we do not implement this here.

\subsection{Zero boundary pressure}

First, we consider the Examples~\ref{Example1} and \ref{Example1b}.
Recall that the solution of a BIP is the posterior measure.  
The mean of the posterior measure is the best unbiased point estimator of the true underlying parameter in the $\mathcal{L}^2$-sense. 
See \cite{Matthies2016InverseSetting} for details on conditional expectations and their properties.
It is important to note that unbiasedness refers only to the stochastic approximation.
The discretised PDE solution introduces a bias compared to the exact PDE solution.
For this reason we measure the approximation accuracy of the posterior measure and also the accuracy of the  posterior mean when used as point estimator. 
In addition, for each sequential sampler we compare the estimated model evidences and the associated computational costs.

\subsubsection{Posterior mean} \label{Subsec:Posteriormean_asapoint_estimator}
We consider synthetic data and thus the true (spatially varying) parameter $\theta_{\mathrm{true}}$ is known.
$\theta_{\mathrm{true}}$ is identical in Examples~\ref{Example1} and \ref{Example1b} (their setup differs only in the noise covariances).
Note that $\theta_{\mathrm{true}}$ is generated using the truncated KL expansion in \eqref{truncKL}. 
Hence the KL truncation error is not included in our experiments.
In the top row of Figure~\ref{Figure_Mean_RandomFields} we plot $\theta_{\mathrm{true}}$ together with typical posterior means estimated with SMC, MLB, and MLS$^{2}$MC, respectively.
In the bottom row of Figure~\ref{Figure_Mean_RandomFields} we plot the corresponding hydrostatic pressure. 
We observe that SMC and MLS${}^2$MC give similar results.
In contrast, the estimate delivered by MLB differs (visually) from the SMC estimate. 
We discuss this below.

\begin{figure}[htb]
\centering
\includegraphics[width=0.99\textwidth]{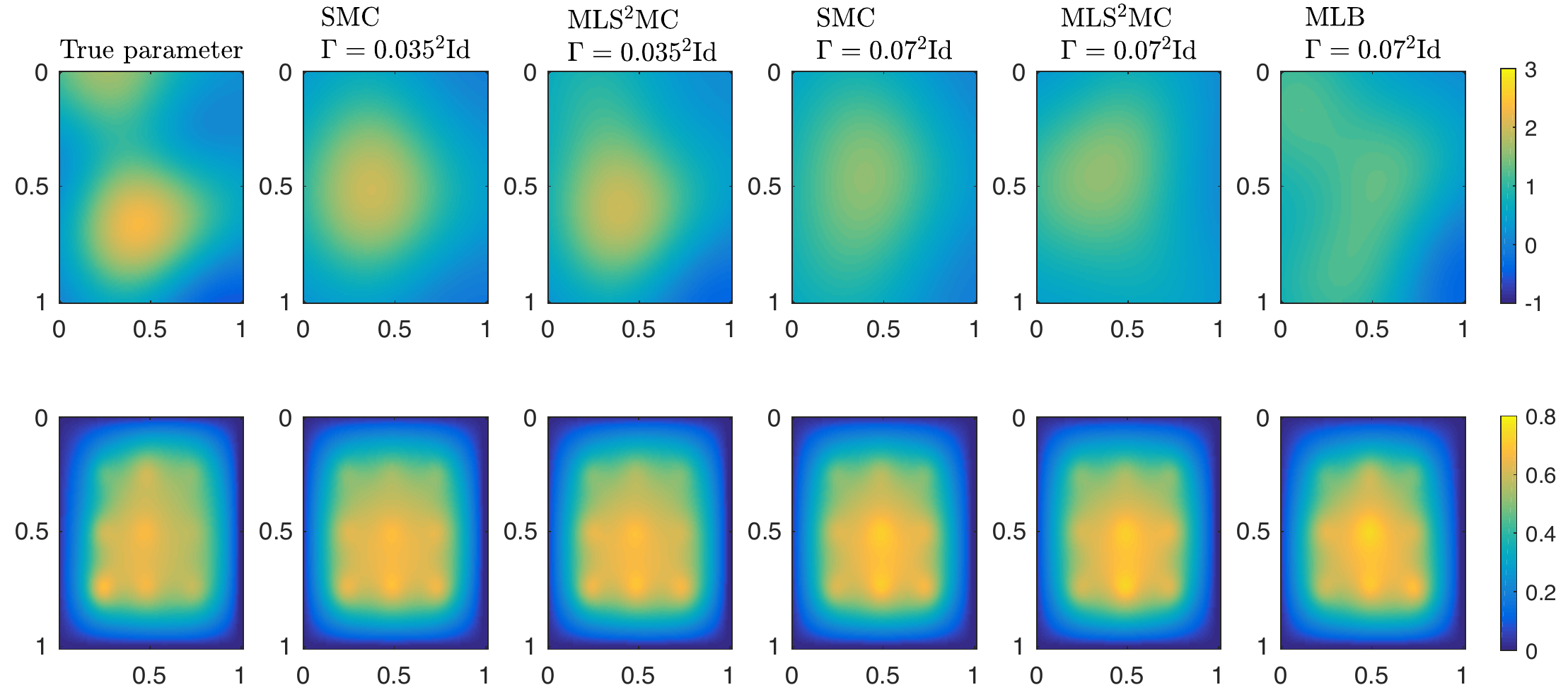}
\caption{Top row: The true underlying log-permeability and various posterior mean estimates based on $J = 1250$ particles and $\tau = 0.5$ for Examples~\ref{Example1} and \ref{Example1b}.
Bottom row: The hydrostatic pressure corresponding to the log-permeability in the top row.}
\label{Figure_Mean_RandomFields}
\end{figure}

Now we evaluate the posterior mean estimates more systematically, and quantitatively. 
We use the following error metric:
\begin{equation}\label{RelErr}
\mathrm{RelErr}(\widehat{\boldsymbol{\theta}}^{\mathrm{KL}},\boldsymbol{\theta}_{\mathrm{true}}^{\mathrm{KL}}) 
:=  
\|\Lambda^{1/2}(\widehat{\boldsymbol{\theta}}^{\mathrm{KL}}-\boldsymbol{\theta}_{\mathrm{true}}^{\mathrm{KL}})\|_1/\|\Lambda^{1/2} \boldsymbol{\theta}_{\mathrm{true}}^{\mathrm{KL}}\|_1,
\end{equation}
where $\widehat{\boldsymbol{\theta}}^{\mathrm{KL}} \in \mathbb{R}^{10}$ is the estimate of the posterior mean (column) vector and  $\boldsymbol{\theta}_{\mathrm{true}}^{\mathrm{KL}} \in \mathbb{R}^{10}$ is the (column) vector of the true parameter values.  
The (row) vector  $\Lambda^{1/2} := (\lambda_1^{1/2},\dots,\lambda_{10}^{1/2})$ contains the square roots of the 10 leading KL eigenvalues. 
Hence, the error measure is a weighted $\mathcal{\ell}^1$ distance, where we weigh the particles according to their contribution in the KL expansion.
We plot the results in Figure~\ref{Figure:rel_error:poster_mean_true}. 
As expected the estimation quality is better for a smaller noise level, consistently for all methods.
We see that SMC is the most accurate method, while MLS${}^2$MC performs slightly worse than SMC, and MLB performs slightly worse than  MLS${}^2$MC.  
This is more pronounced for small numbers of particles $J$ and a relatively large coefficient of variation $\tau^\ast=1$.  
The results are consistent with the fact that in every importance sampling update we introduce a sampling error. 
A large number of updates gives a large sampling error. 
The number of updates is minimal in SMC and maximal in MLB.
Hence we expect SMC to give a better estimation result compared to MLB.
The estimates obtained with MLS${}^2$MC are similar to the estimation results of SMC.
Overall, these experiments confirm our motivation for MLS${}^2$MC given in \S\ref{sec:ML}.

\begin{figure}[htb]
\centering
\subfloat[Example \ref{Example1}: $\Gamma = 0.07^2\mathrm{Id}$]{
\includegraphics[width=0.495\textwidth]{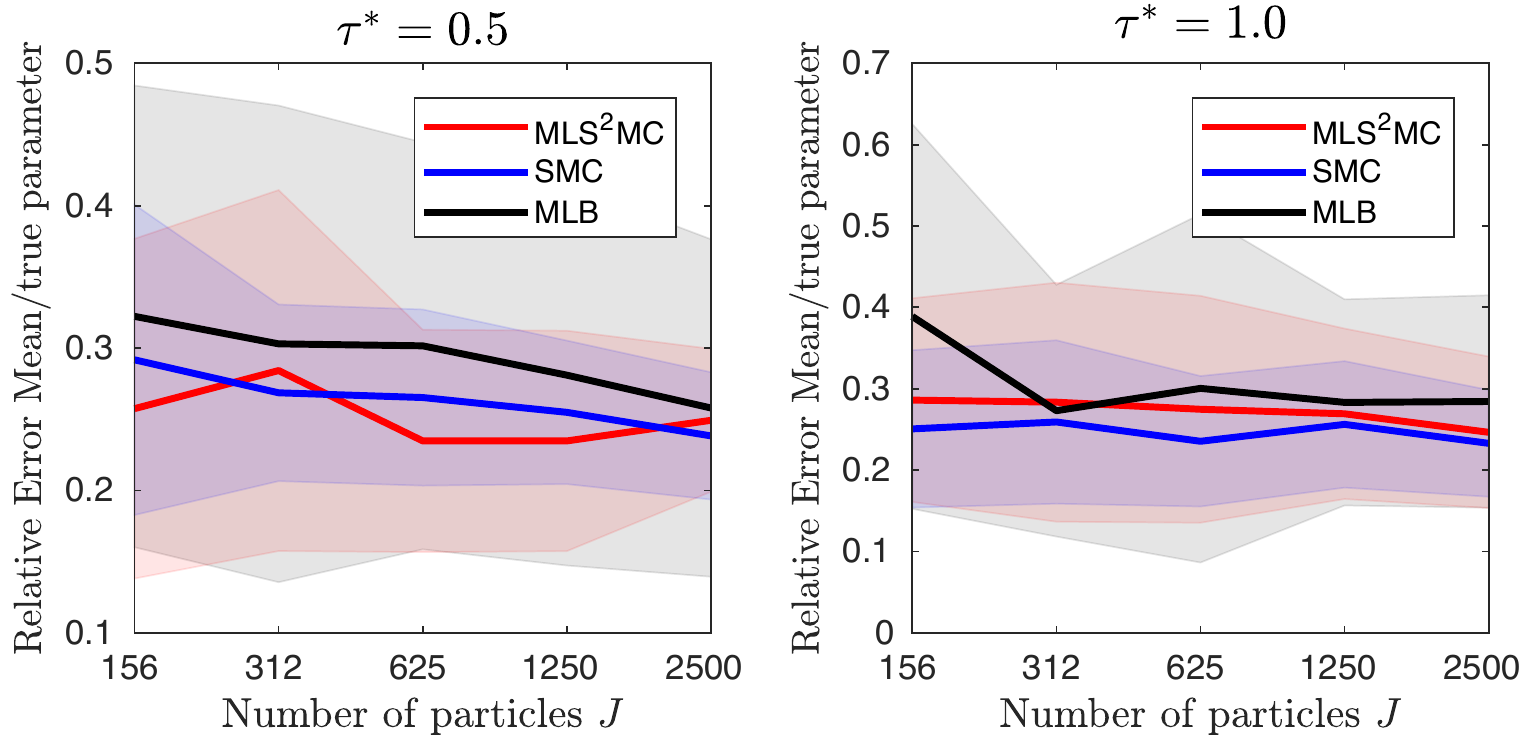}}
\subfloat[Example \ref{Example1b}:  $\Gamma = 0.035^2\mathrm{Id}$]{
\includegraphics[width=0.495\textwidth]{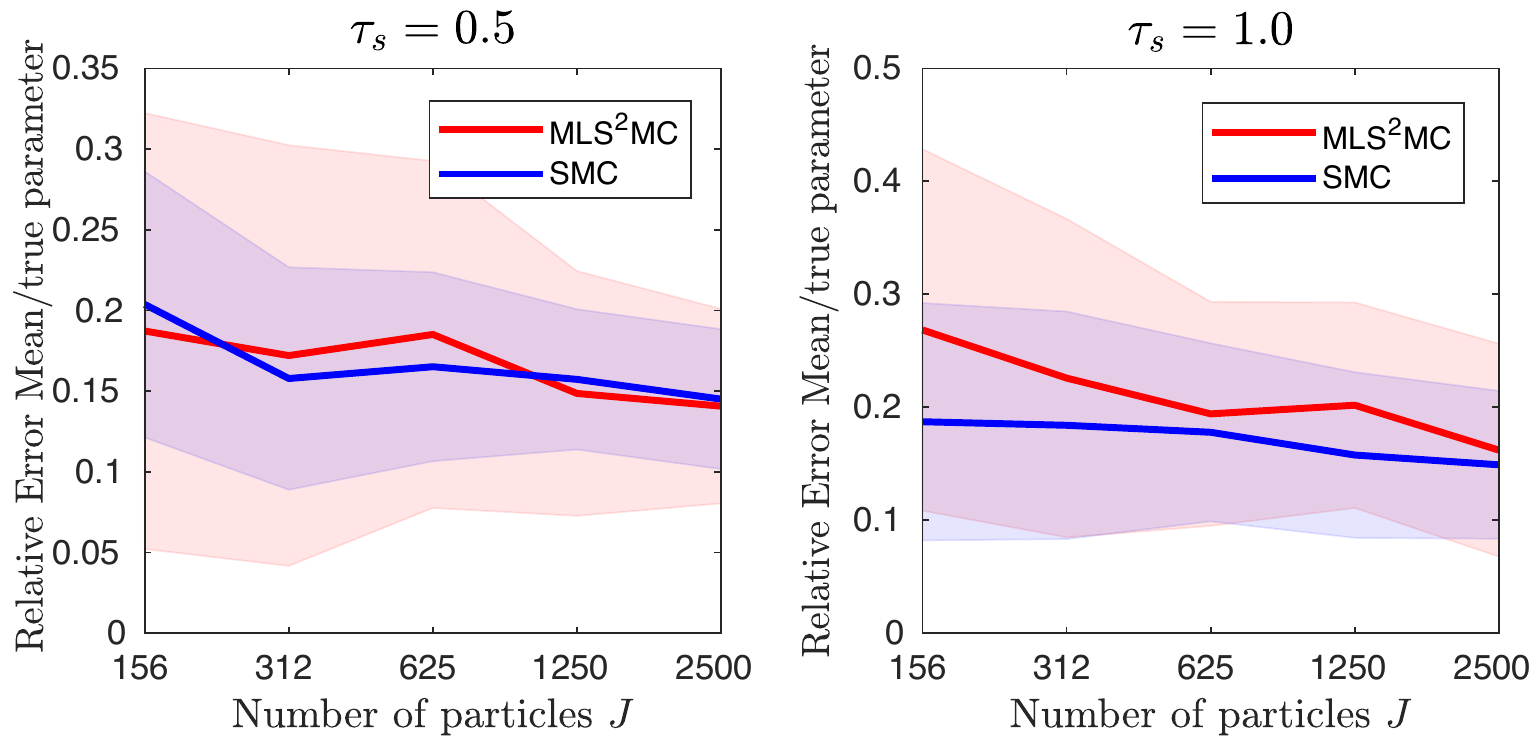}}
\caption{$\mathrm{RelErr}$ of the posterior mean estimate compared to the true parameter. 
The bold lines show the sample mean of the error taken over 50 runs.
The shaded areas show the associated standard deviation, again taken over 50 runs.} 
\label{Figure:rel_error:poster_mean_true}
\end{figure}

Next we consider the misfit of the (discretised) model output $\mathcal{G}(\widehat{\boldsymbol{\theta}}^{\mathrm{KL}})$ and the observed data:
\begin{equation}\label{RelMisfit}
\mathrm{RelMisfit}({\widehat{\boldsymbol{\theta}}^{\mathrm{KL}}}) := \|\Gamma^{-1/2}(y-\mathcal{G}_{h_{N_{\mathrm{L}}}}({\widehat{\boldsymbol\theta}}^{\mathrm{KL}}))\|^2_2/\|\Gamma^{-1/2}y\|^2_2.
\end{equation}
We plot the relative misfit in Figure~\ref{Figure:rel_error:rel_misfit}. 
As expected we do not observe significant differences for the two noise levels since the noise precision $\Gamma^{-1}$ cancels in the relative expression.
For all methods we see that the misfit is reasonably small.
Hence the posterior mean estimate is a good approximation to the maximum a posterior (MAP) estimator.

\begin{figure}[htb]
\centering\subfloat[Example \ref{Example1}: $\Gamma = 0.07^2\mathrm{Id}$]{
\includegraphics[width=0.495\textwidth]{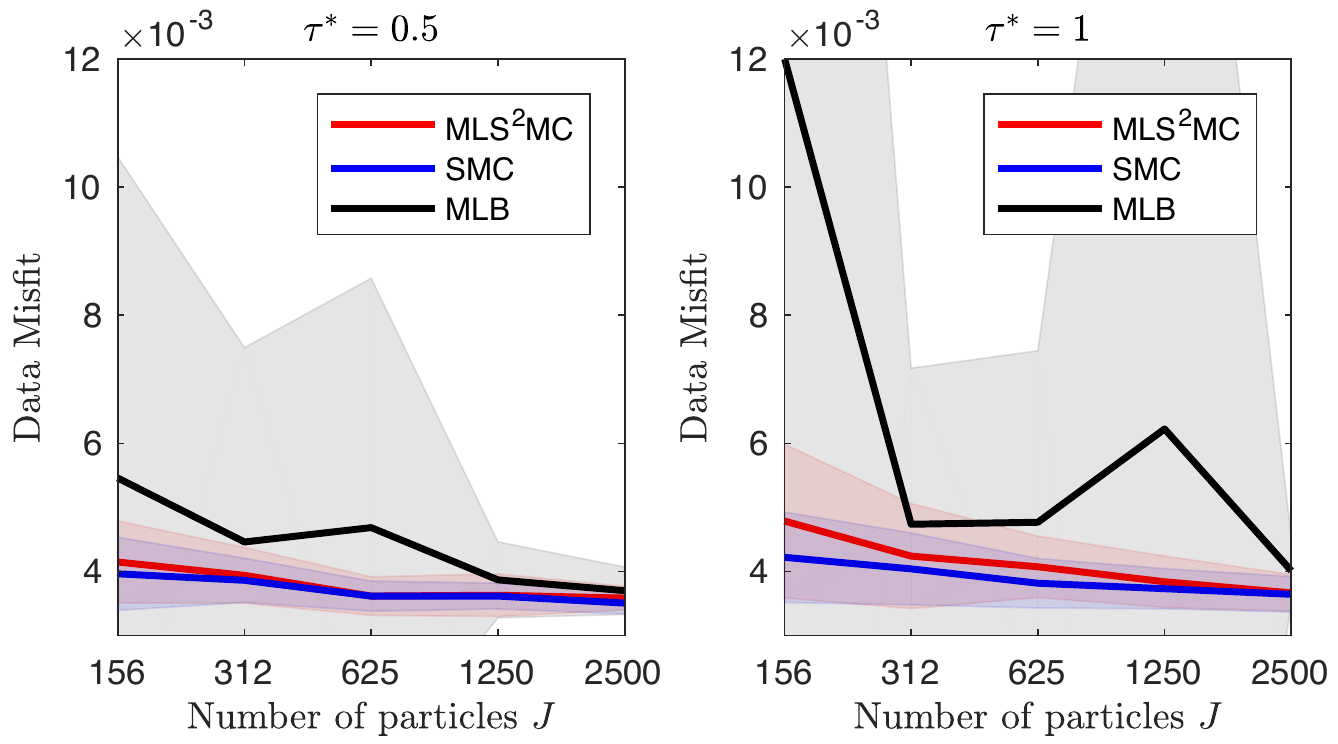}}
\subfloat[Example \ref{Example1b}:  $\Gamma = 0.035^2\mathrm{Id}$]{
\includegraphics[width=0.495\textwidth]{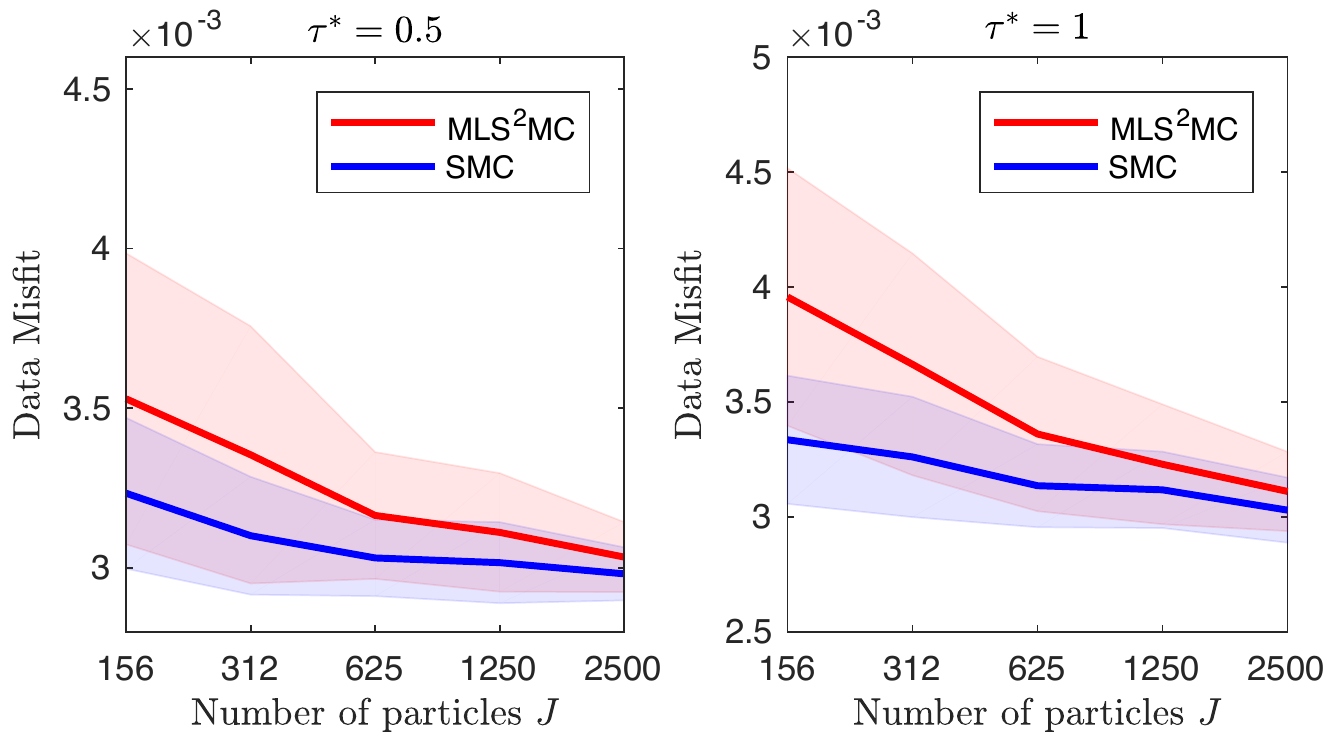}}
\caption{$\mathrm{RelMisfit}$ of the posterior mean estimate compared to the observations $y$.
The bold lines show the sample mean of the error taken over 50 runs.
The shaded areas show the associated standard deviation, again taken over 50 runs.} 
\label{Figure:rel_error:rel_misfit}
\end{figure}

\subsubsection{Posterior measure} \label{Subsubsec:Approximating_Post_measure}
Now we only consider the leading three KL random variables $\theta_1^{\mathrm{KL}}, \theta_2^{\mathrm{KL}}$ and $\theta_3^{\mathrm{KL}}$. 
These parameters capture 76\% of the variance of the prior random field. 
In Figure \ref{Figure:ecdf_examples_firstKLmode} we plot the empirical cumulative distribution functions (ecdfs) of $\theta_1^{\mathrm{KL}}$ for representative simulations in Example \ref{Example1} and \ref{Example1b}.

\begin{figure}[htb]
\centering
\subfloat[Example \ref{Example1}]{\includegraphics[width=0.495\textwidth]{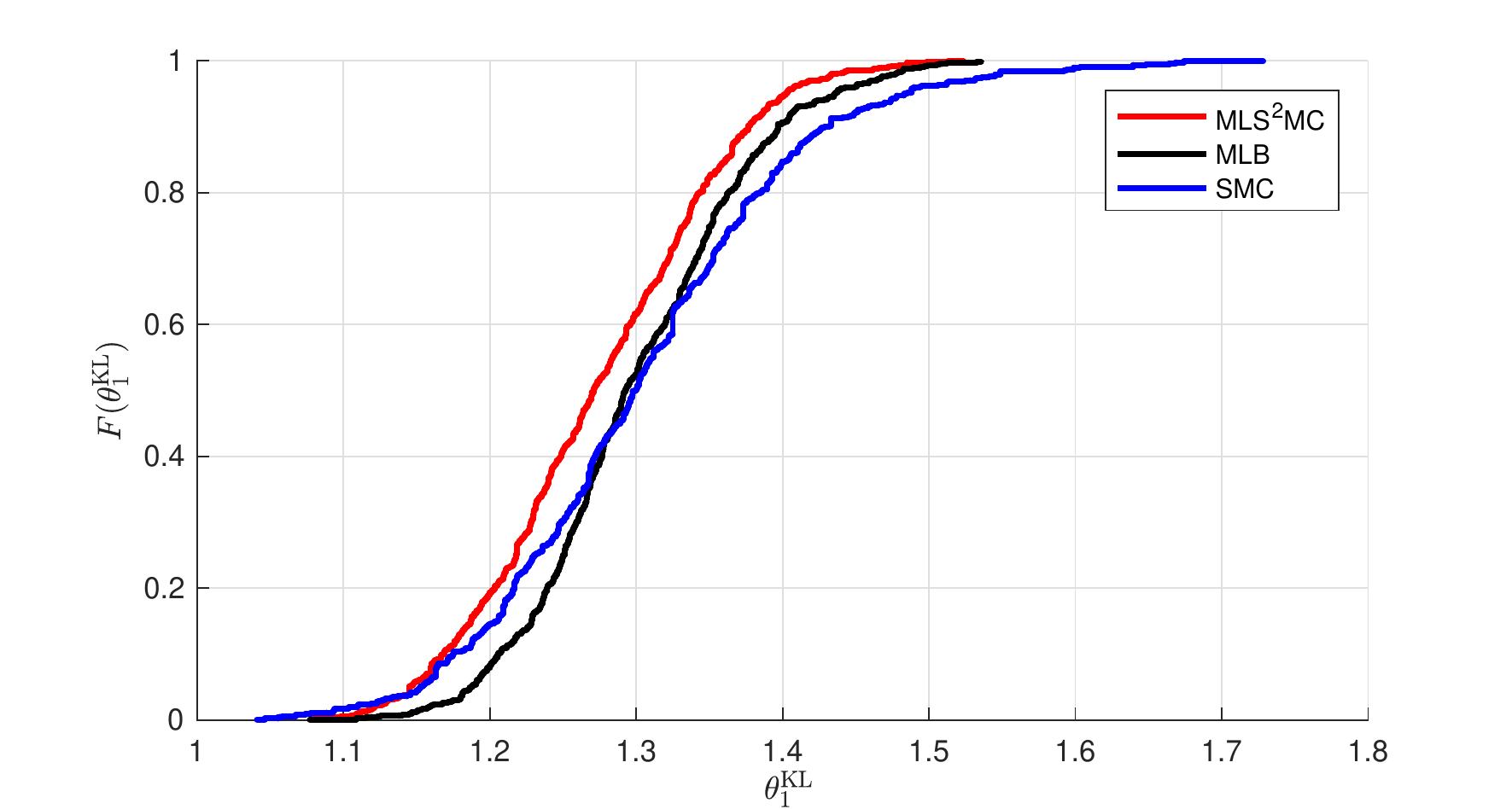}} 
\subfloat[Example \ref{Example1b}]{\includegraphics[width=0.495\textwidth]{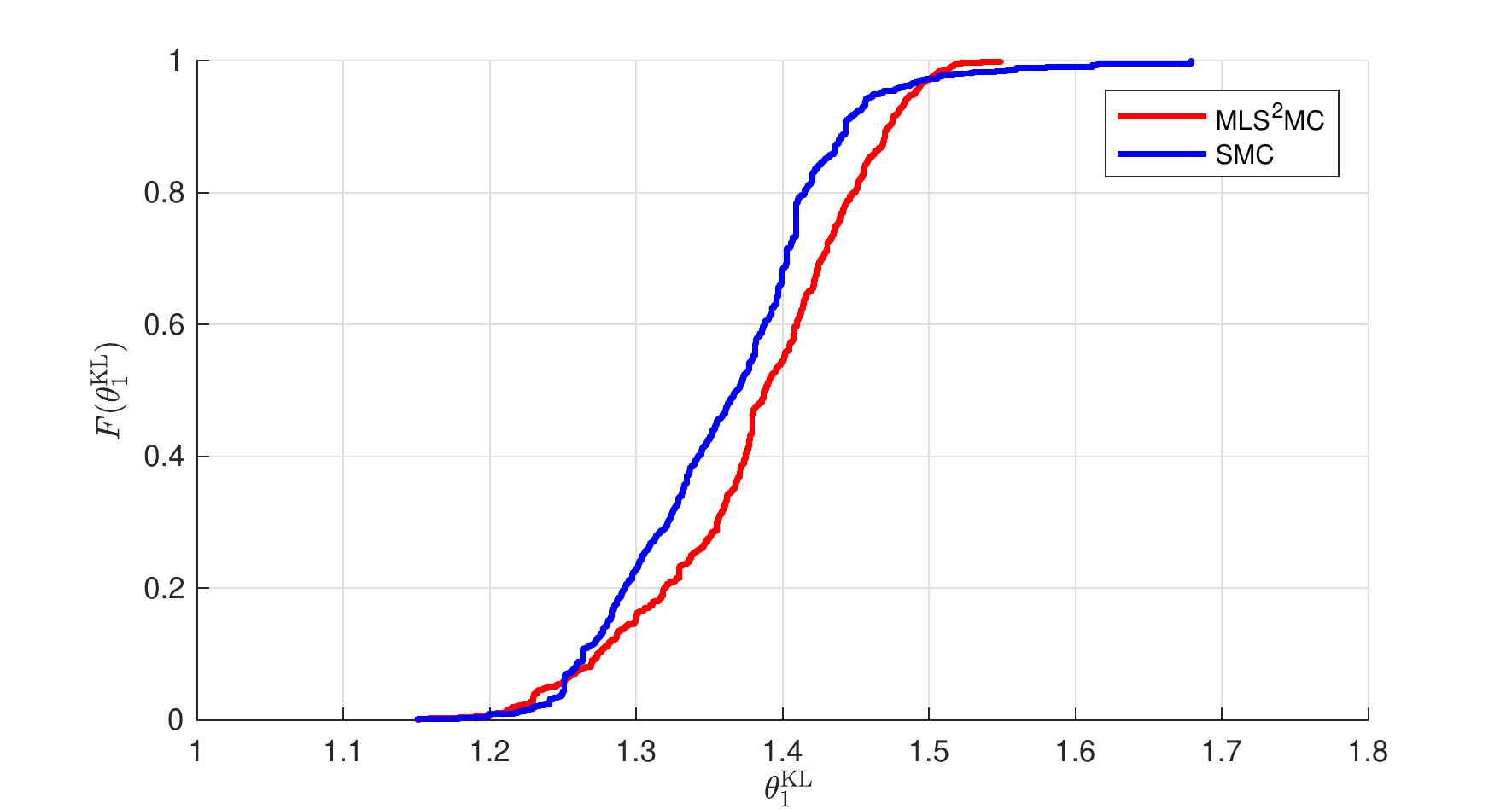}}
\caption{Empirical cumulative distribution function of the posterior measure of the leading KL random variable estimated with $J=1250$ particles and $\tau^\ast = 0.5$.} \label{Figure:ecdf_examples_firstKLmode}
\end{figure} 

We assess the accuracy of the posterior measure approximations produced by MLS${}^2$MC and MLB by comparing it with the associated (single-level) SMC method, using the same values for $J$ and  $\tau^\ast$.  
We compute the \textit{Kolmogorov-Smirnoff (KS) distance}\footnote{
The KS distance has several applications in statistics.
It is often used to compare two discrete probability  measures or a continuous and a discrete probability  measure. 
For example, the KS distance is the test statistic used in the Kolmogorov-Smirnoff test. 
See \cite{Daniel1990AppliedStatistics} for details.
}  of all $50 \cdot 50 = 2500$ pairs of simulations of (MLS${}^2$MC, SMC) and (MLB, SMC), respectively. 

We plot the sample means and standard deviations of the 2500 KS distances of the leading three KL random variables in Figure~\ref{Figure_KS-distances}. 
Since we expect some scattering within the reference SMC approximation itself we also show the $2500$ KS distances within the SMC simulations. 
This line can be used as base line to account for the intrinsic scattering within the stochastic methods.
The results are similar to the observations we made for the posterior mean approximation in the previous subsection.
In Example~\ref{Example1} there is no significant difference between SMC and MLS${}^2$MC. 
MLB performs slightly worse; we suspect that this is again caused by the larger number of intermediate importance sampling updates.
In Example~\ref{Example1b} we observe a larger discrepancy of the approximate posterior measures compared to Example~\ref{Example1}.

\begin{figure}[hbt]
\centering
\subfloat[Example \ref{Example1}: $\Gamma = 0.07^2\mathrm{Id}$]{
\includegraphics[width = 0.495\textwidth]{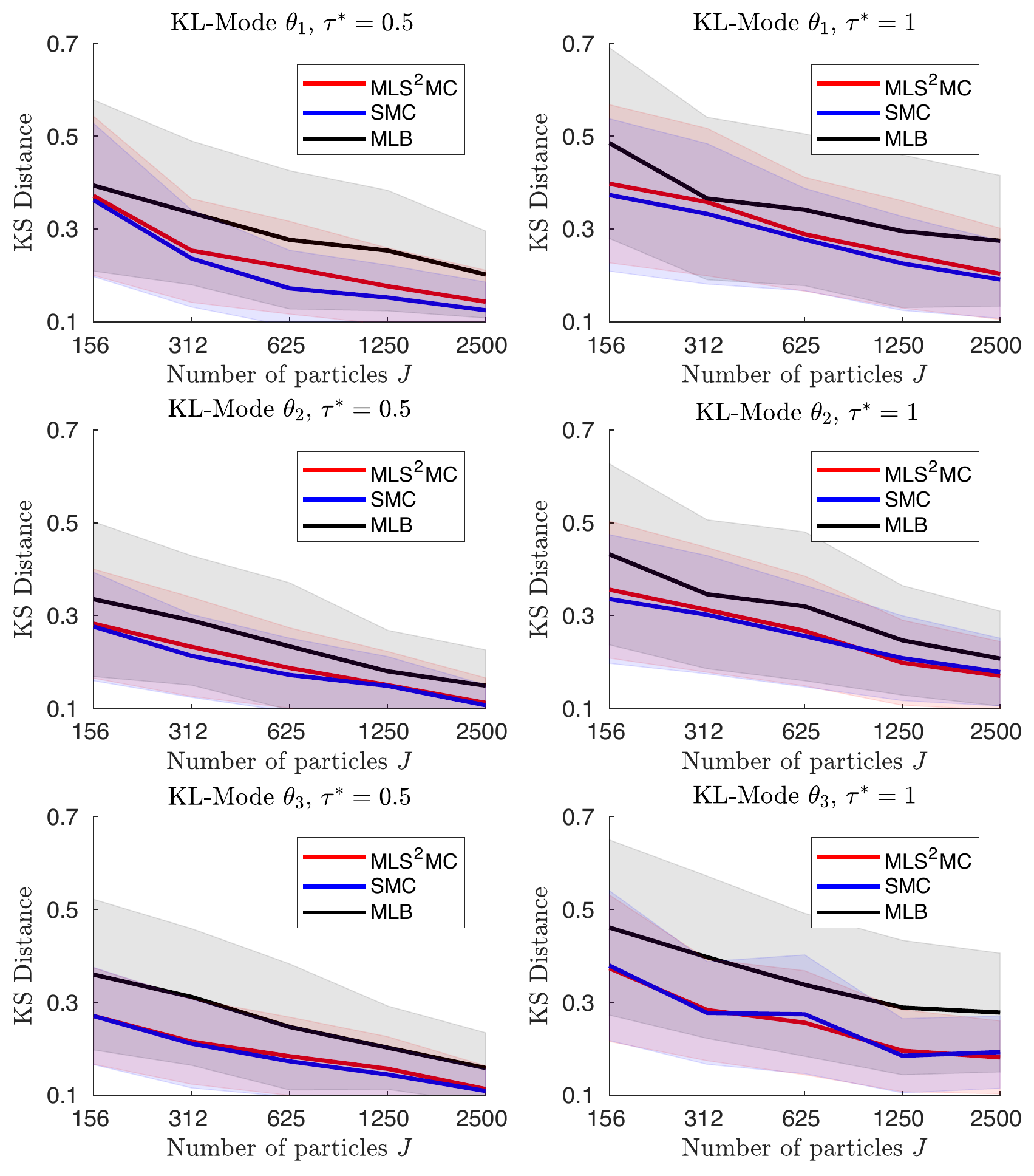}} 
\subfloat[Example \ref{Example1b}: $\Gamma = 0.035^2\mathrm{Id}$]{
\includegraphics[width = 0.495\textwidth]{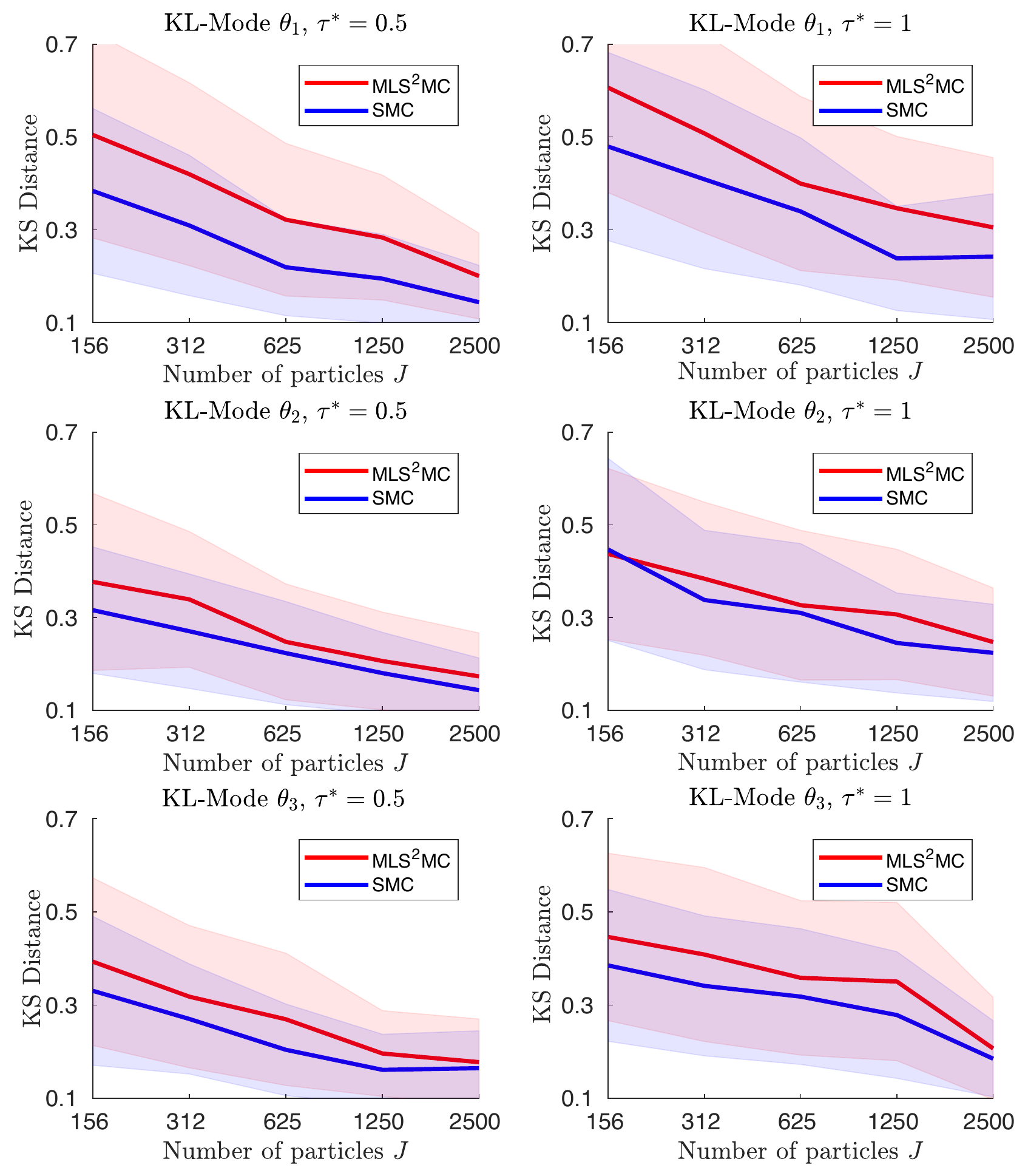}}
\caption{KS distances of posterior measure approximations. 
The bold lines show the sample means of KS distances of $50 \cdot 50$ combinations of SMC and either MLS${}^2$MC, MLB or SMC.
The shaded areas show the associated standard deviations.} 
\label{Figure_KS-distances}
\end{figure}

\subsubsection{Model evidence}

Every SMC-type method delivers automatically an estimate $\widehat{Z}$ of the model evidence $Z_y$ in \eqref{evidence}. This is the normalising constant of the  Radon-Nikodym derivative of the posterior w.r.t. the prior. See \cite{DelMoral2006SequentialSamplers,Gelman1998SimulatingSampling,Neal2005EstimatingSampling} for details.

$\widehat{Z}$ is a random variable, and in each simulation run of SMC, MLB or MLS$^2$MC we obtain a realisation of it.
We plot the ecdfs for 50 runs of SMC, MLB and MLS$^2$MC each in Figure \ref{Figure_ECDF_Evidence}. 
Note that the random variable $\widehat{Z}$ is a biased estimator for the model evidence, due to the adaptivity of the algorithm, see \cite{Beskos2016OnMethods}. 

In addition, we compute the distance of $\widehat{Z}$ to a reference solution $Z^{\mathrm{ref}}$ given by the geometric mean of 50 estimates produced by single-level SMC. 
We consider the geometric mean since the model evidence is a prefactor. 
For the same reason we consider the $\log$ of the model evidence rather than the model evidence itself from now on.  
We use the error metric 
\begin{equation*}
\mathrm{RelErrEvid}(\widehat{Z}, Z^{\mathrm{ref}}) 
:= 
\|\log(\widehat{Z})-\log(Z^{\mathrm{ref}})\|_1/\|\log(Z^{\mathrm{ref}})\|_1.
\end{equation*}
Again we compare the SMC  estimates with the reference solution to obtain a base value for the dispersion within the stochastic algorithms. 
The results are given in Figure~\ref{Figure_Relerror_Evidence}. 
We see that MLB gives poor estimates of the model evidence compared to SMC and MLS${}^2$MC. 
This is consistent with the results for the KS distances of the posterior measures of $\theta_1^{\mathrm{KL}},\theta_2^{\mathrm{KL}}$, and $\theta_3^{\mathrm{KL}}$ where MLB produced significantly different approximations compared to SMC and MLS${}^2$MC.

\begin{figure}[hbt]
\centering
\subfloat[Example \ref{Example1}]{\includegraphics[width=0.495\textwidth]{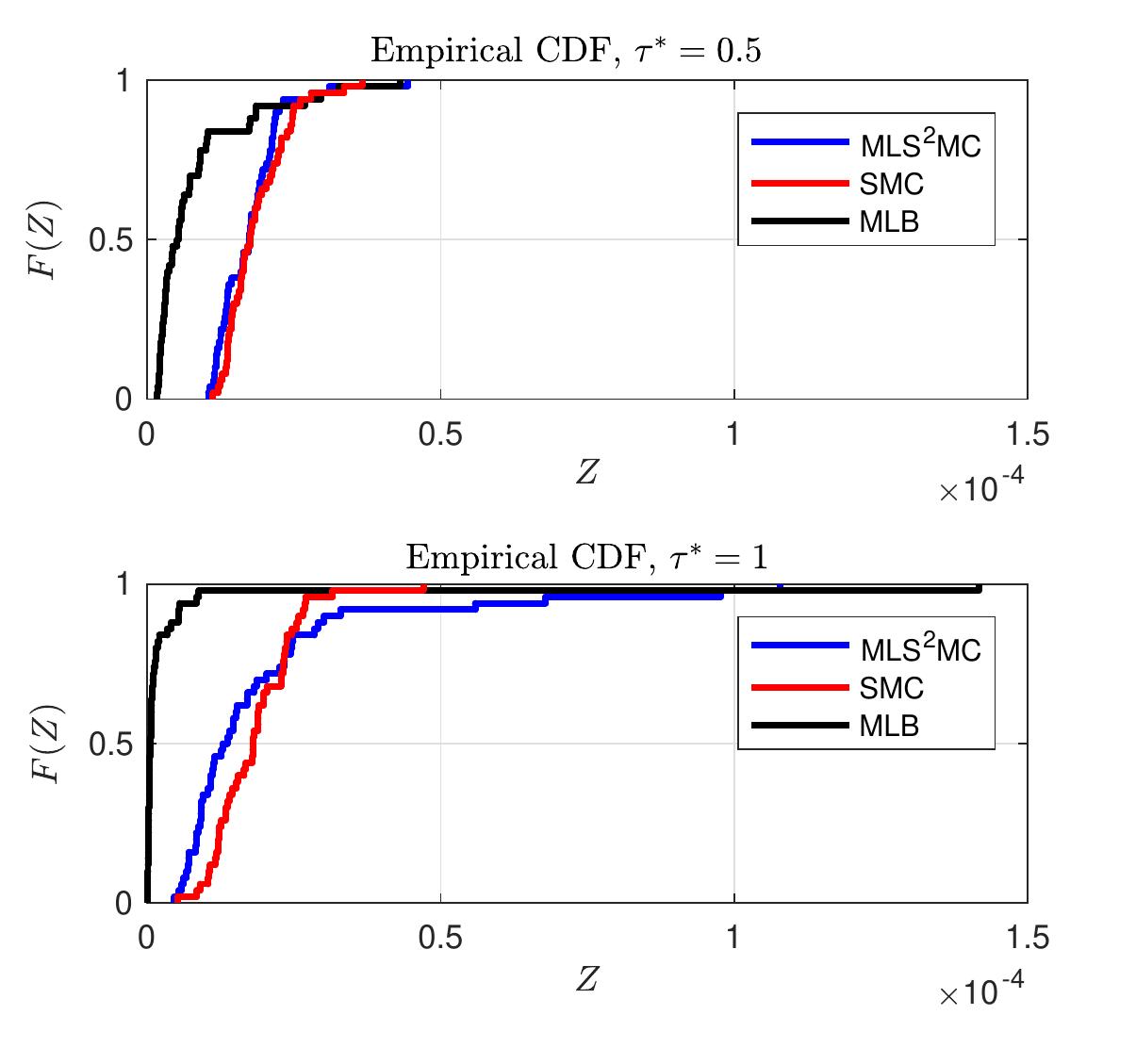}}
\subfloat[Example \ref{Example1b}]{\includegraphics[width=0.495\textwidth]{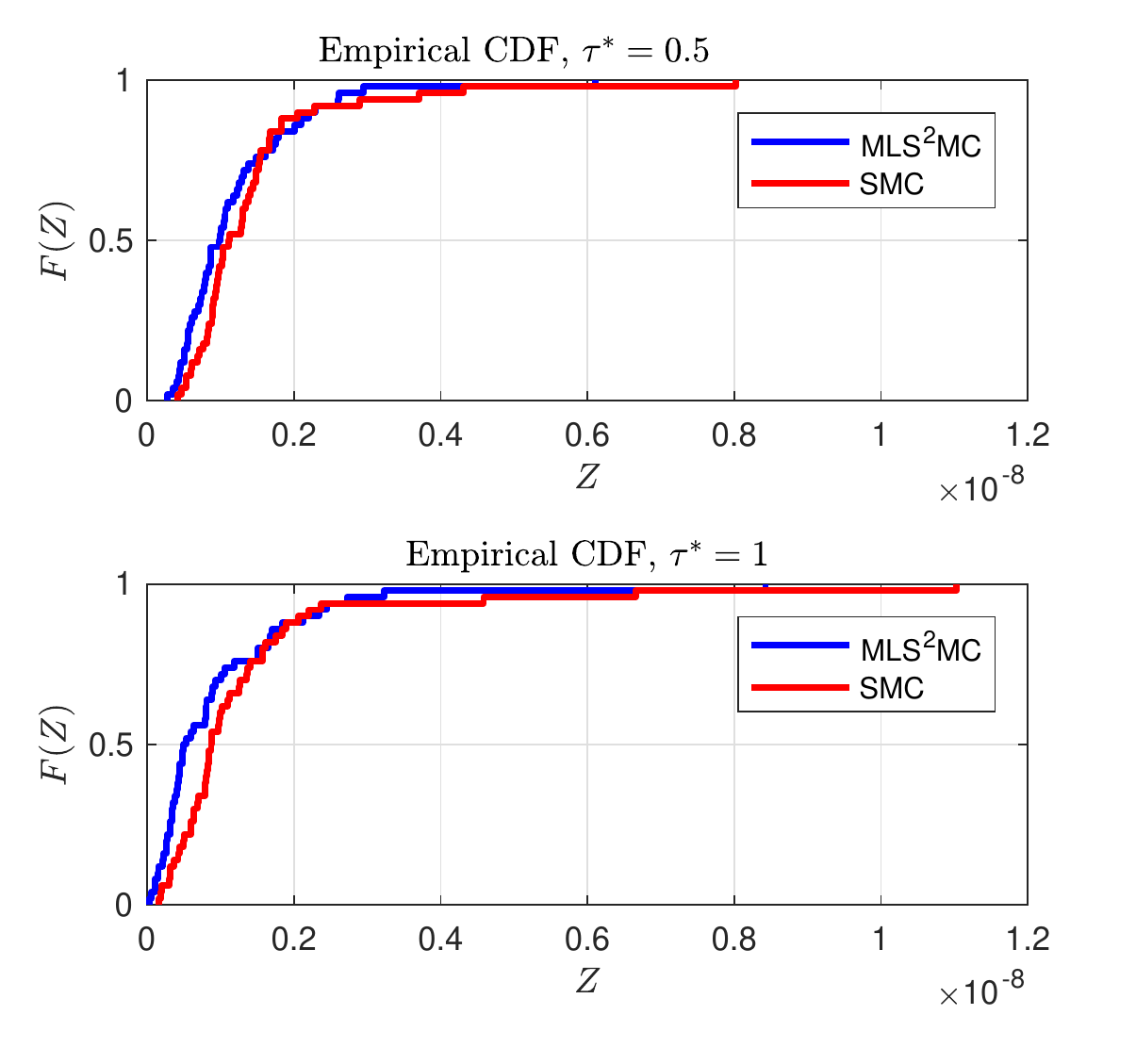}}
\caption{Empirical cumulative distribution functions the model evidences of the 50 posterior measures, each computed with $J=2500$ particles.} 
\label{Figure_ECDF_Evidence}
\end{figure}

\begin{figure}[htb]
\centering
\subfloat[Example \ref{Example1}]{\includegraphics[width=0.495\textwidth]{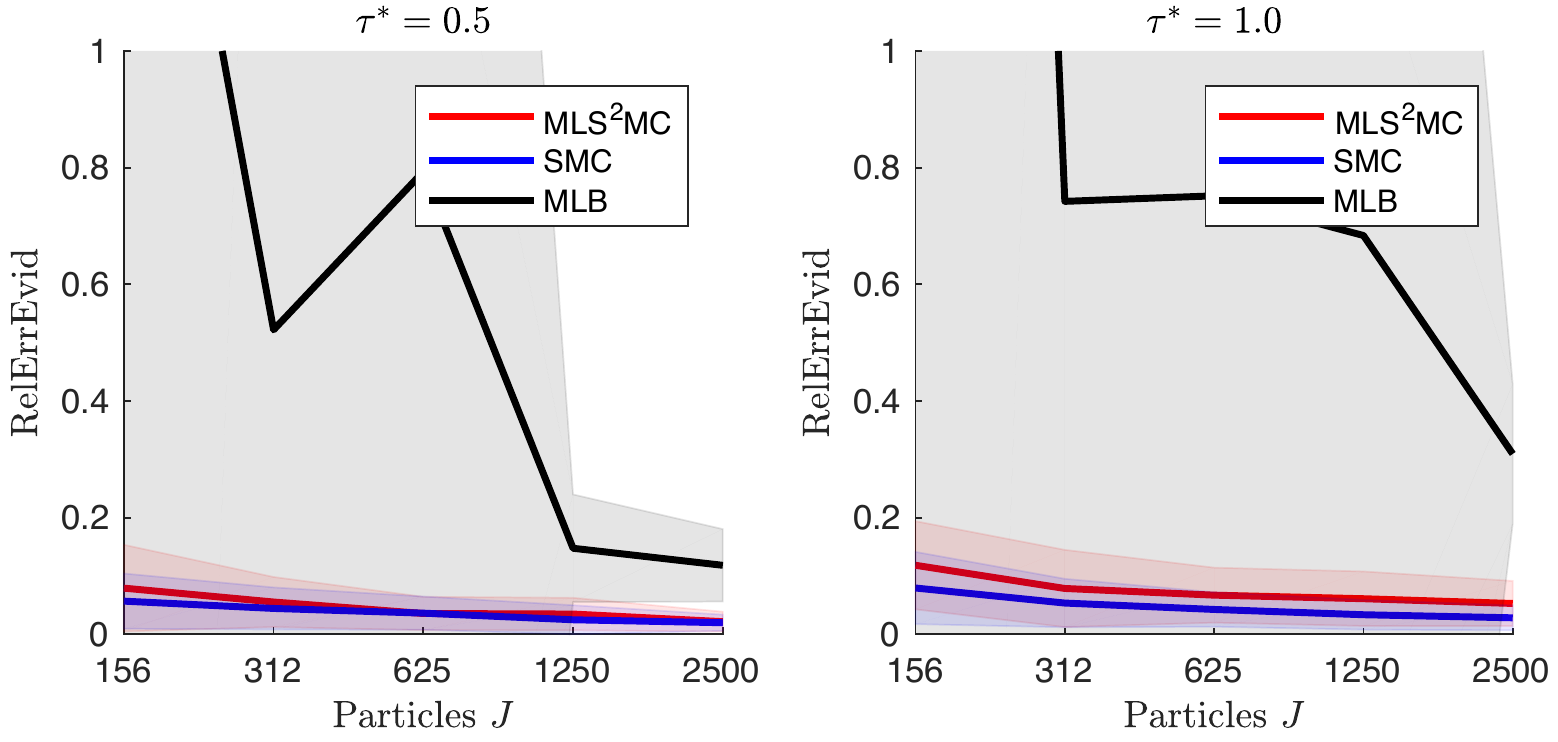}}
\subfloat[Example \ref{Example1b}]{\includegraphics[width=0.495\textwidth]{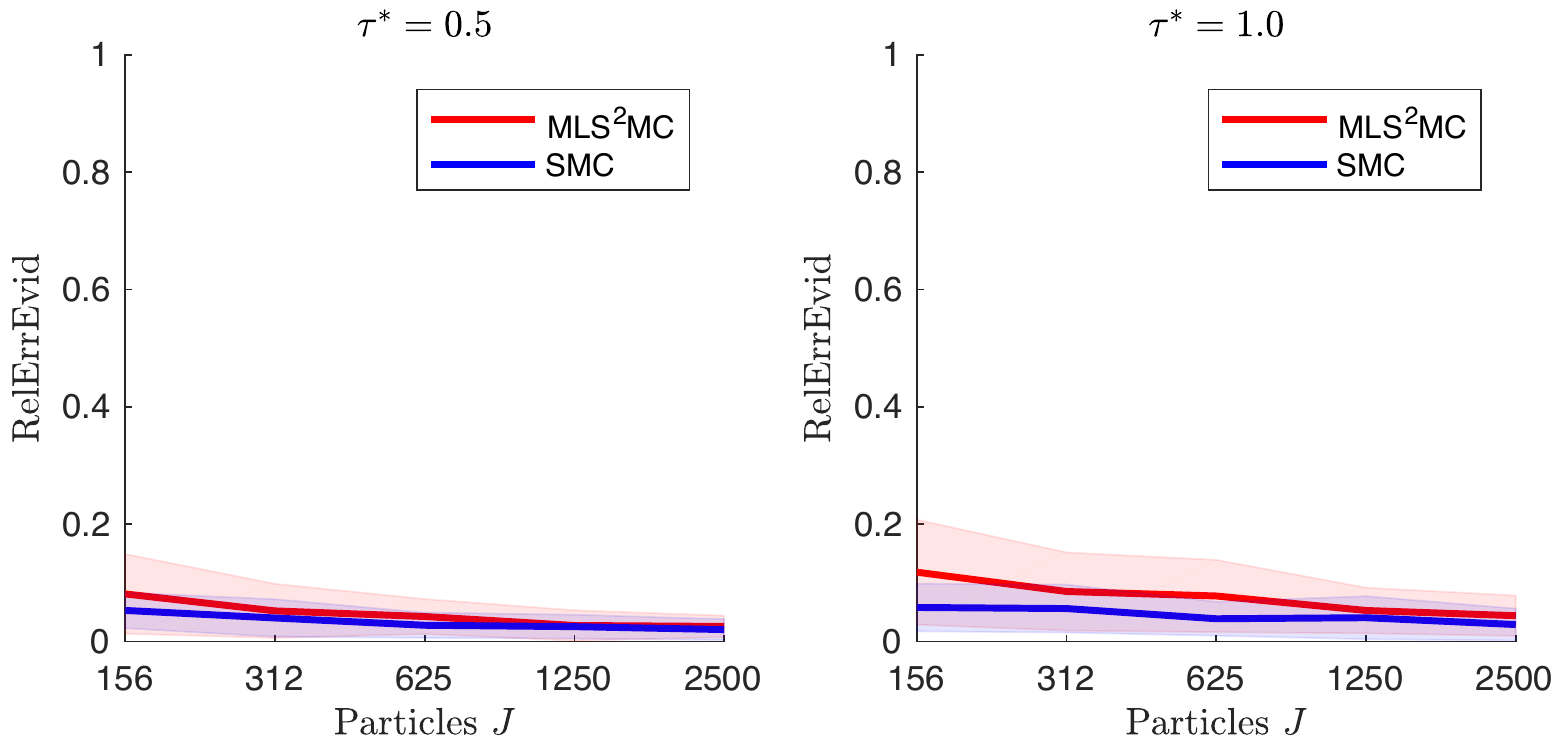}}
\caption{Relative error of the estimated model evidences. The bold lines show the sample mean of the error taken over 50 experiments.
The shaded areas show the associated standard deviation, again taken over 50 runs.} 
\label{Figure_Relerror_Evidence}
\end{figure}

\subsubsection{Adaptive Update Scheme}
In MLS${}^2$MC we apply the adaptive update scheme introduced in \S\ref{Subsec:Adaptive_Update_Schemes}. 
We always use $\tilde{J} := 100$ particles to predict the number of intermediate bridging steps.
In Figure~\ref{Figure_updateschemes} we present realisations of the adaptive update scheme.
Note that these are realisations of the schematic sketch in Figure~\ref{Figure_ComparisonPlot}. 

We observe that the first discretisation level $\ell=1$ is very inaccurate. 
In all realisations the update scheme leaves this level with a very small inverse temperature. 
This might be the reason why Multilevel Bridging performs poorly here. 
Indeed, given the inverse temperature $\beta = 1$, the bridging from $\ell = 1$ to $\ell = 2$ requires many intermediate bridging steps.
This in turn induces a large sample error in MLB as observed throughout this section. 
Since the evaluation of $\mathcal{G}_{h_1}$ and $\mathcal{G}_{h_2}$ is cheap the influence on the computational cost of MLB is negligible.

Observe that for $\tau^\ast = 0.5$ the algorithm might choose to go to $\ell = 3$ before arriving at the maximal inverse temperature $\beta = 1$. 
For $\tau^\ast = 1.0$ the algorithm goes to $\beta = 1$ first, before moving to the discretisation level $\ell = 3$. 
We anticipated this situation.
In the first case, for a small value of $\tau^\ast$, the algorithm is more conservative, meaning that the level updates are performed early.
This strategy increases the accuracy but also the computational cost of the method. 
The path selected for the larger value $\tau^\ast = 1.0$ is computationally cheaper, however, it might give a larger sampling error. 
Note that we do in fact observe a larger error in the examples where $\tau^{\ast} = 1.0$. 

\begin{figure}[htb]
\centering
\subfloat[Example \ref{Example1}]{\includegraphics[width=0.495\textwidth]{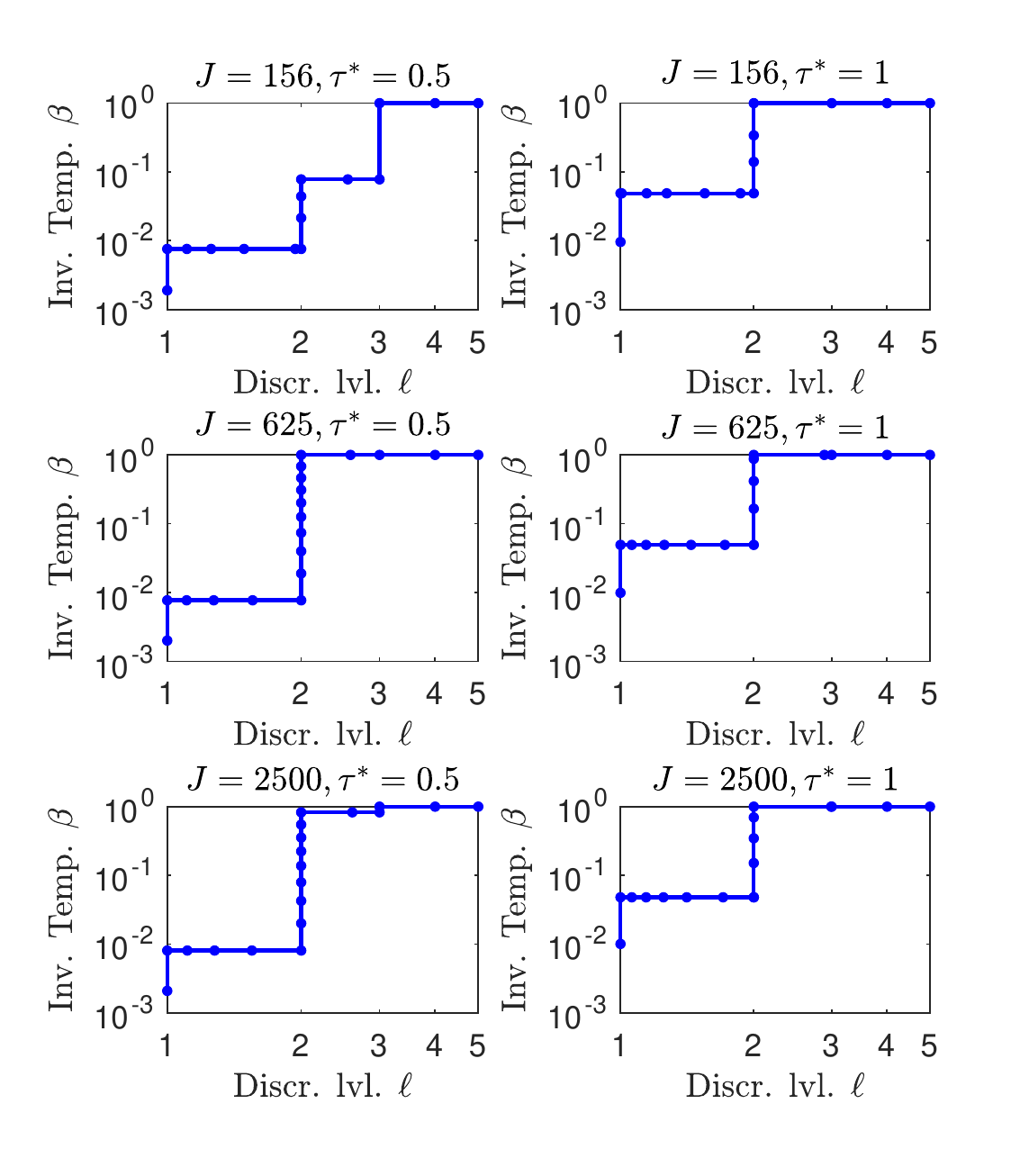}}
\subfloat[Example \ref{Example1b}]{\includegraphics[width=0.495\textwidth]{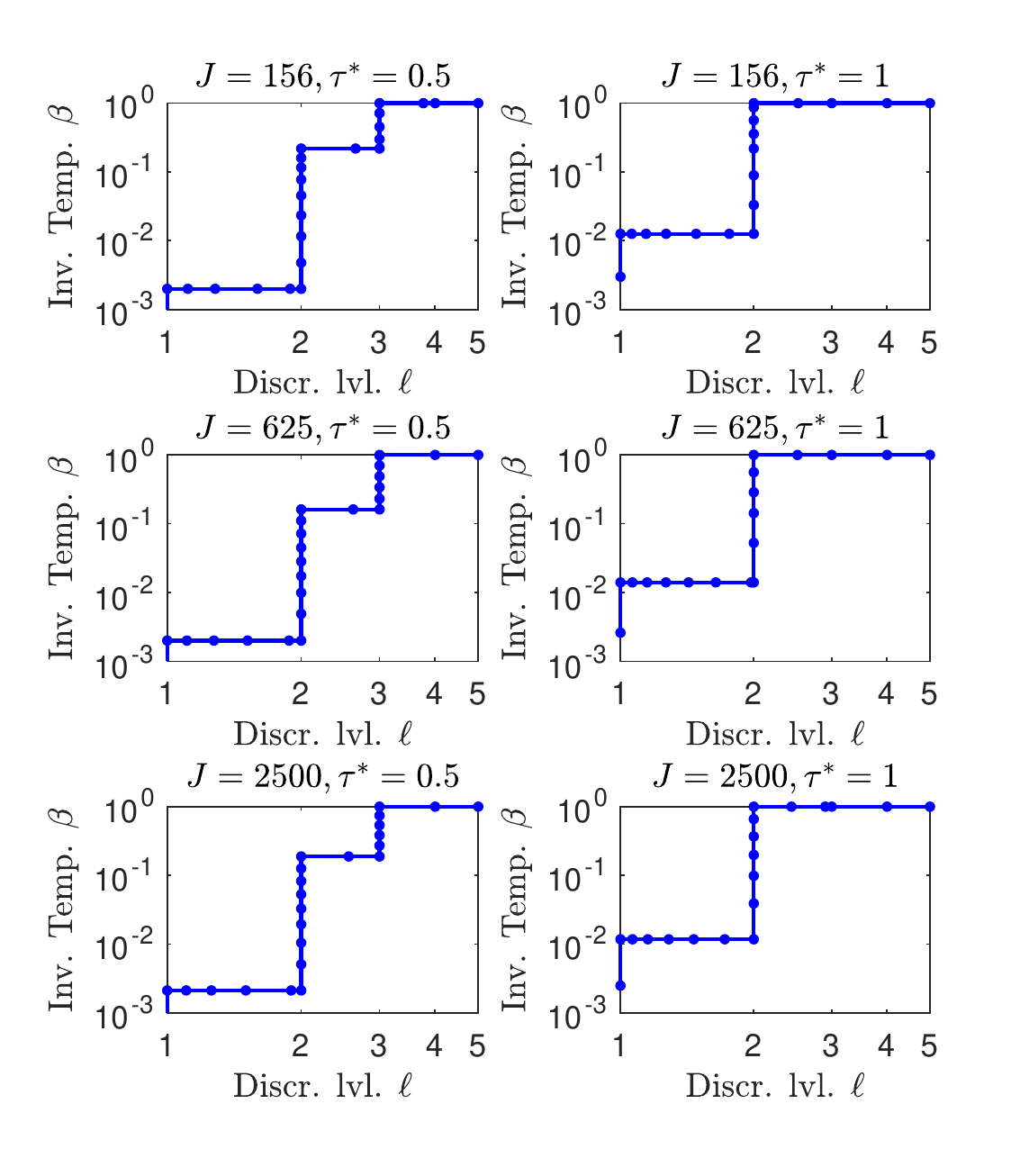}}
\caption{Realisations of the adaptive update scheme \eqref{eq_updatescheme_final} within the MLS${}^2$MC algorithm. 
Each dot corresponds to one intermediate probability  measure.}
\label{Figure_updateschemes}
\end{figure}

\subsubsection{Computational Cost} \label{Sec:Compu_Cost_Ex1.1b}
Our implementation of the SMC-type samplers and the finite element approximation is not optimized.
For these reasons we compare the computational cost in terms of floating point operations, and not in terms of the elapsed time. 
The cost of a single evaluation of $\mathcal{G}_{h_{\ell}}$ is
\begin{equation*}
\mathcal{C}_\ell := 2^{2(\ell-5)}, \quad \ell = 1,\dots,5.
\end{equation*}

This is motivated in Example~\ref{Examples_CostEvaluations} where we take $d=2$.
In Figure~\ref{Figure_comput_cost} we plot $\mathcal{C}_\ell$ against the number of particles $J$.
As expected, the cost scales linearly in $J$. 
If $J$ is fixed, then we observe a speed-up of factor 4 for both MLB and MLS${}^2$MC compared to single-level SMC.
Increasing the discretisation level by one unit increases the cost by a factor of 4 in single-level SMC.
Hence, using either of the multilevel methods gives us one discretisation level more for the same computational cost as single-level SMC. 
However, in the preceding sections we observed that the MLS${}^2$MC samplers are more accurate compared to MLB.
In Figure \ref{Figure_cost_vs_accuracy} we compare computational cost and accuracy directly.
{We measure the accuracy in terms of the relative error of the model evidence.}
Given the relatively large $\tau^* = 1.0$, the additional stochastic error that is introduced in MLS${}^2$MC outweighs the advantages in terms of computational cost.
For $\tau^* = 1.0$ we see that MLS${}^2$MC is not as accurate as SMC, because in MLS${}^2$MC we perform a much larger number of intermediate update steps.
On the contrary, for the smaller value $\tau^\ast = 0.5$ and a fixed accuracy of the estimator,  MLS${}^2$MC is strictly cheaper than SMC.
Overall, this demonstrates the advantages of MLS${}^2$MC in terms of both cost and accuracy.

\begin{figure}[htb]
\centering
\subfloat[Example \ref{Example1}]{\includegraphics[width=0.495\textwidth]{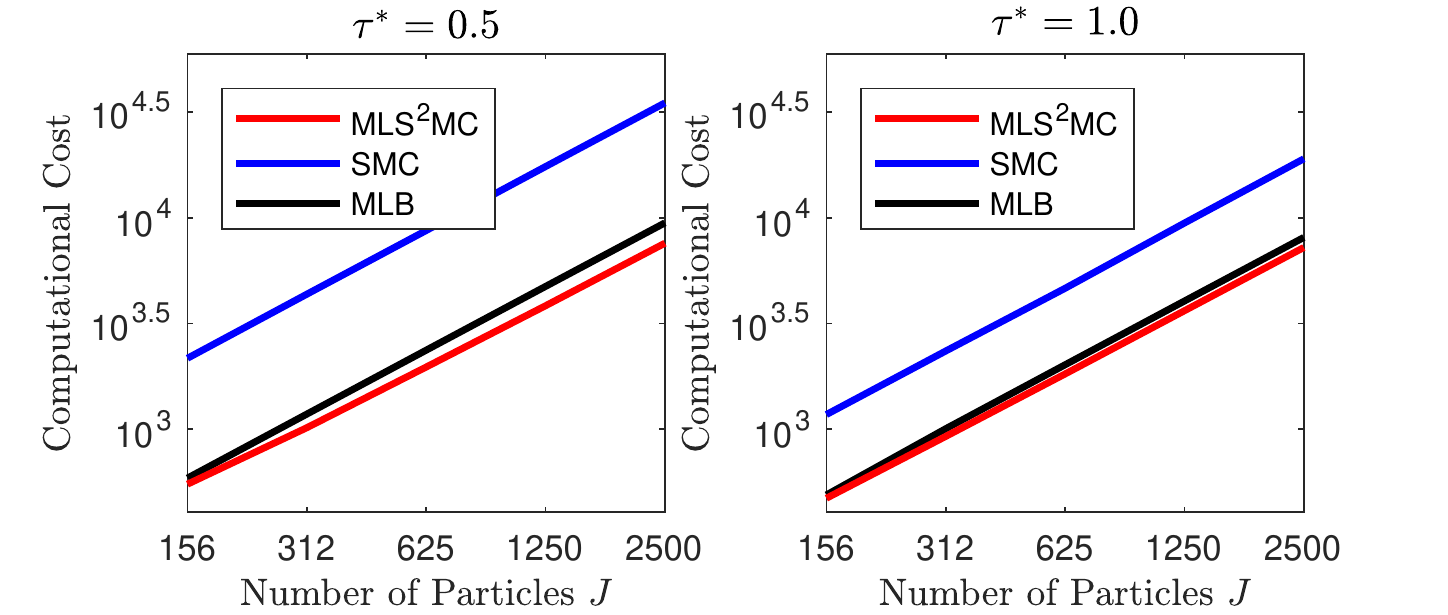}}
\subfloat[Example \ref{Example1b}]{\includegraphics[width=0.495\textwidth]{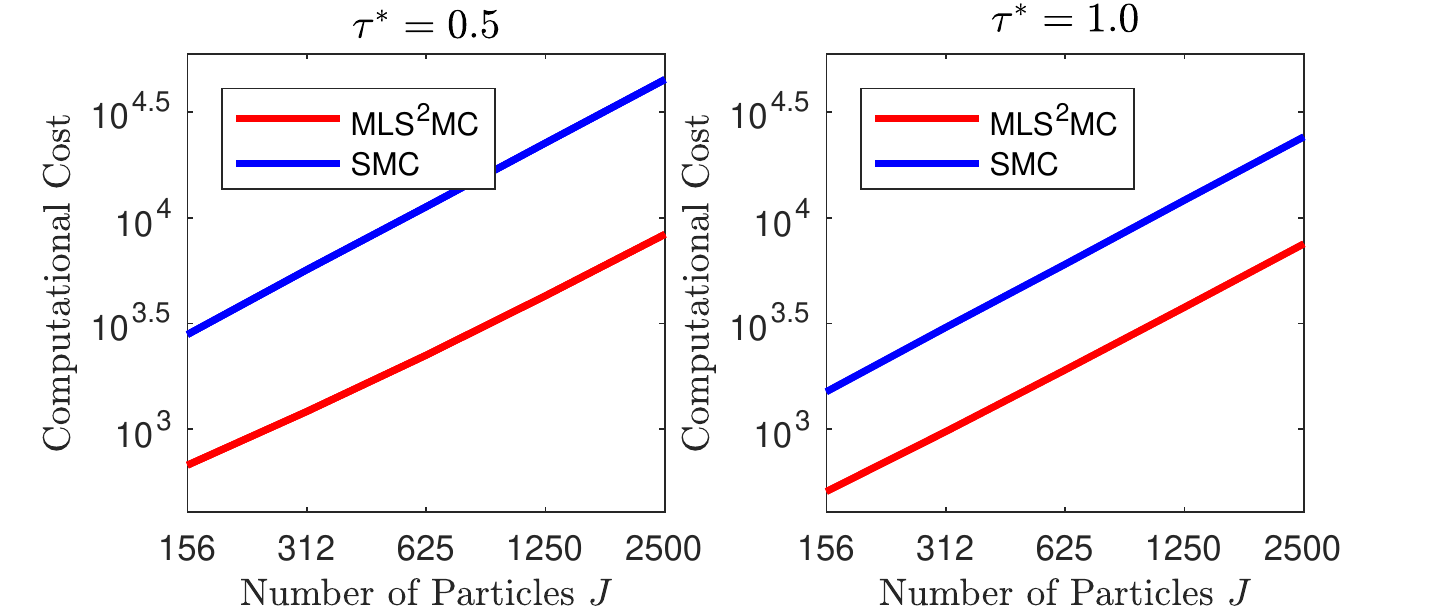}}
\caption{Computational cost of the SMC-type samplers. Each of the bold lines represents the mean computational cost throughout 50 simulations. The costs are measured in terms of the theoretical number of floating point operations per PDE solve on the given discretisation level. These costs are normalised such that $\mathcal{C}_{{N_{\mathrm{L}}}} = 1$.}  
\label{Figure_comput_cost}
\end{figure}

\begin{figure}[htb]
\centering
\subfloat[Example \ref{Example1}]{\includegraphics[width=0.4\textwidth]{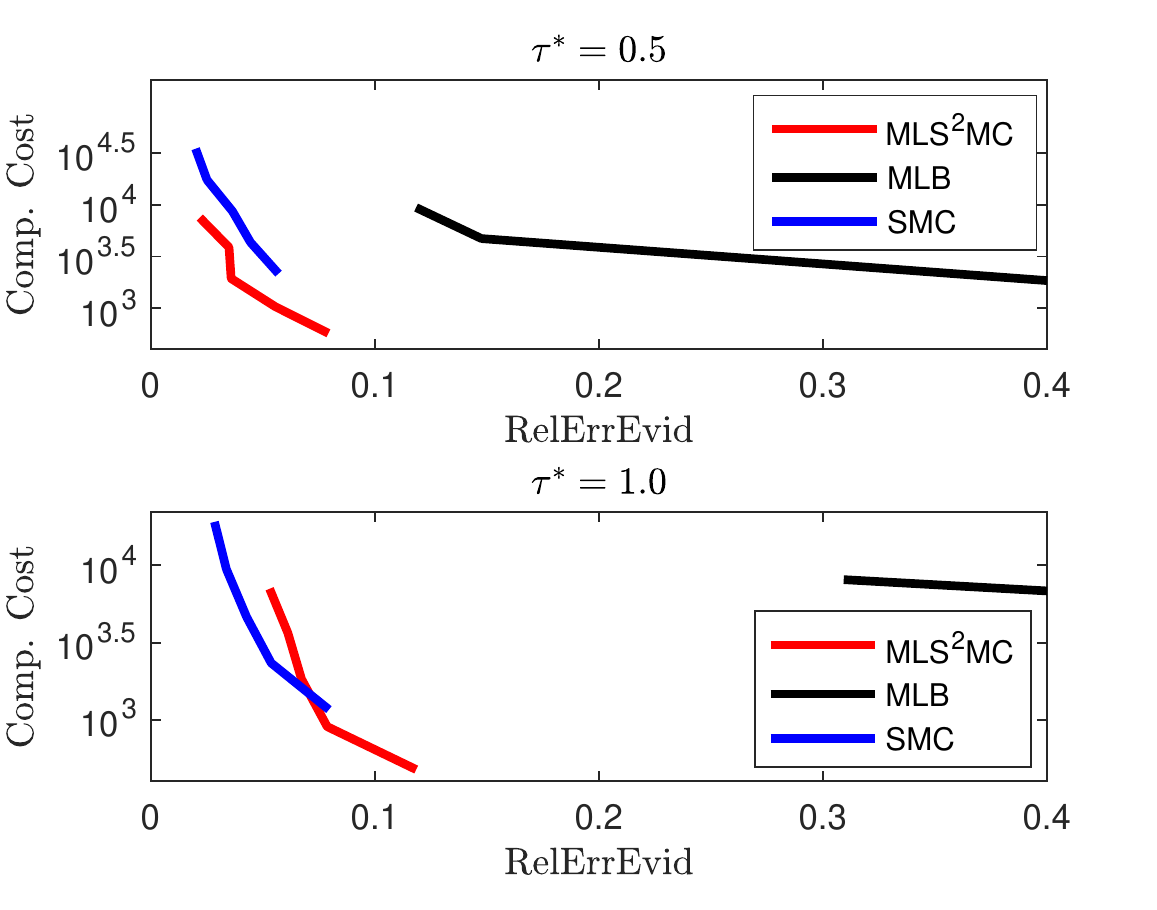}}
\subfloat[Example \ref{Example1b}]{\includegraphics[width=0.4\textwidth]{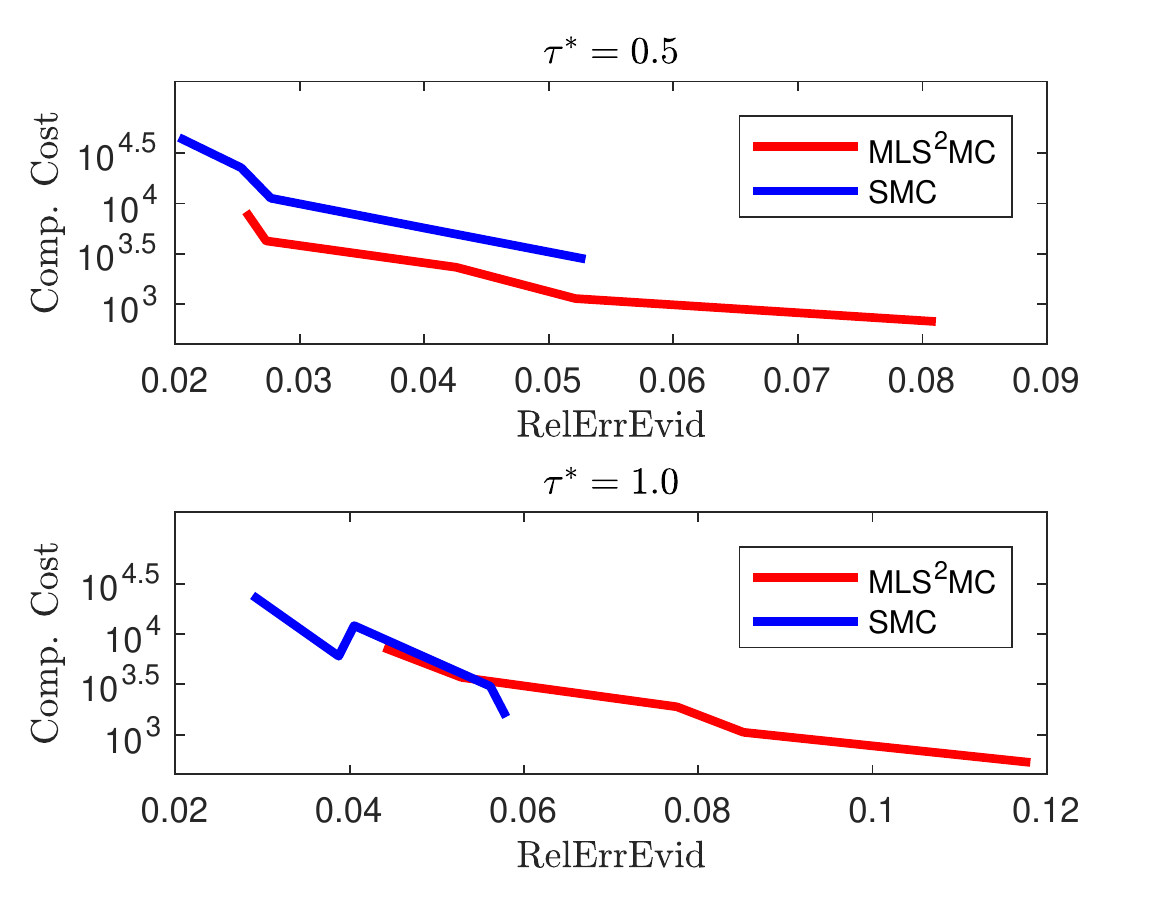}}
\caption{Comparison of the computational cost and accuracy of MLS$^2$MC, SMC and MLB for $\tau^{\ast} \in \{0.5, 1.0\}$. 
The different levels of accuracy are associated with different numbers of samples $J$. 
This combines Figures \ref{Figure_Relerror_Evidence} and \ref{Figure_comput_cost}.}
\label{Figure_cost_vs_accuracy}
\end{figure}

\subsection{Flow Cell}

Now we consider Example~\ref{Example2}.
We are particularly interested in the performance of MLS${}^2$MC in high dimensions. 
We compare only MLS${}^2$MC and single-level SMC since the adaptive update scheme in MLS${}^2$MC delivers the same sequence of intermediate probability  measures as MLB.
In addition, we also choose the maximal discretisation level adaptively within MLS${}^2$MC.
See \S \ref{Final level updates} for a discussion. 
Note that we use 16 rather than 8 finite elements in each spatial direction on the coarsest level. 

\subsubsection{Posterior approximation in high dimensions}
We present the posterior mean estimates and the true underlying parameter in Figure~\ref{Figure_Flow_Randomfield}.
We see that the estimation results are visually not as informative as the previous examples. 
Indeed, one can only recognize the coarse-scale structure of the true parameter.  
Recall that in \S\ref{Subsubsec:Approximating_Post_measure}, we considered the three leading KL terms.
In this example however, the three leading KL terms capture only about 8\% of the prior variance.
Informative results would require the consideration of a large number of marginal distributions.
However, since this is not illustrative for the reader we consider the random field at two fixed points in the spatial domain;
these points are $x^{(1)} = (0.5, 0.5)$ and $x^{(2)} = (0.75, 0.25)$. 

Before looking at the KS distances of the distributions of $\theta_{N_{\mathrm{sto}}}(x^{(1)})$ and $\theta_{N_{\mathrm{sto}}}(x^{(2)})$ we assess their posterior mean estimates.
The relative error of the posterior means in these points compared to the true values $\theta_{{\mathrm{true}}}(x^{(1)})$ and $\theta_{{\mathrm{true}}}(x^{(2)})$ is given in Figure \ref{Figure_RelErr_twopoints_flow}. 
While the estimate of $\theta_{N_{\mathrm{sto}}}(x^{(1)})$ is quite accurate, the estimate of  $\theta_{N_{\mathrm{sto}}}(x^{(2)})$ is very inaccurate -- consistently in both methods.
This is consistent with the plots of the posterior means in Figure \ref{Figure_Flow_Randomfield}.

Next we consider the relative misfit defined in \eqref{RelMisfit}.
We plot this error metric in Figure~\ref{Figure_RelMisfit_flow}.   
Even though the parameters are approximated quite poorly the relative misfit is fairly small. 
Hence, the data might be not sufficient to identify the underlying parameter more precisely.

\begin{figure}[htb]
\centering
\includegraphics[width=0.7\textwidth]{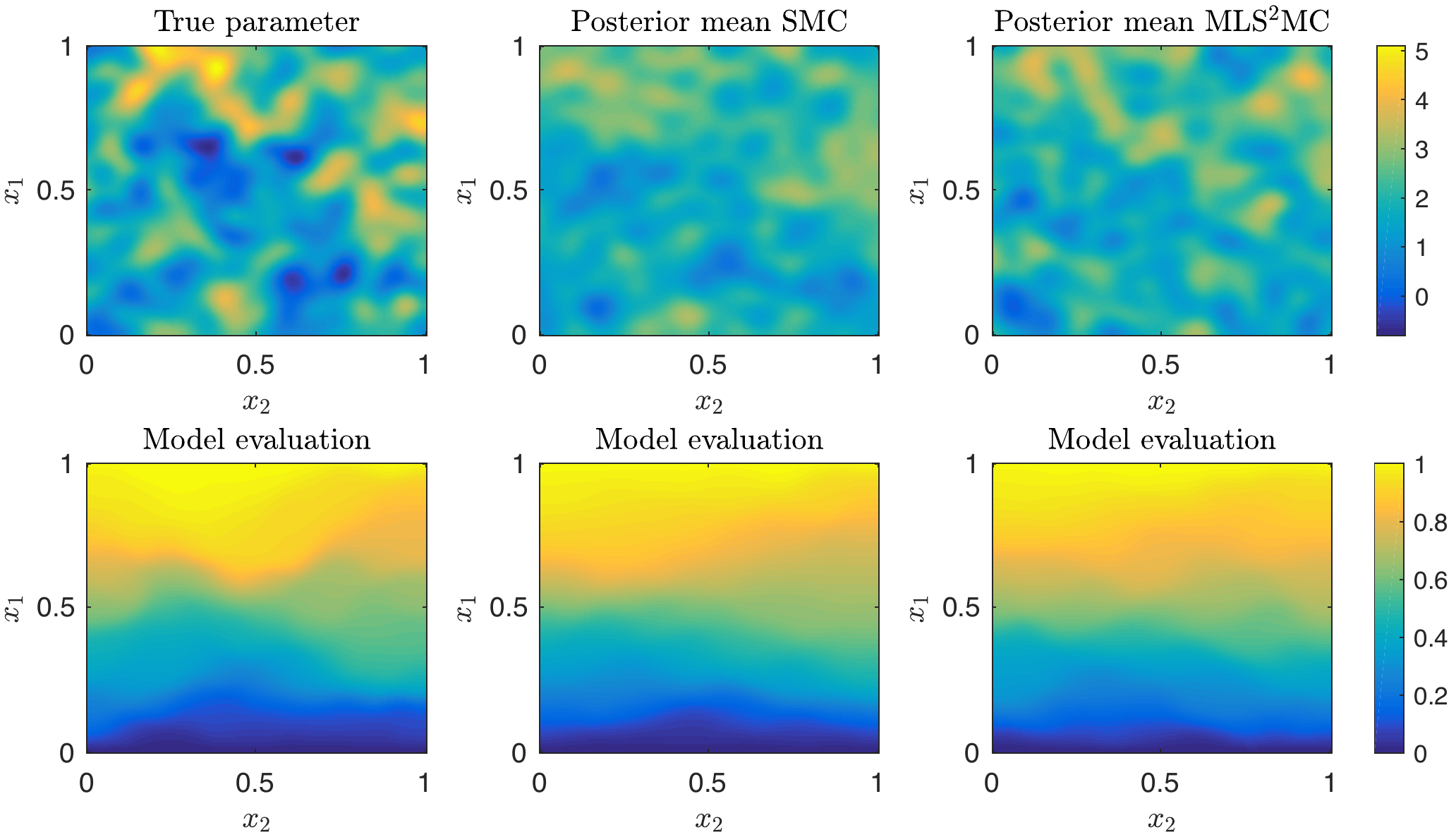}
\caption{Top row: True underlying parameter (left) and posterior mean estimates of SMC (center) and MLS${}^2$MC (right) in Example~\ref{Example2}. 
The estimations are based on $J=1000$ particles. 
Bottom row: Hydrostatic pressure corresponding to the log-permeability in the top row.} 
\label{Figure_Flow_Randomfield}
\end{figure}
  
 \begin{figure}
 \centering
 \includegraphics[width=0.6\textwidth]{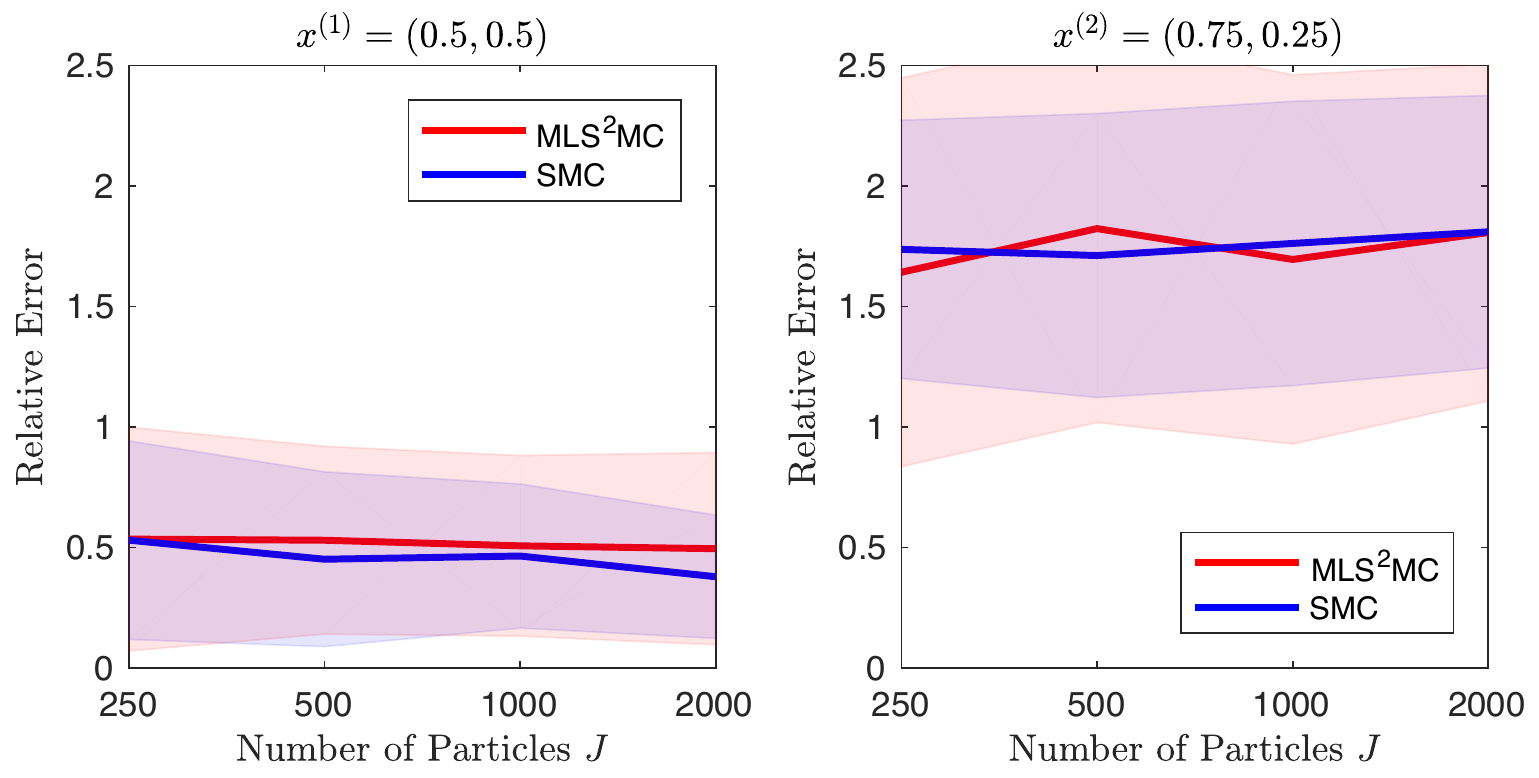}
 \caption{Relative error of posterior mean estimates compared to the true parameter in $x^{(1)}$ (left) and $x^{(2)}$(right) in Example~\ref{Example2}. 
 The bold lines show the sample mean of the error taken over 50 experiments.
The shaded areas show the associated standard deviation, again taken over 50 runs.} 
 \label{Figure_RelErr_twopoints_flow}
 \end{figure}
 
\begin{figure}
 \centering
 \includegraphics[width=0.3\textwidth]{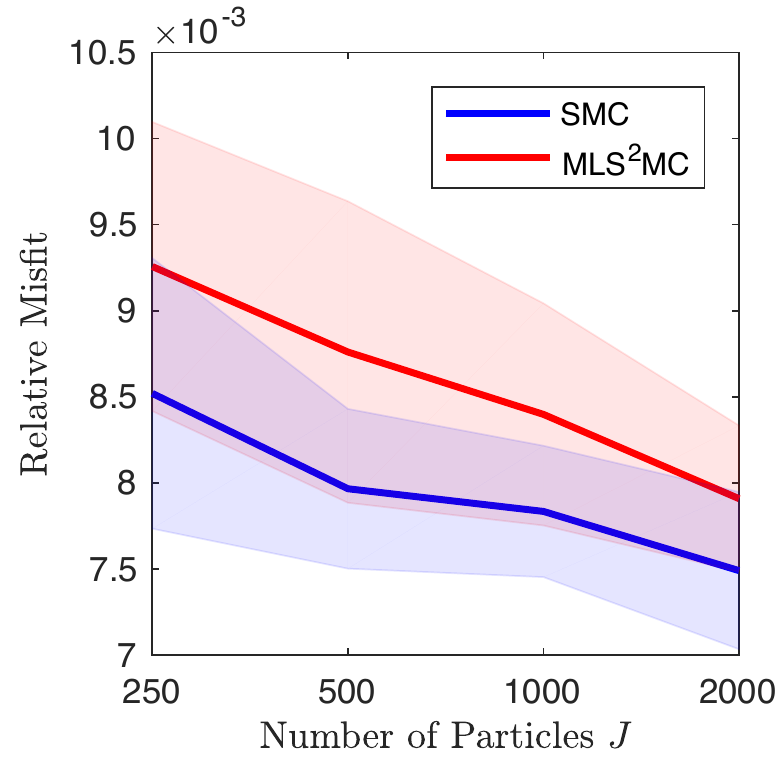}
 \caption{Relative misfit of the posterior mean estimates compared to the observations in Example~\ref{Example2}.  
 The bold lines show the sample mean of the error taken over 50 experiments.
The shaded areas show the associated standard deviation, again taken over 50 runs.} 
 \label{Figure_RelMisfit_flow}
 \end{figure}
 
We now move on to assess the approximation accuracy of the posterior measures. 
To this end we consider again the random variables  $\theta_{N_{\mathrm{sto}}}(x^{(1)})$ and $\theta_{N_{\mathrm{sto}}}(x^{(2)})$. 
We compute the KS distances of their posterior measures as discussed in \S\ref{Subsubsec:Approximating_Post_measure}. 
That is, we compare 50 MLS${}^2$MC approximations with 50 SMC approximations, using the identical number of particles. 
To obtain a base value for the KS distance we again compare also the SMC approximations to one another.
The results are presented in Figure~\ref{Figure_Flow_KS_distances}.
As in Examples \ref{Example1} and \ref{Example1b} we see that MLS${}^2$MC approximates the SMC reference solution very well. 

\begin{figure}
 \centering
 \includegraphics[width=0.6\textwidth]{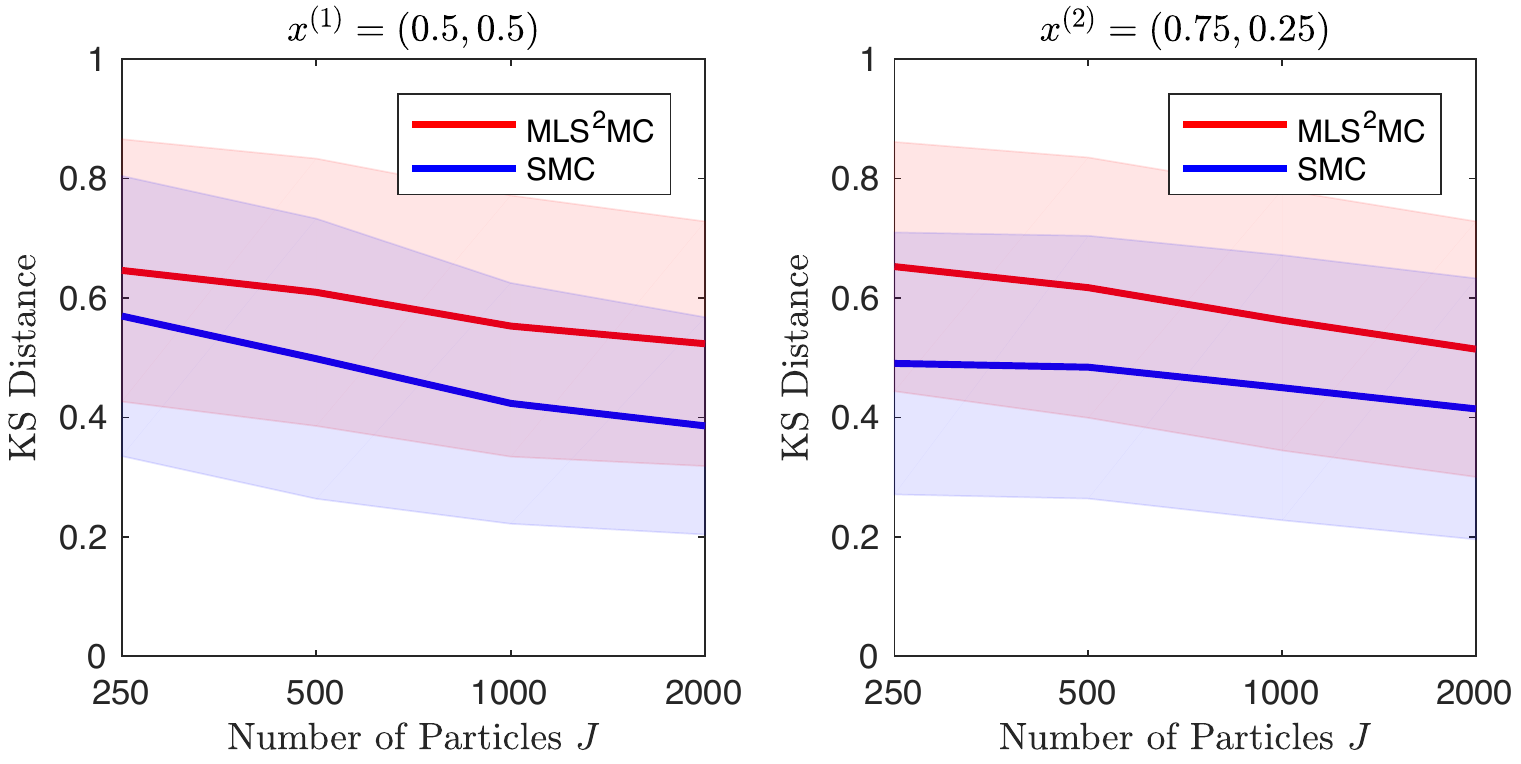}
 \caption{KS distances of the marginal posterior distributions of $\theta_{N_{\mathrm{sto}}}(x^{(1)})$ (left) and $\theta_{N_{\mathrm{sto}}}(x^{(2)})$ (right) in Example \ref{Example2}. 
 We compare the MLS${}^2$MC approximation with the SMC approximations and also the SMC approximations to one another.
 The bold lines show the sample means of KS distances of $50 \cdot 50$ combinations of SMC and either MLS${}^2$MC, or SMC.
The shaded areas show the associated standard deviations.} 
 \label{Figure_Flow_KS_distances}
 \end{figure}
 
\subsubsection{Adaptive Update Scheme}
 We present again some representative update schemes in Figure~\ref{Figure_Flow_update Schemes}. 
 We see that MLS${}^2$MC chooses the same updates as MLB. 
 This can be justified as follows: 
 First of all, we started with a finer PDE discretisation on the initial level. 
 Hence, the Bridging with large inverse temperatures should be genuinely easier. 
 Moreover, the noise level in this Example~\ref{Example2} is not as small as in Example~\ref{Example1b}. 
 In such a setting, MLB is optimal. 
 
Recall that the maximal discretisation level is chosen adaptively. 
The samplers using $J \in \{250, 500, 1000\}$ particles stop on level 4, whereas the samplers using $J = 2000$ particles continue to level 5. 
Hence it might not be possible to capture the  difference between the discretisations $\mathcal{G}_{h_4}$ and  $\mathcal{G}_{h_5}$ using a small number of particles.
In Figure~\ref{Figure_Flow_KS_distances} we do not see a significant difference between the MLS${}^2$MC approximations using $J \in \{250, 500, 1000\}$ particles and the respective SMC approximations.  
This might be surprising, since the posterior approximations are based on different PDE discretisations. 
However,  SMC also uses  $J \in \{250, 500, 1000\}$ particles for its approximation.  
If the $J$ particles were not able to capture the difference  between the models  $\mathcal{G}_{h_4}$ and  $\mathcal{G}_{h_5}$ in MLS${}^2$MC, this should also be the case in SMC. 
Hence, by using the adaptive update scheme, we can reduce the final discretisation level without losing accuracy.

\begin{figure}
\centering
 \includegraphics[width=0.7\textwidth]{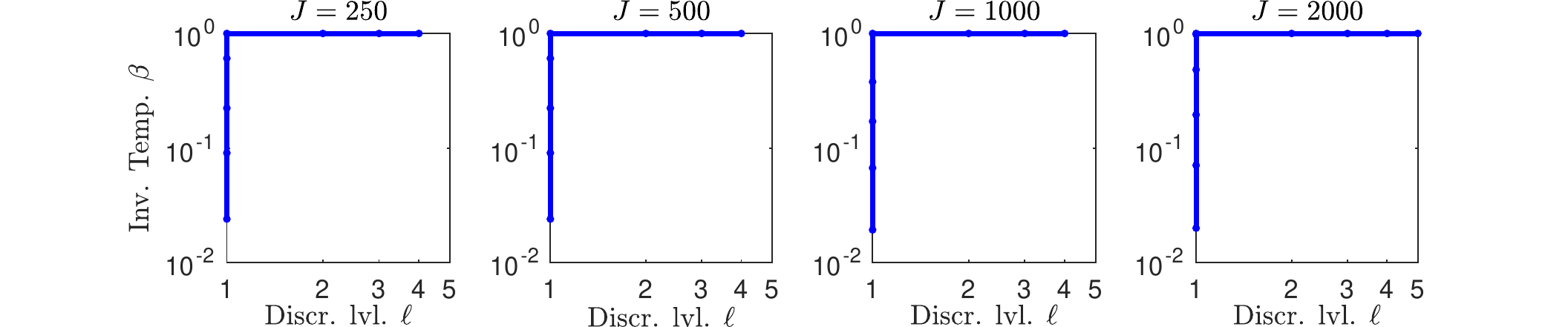}
 \caption{Realisations of the adaptive update scheme \eqref{eq_updatescheme_final_Lmax} in MLS${}^2$MC applied to Example~\ref{Example2}. 
 Each dot represents one intermediate probability measure.} 
 \label{Figure_Flow_update Schemes}
\end{figure}

\subsubsection{Computational Cost}
We give the computational cost again in terms of number of PDE evaluations with their respective theoretical number of floating point operations. 
Furthermore, we normalize $\mathcal{C}_4 = 1 $ to be consistent with Examples~\ref{Example1} and \ref{Example1b}.
Hence, $\mathcal{C}_\ell = 2^{2(4-\ell)}$.  
We present the cost of the simulations in Figure~\ref{Figure_Flow_cost}. 
We observe a speed-up of a factor 4 compared to single-level SMC, considering the number of particles.
This is similar to the results in Example~\ref{Example1} and \ref{Example1b}. 
Furthermore, in this figure we see a kink at $J = 1000$ in the graph representing the MLS${}^2$MC method. 
This corresponds to a disproportional increment in logarithmic computational cost we observe when using  $J=2000$ particles. 
It is caused by  the larger maximal discretisation level our algorithm chooses adaptively.
\begin{figure}
\centering
 \includegraphics[width=0.48\textwidth]{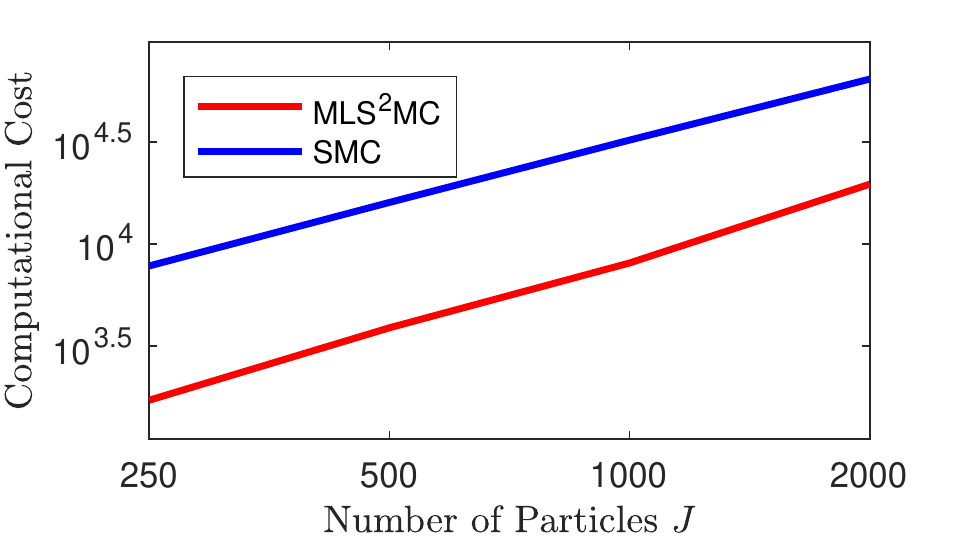} \includegraphics[width=0.37\textwidth]{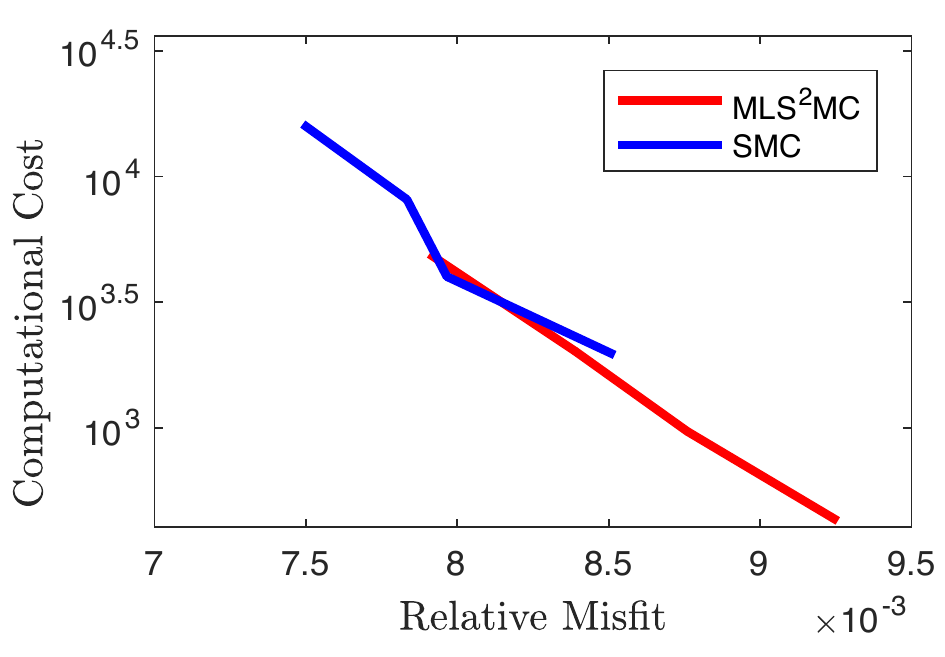}
 \caption{Cost of SMC and MLS${}^2$MC evaluations in Example~\ref{Example2}. 
 Each of the bold lines represents the mean computational cost throughout 50 simulations. 
 The y-axes represents the costs in terms of the theoretical number of floating point operations per PDE solve on the given discretisation level. 
 These costs are normalised such that $\mathcal{C}_{{N_{\mathrm{L}}}} = 1$.
 The x-axes represents either the number of particles $J$ (left) or the relative misfits that are also given in Figure \ref{Figure_RelMisfit_flow} (right).}
 \label{Figure_Flow_cost}
\end{figure}
In Figure \ref{Figure_comput_cost}, we also compare computational cost and accuracy of the posterior mean estimates {in terms of the relative misfit.}
We see that MLS$^2$MC is less accurate than SMC.
This is consistent with the numerical results in Examples \ref{Example1} and \ref{Example1b}; see \S \ref{Sec:Compu_Cost_Ex1.1b}.
There we have noticed that the large $\tau^\ast = 1.0$ leads to a large stochastic error in MLS$^2$MC, but not in SMC.
We expect that this problem can be solved by choosing a small $\tau^\ast$.

\section{Conclusion and outlook}\label{sec:concl}
We introduce a novel Sequential Monte Carlo method to approximate a posterior measure, the solution of a Bayesian inverse problem.
The posterior measure is associated with the solution of a discretised PDE, and thus every Monte Carlo sample is expensive.
We suggest an efficient, adaptive SMC sampler termed MLS$^2$MC.
The new sampler combines tempering on a fixed PDE discretisation as in single-level SMC, and a bridging scheme to transfer samples from coarse to fine discretisation levels. 
MLS$^2$MC is based on a heuristic choice between tempering and bridging, {   and does not require parameter tuning}. 
It can be used consistently with black box models, and also in the small noise limit.

MLS$^2$MC is a generalisation of multilevel bridging introduced by Koutsourelakis in \cite{Koutsourelakis2009AParameters}.
Numerical experiments show that MLS$^2$MC is as accurate as single-level SMC with tempering, and more accurate compared to multilevel bridging.
Both MLS$^2$MC and multilevel bridging are four times cheaper than the associated single-level SMC sampler for PDE problems in 2D space.
In some situations our adaptive choice between tempering and bridging recovers the multilevel bridging algorithm.

MLS$^2$MC is a particle filter which is known to perform well in high-dimensional parameter spaces.
We confirm this in numerical experiments where we work in parameter spaces of dimension up to 320.
Moreover, a by-product of SMC samplers is an estimate for the model evidence which is important in Bayesian Model Selection. 
The model evidence estimates of MLS$^2$MC are as accurate as those delivered by the associated single-level SMC sampler.

In future works we plan to analyze the convergence of MLS$^2$MC and give mathematical arguments for its efficiency in terms of accuracy and computational cost. 
We will also combine MLS$^2$MC with approximate particle filters, e.g. the Ensemble Kalman Filter.
Moreover, we are interested in applying MLS$^2$MC to real-world problems and time-dependent settings. 
Importantly, MLS$^2$MC can handle black box models, and does not rely on model hierarchies built from finite element meshes.
Thus it could be used with models of different fidelities where the fidelity is not associated with the mesh size.

Alternatively, MLS$^2$MC could be used to estimate expected values of quantities of interest with respect to the posterior measure.
To this end, it could be combined with multilevel SMC within the final level updates.
Multilevel SMC relies on variance reduction, and it would be interesting to study the possible reduction of the number of particles associated with fine discretisation levels.

\section*{Acknowledgements}

This work was supported by Deutsche Forschungsgemeinschaft (DFG) through the TUM International Graduate School of Science and Engineering (IGSSE) within the project 10.02 BAYES. 
The numerical experiments were performed on the Linux clusters of the Leibniz Rechenzentrum at the Bayerische Akademie der Wissenschaften.

{ \small
  \bibliographystyle{plain} 
 \bibliography{Mendeley}
}
\end{document}